\documentclass{svjour3}                     
\smartqed  
\usepackage{graphicx}
\usepackage{mathptmx}      
\usepackage{latexsym}
\journalname{Astronomy \& Astrophysics Review}
\begin{document}

\title{Astronomy in Antarctica
}

\titlerunning{Astronomy in Antarctica}        

\author{Michael G. Burton}


\institute{M.G. Burton \at
              School of Physics, University of New South Wales, Sydney, NSW 2052, Australia \\
              Tel.: +61-2-9385-4553\\
              Fax: +61-2-9385-6060\\
              \email{m.burton@unsw.edu.au}           
}

\date{Received: 01/05/10 / Accepted: 10/06/10}

\maketitle

\begin{abstract}
  Antarctica provides a unique environment for astronomers to
  practice their trade.  The cold, dry and stable air found above the
  high Antarctic plateau, as well as the pure ice below, offers new
  opportunities for the conduct of observational astronomy across both
  the photon and the particle spectrum.  The summits of the Antarctic
  plateau provide the best seeing conditions, the darkest skies and
  the most transparent atmosphere of any earth-based observing site.
  Astronomical activities are now underway at four plateau sites: the
  Amundsen-Scott South Pole Station, Concordia Station at Dome C,
  Kunlun Station at Dome A and Fuji Station at Dome F, in addition to
  long duration ballooning from the coastal station of McMurdo, at
  stations run by the USA, France / Italy, China, Japan and the USA,
  respectively.  The astronomy conducted from Antarctica includes
  optical, infrared, terahertz and sub-millimetre astronomy,
  measurements of cosmic microwave background anisotropies, solar
  astronomy, as well as high energy astrophysics involving the
  measurement of cosmic rays, gamma rays and neutrinos.  Antarctica is
  also the richest source of meteorites on our planet.

  An extensive range of site testing measurements have been made over
  the high plateau sites.  In this paper we summarise the facets of
  Antarctica that are driving developments in astronomy there, and
  review the results of the site testing experiments undertaken to
  quantify those characteristics of the Antarctic plateau relevant for
  astronomical observation.  We also outline the historical
  development of the astronomy on the continent, and then review the
  principal scientific results to have emerged over the past three
  decades of activity in the discipline.  These range from
  determination of the dominant frequencies of the 5 minute solar
  oscillation in 1979 to the highest angular scale measurements yet
  made of the power spectrum of the CMBR anisotropies in 2010.  They
  span through infrared views of the galactic ecology in star
  formation complexes in 1999, the first clear demonstration that the
  Universe was flat in 2000, the first detection of polarization in
  the CMBR in 2002, the mapping of the warm molecular gas across the
  $\sim 300$\,pc extent of the Central Molecular Zone of our Galaxy in
  2003, the measurement of cosmic neutrinos in 2005, and imaging of
  the thermal Sunyaev Zel'dovich effect in galaxy clusters in 2008.

  This review also discusses how science is conducted in Antarctica,
  and in particular the difficulties, as well as the advantages, faced
  by astronomers seeking to bring their experiments there.  It also
  reviews some of the political issues that will be encountered, both
  at national and international level.  Finally, the review discusses
  where Antarctic astronomy may be heading in the coming decade, in
  particular plans for infrared and terahertz astronomy, including the
  new facilities being considered for these wavebands at the high
  plateau stations.


\end{abstract}


\section{Why Astronomy in Antarctica?}
\label{sec:why}
\subsection{The Antarctic Continent and its high ice plateau}
The Antarctic continent is the highest, driest and coldest of the
continents on the Earth.  It is the end of the Earth, literally as
well metaphorically.  As with all endeavours in Antarctica, it was the
last continent where humans began to conduct astronomical observations
from.  The first astronomical discovery was made less than a century
ago, and it has only been in the past two decades that major ventures
in the discipline have taken place. Yet, with current technology,
astronomy is no longer difficult to undertake in Antarctica, given
appropriate resources and fore-planning.  Moreover, the potential
Antarctica offers for furthering a wide and diverse range of frontier
investigations in astronomy is unmatched, in comparison with any other
location on our planet.

Antarctica is the fifth largest continent, with a land area of 14
square million kilometres.  The amount of exposed land is, however,
tiny, about 2\% of the total and confined almost entirely to the
coastal fringe.  As a continent, Antarctica is dominated by ice, with
the area covered virtually doubling between the summer and winter
extremes; the ice sheets extend up to one thousand kilometres over the
Southern Ocean from the coast at their September peak.  From an
astronomer's perspective it is the ice mass of the Antarctic plateau
that draws the attention.  For, while the continent is crossed by one of
the world's great mountain ranges, the Trans Antarctic Mountains that
stretch nearly 5,000\,km from the Weddell Sea to the Ross Sea, all but
its highest peaks (the nunataks) are obscured from view, buried
under the ice sheet that makes up the Antarctic plateau.

The land itself lies under up to four kilometres of ice, with the ice
surface very gradually rising from the coast, over a distance of
several hundred kilometres, to reach over 4,000\,m at Dome A (though
the peaks of some of the mountains rise higher than this, for instance
the Vinson Massif in West Antarctica is 4,897\,m high).  The area of
ice over 3,000\,m elevation is almost as large as the continent of
Australia. This is the Antarctic plateau.  Its great extent makes
Antarctica the highest continent, as measured by average elevation.
It also contains the coldest and driest regions of our planet. A
temperature of $-90^{\circ}$C was once measured at the Russian Vostok
station, the lowest ever recorded, and winters average $-60^{\circ}$C
over the plateau.  Typical precipitable water vapour levels of
250$\mu$m exist for much of the year and levels can fall below
100$\mu$m in places at times, the driest air on Earth.

Of crucial importance for the conduct of astronomy is that there is
little wind on top of the plateau.  The Antarctic atmospheric
circulation pattern centres about the South Pole, and the dominant
airflow is a slow settling from the stratosphere, to feed a steady,
downward-flowing wind off the plateau.  This wind is katabatic in
origin, starting from the highest points and picking up speed as it
falls towards the coast, under gravity.  With an average slope of
about one tenth of a degree, the wind is also gentle.  Over the
highest parts of the plateau typical wind speeds are only 1-2\,m/s.
Though, as it nears the coast and the gradient increases, the wind
speed picks up -- and can lead to the ferocious storms that are a part
of Antarctic folklore.

The debilitating effects of the atmosphere are largely confined to a
thin turbulent surface boundary layer.  In winter this can be only a
few metres thick over the summits of the plateau.  Above it, the
seeing is exceptionally low.  So too is the level of scintillation
noise, and both the isoplanatic angle and coherence time are large, in
comparison with the best temperate latitude sites.  These facets
provide conditions than can yield extraordinary image clarity and
stability for a wide range astronomical observations.  As these
conditions have been quantified over the past decade, and their
implications understood, they have drawn the interest of astronomers
to the continent for its potential to provide superlative observatory
sites.

\begin{figure*}
\begin{center}
\includegraphics[width=1.0\textwidth]{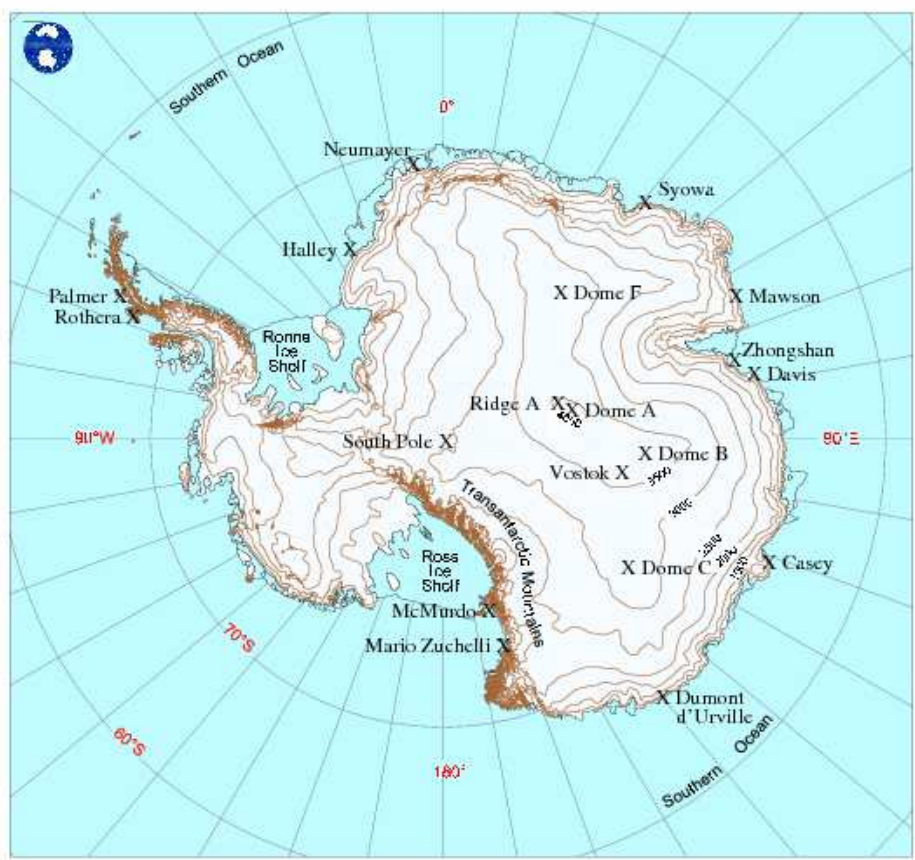}
\end{center}
\caption{Topographic map of Antarctica, with the location of the
  principal research stations discussed in the text indicated.  The
  high Antarctic plateau runs along the ridge from Dome F to Dome C,
  through Dome A, Dome B and Vostok.  Ridge A lies 400\,km SW of Dome
  A\@. The South Pole lies on the flank of the Antarctic plateau.
  Coastal stations supporting high plateau operations are also marked;
  McMurdo (USA), Mario Zuchelli (Italy), Dumont d'Urville (France),
  Zhongshan (China) and Syowa (Japan).  In addition, the locations of
  major coastal stations at Casey, Davis \& Mawson (Australia), Halley
  \& Rothera (UK), Palmer (USA) and Neumayer (Germany) are shown.  Map
  adapted from a figure supplied by the Australian Antarctic Division,
  with acknowledgment.}
\label{fig:map}
\end{figure*}

\subsection{This Review}
\label{sec:review}
This paper provides an overview of the subject of astronomy in
Antarctica.  We describe the features of the continent of special
interest to astronomers, and the sites under consideration for
observatories.  We also summarise the principal results obtained from
the extensive program of site testing of the Antarctic plateau that
has taken place over the past two decades.  A brief history of the
development of astronomy in Antarctica is given, followed by examples
of some of the science obtained.  As will be apparent, this includes a
diverse range of fields as well as techniques. A section discusses how
science is conducted in Antarctica today, and what the principal
difficulties are for the investigator, in comparison to undertaking
temperate-latitude astronomy.  The review finishes with some personal
thoughts of where astronomy will develop on the continent over the
coming years.  Many sources have been drawn upon for this review, and
in particular it extends the earlier reviews of Storey (2005), Burton
(2005) and Storey (2009).  Several books have been devoted entirely to
Astronomy in Antarctica, the proceedings of conferences held on the
subject.  They include the proceedings of the American Institute of
Physics conference on Astronomy in Antarctica in Newark, USA in 1989
(Pomerantz 1990), the Astronomical Society of the Pacific symposium on
Astronomy in Antarctica in Chicago, USA in 1997 (Novak \& Landsberg,
1998), the Concordia station workshop in Capri, Italy in 2003 (Fossat
\& Candidi 2003) and the three European ARENA conferences, held in
Roscoff, France in 2006 (Epchtein \& Candidi 2007), Potsdam, Germany
in 2007 (Zinnecker, Epchtein \& Rauer 2008) and Frascati, Italy in
2009 (Spinoglio \& Epchtein 2010).

\subsection{Sites for Astronomy in Antarctica}
\label{sec:sites}
\begin{table}
\caption{Astronomical Sites in Antarctica}
\label{table:sites}       

\begin{tabular}{lccclcll}
\hline\noalign{\smallskip}
Location & Elevation & Latitude & Longitude & National & Established & Astronomical Applications & Comment \\
& \multicolumn{1}{c}{m} & \multicolumn{1}{c}{$^{\circ}$} & \multicolumn{1}{c}{$^{\circ}$} & Operator && 
of Particular Interest \\
\noalign{\smallskip}\hline\noalign{\smallskip}
South Pole & 2,835 & -90 & $\ldots$ & USA & 1957 & CMBR, Neutrinos, Sub-mm & Amundsen-Scott station. \\
&&&&&&& On flank of Antarctic plateau. \\
Dome A & 4,083 & -80 & +78 & China & 2009 & Optical, THz, IR & First visited 2005. Construction  \\
&&&&&& Time-series & of Kunlun station started 2009. \\
Dome B & 3,809 & -76 & +95 & $\ldots$ & $\ldots$ & THz, IR & Not yet visited by humans. \\
& \\
Dome C & 3,268 & -75 & +123 & France & 2005 & Optical, Sub-mm, IR, Solar & Concordia station. \\
&&&& \& Italy && Time-series, Interferometry \\
Dome F & 3,810 & -77 & +39 & Japan & 1995 & THz, IR & Fuji station, wintered (ice cores). \\
&&&&&&& Site testing now commencing. \\
Vostok & 3,488 & -78 & +107 & Russia & 1957 & Sub-mm, IR & No site quantification as yet. \\
& \\
Ridge A & 4,053 & -82 & +74 & $\ldots$ & $\ldots$ & THz, IR & Possibly `best' site? \\
&&&&&&& Not yet visited by humans. \\
McMurdo & 0 & -78 & +167 & USA & 1956 & Long Duration Ballooning & Coastal station hosting  \\
&&&&&& (CMBR, THz, sub-mm, Solar, & the LDBF. \\
&&&&&& neutrinos), Cosmic Rays \\
Mawson & 0 & -68 & +63 & Australia & 1954 & Cosmic Rays & Coastal station. \\
\noalign{\smallskip}\hline
\end{tabular}

This table summarises the sites in Antarctica where astronomy is
conducted or is under consideration, including the applications of
particular interest at individual sites.
\end{table}

\begin{figure*}
\begin{center}
\includegraphics[width=1.0\textwidth]{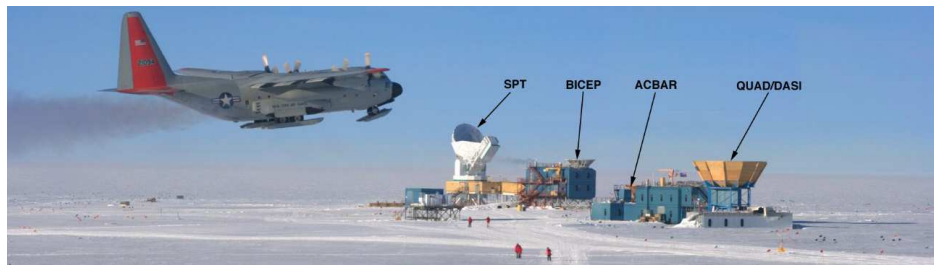}
\end{center}
\caption{The Martin A Pomerantz Observatory (MAPO) in the ``Dark
  Sector'' of the US Amundsen-Scott South Pole Station.  From left to
  right, the telescopes are the 10\,m South Pole Telescope (SPT), the
  25\,cm BICEP, the 2.1\,m Viper with ACBAR and the Degree Angular
  Scale Interferometer (DASI) with QUaD\@. \S\ref{sec:cmbr} provides
  further information on their capabilities. A ski-equipped LC130
  aircraft, as used for supplying the station with all personnel and
  cargo, takes off from the ski-way to left. Several humans seen to
  foreground provide a sense of scale.  They are traversing across the
  ski-way from the main South Pole Station base 1\,km away (behind the
  camera).  Image courtesy of Steffan Richter.}
\label{fig:southpole}
\end{figure*}

\begin{figure*}
\begin{center}
\includegraphics[width=1.0\textwidth]{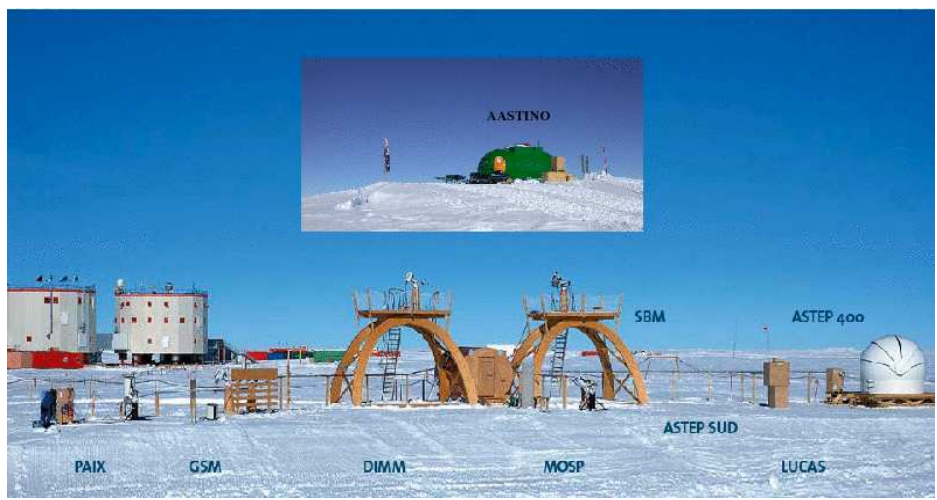}
\end{center}
\caption{Panoramic view of the French--Italian Concordia Station at
  Dome C, seen from the cluster of experiments that make up the
  Concordiastro site characterisation program.  These are labelled
  from left to right: PAIX (photometric extinction), GSM (seeing),
  DIMM (seeing), MOSP (turbulence outer scale), SBM (sky brightness),
  ASTEP (photometry) Sud, ASTEP 400 and LUCAS (spectra of earth-shine
  off the Moon). The inset shows the AASTINO, the Australian automated
  site testing laboratory with a SODAR (sonic radar) and MASS
  (seeing).  The twin towers of the station are seen to rear, rising
  20\,m above the ice. Credit Karim Agabi (main picture) and John
  Storey (for AASTINO insert).
}
\label{fig:domec}
\end{figure*}

\begin{figure*}
\begin{center}
\includegraphics[width=1.0\textwidth]{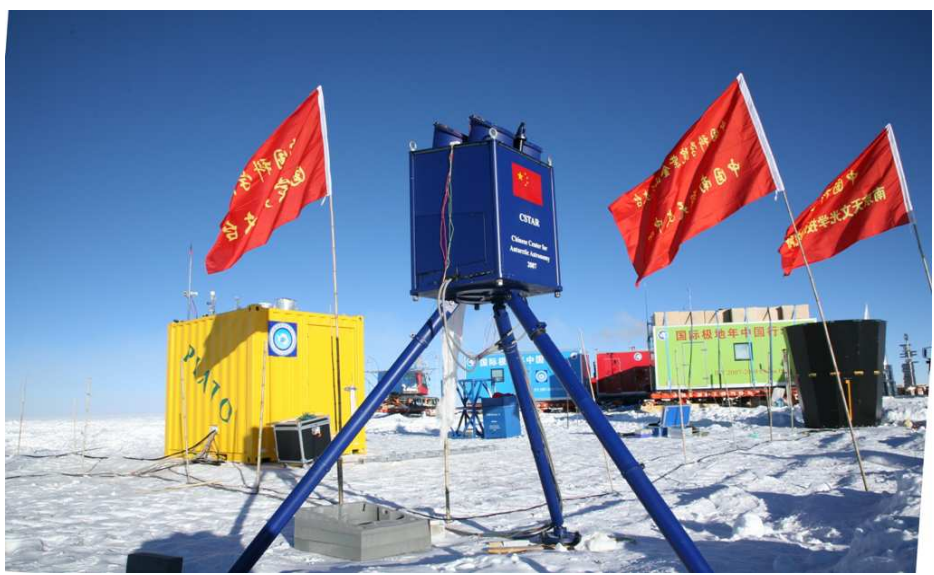}
\end{center}
\caption{The Chinese Kunlun Station at Dome A\@.  To foreground is the
  CSTAR experiment (4 fixed optical telescopes pointed at the South
  Celestial Pole). It is controlled from the instrument module of the
  PLATO autonomous laboratory (yellow cabin to left). On its roof is
  the Gattini camera for monitoring sky brightness and cloud cover.
  The black cone is the SNODAR acoustic radar.  The first modules of
  Kunlun station are to rear. Credit Zhenxi Zhu and Xu Zhou.
}
\label{fig:domea}
\end{figure*}

\begin{figure*}
\begin{center}
\includegraphics[width=1.0\textwidth]{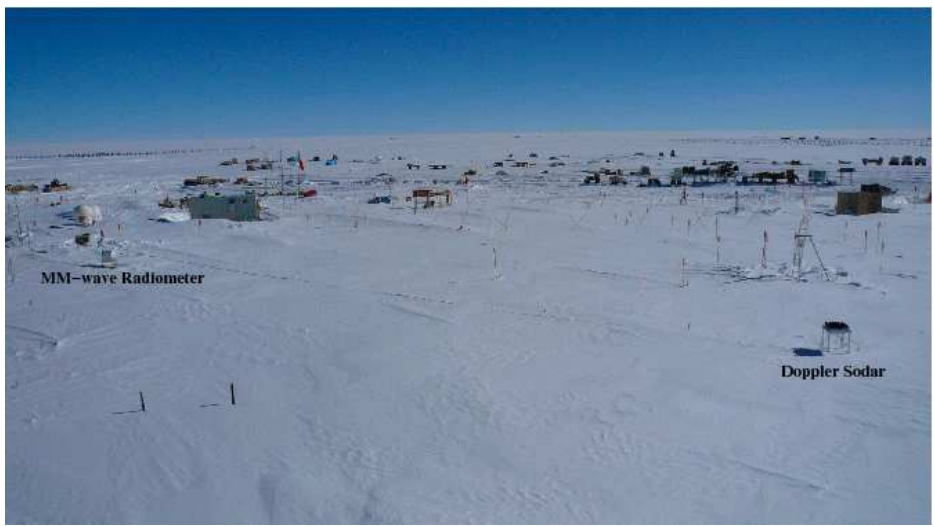}
\end{center}
\caption{The Japanese Dome Fuji station at Dome F in the 2006-07 summer, 
when the primary activity was ice core drilling. Two site testing
experiments are indicated; a mm-wave radiometer for measuring sky
transparency and a Doppler sodar for boundary layer turbulence.
Credit Hideaki Motoyama, National Institute Polar Research, Japan.}
\label{fig:domef}
\end{figure*}

\begin{figure*}
\begin{center}
\includegraphics[width=1.0\textwidth]{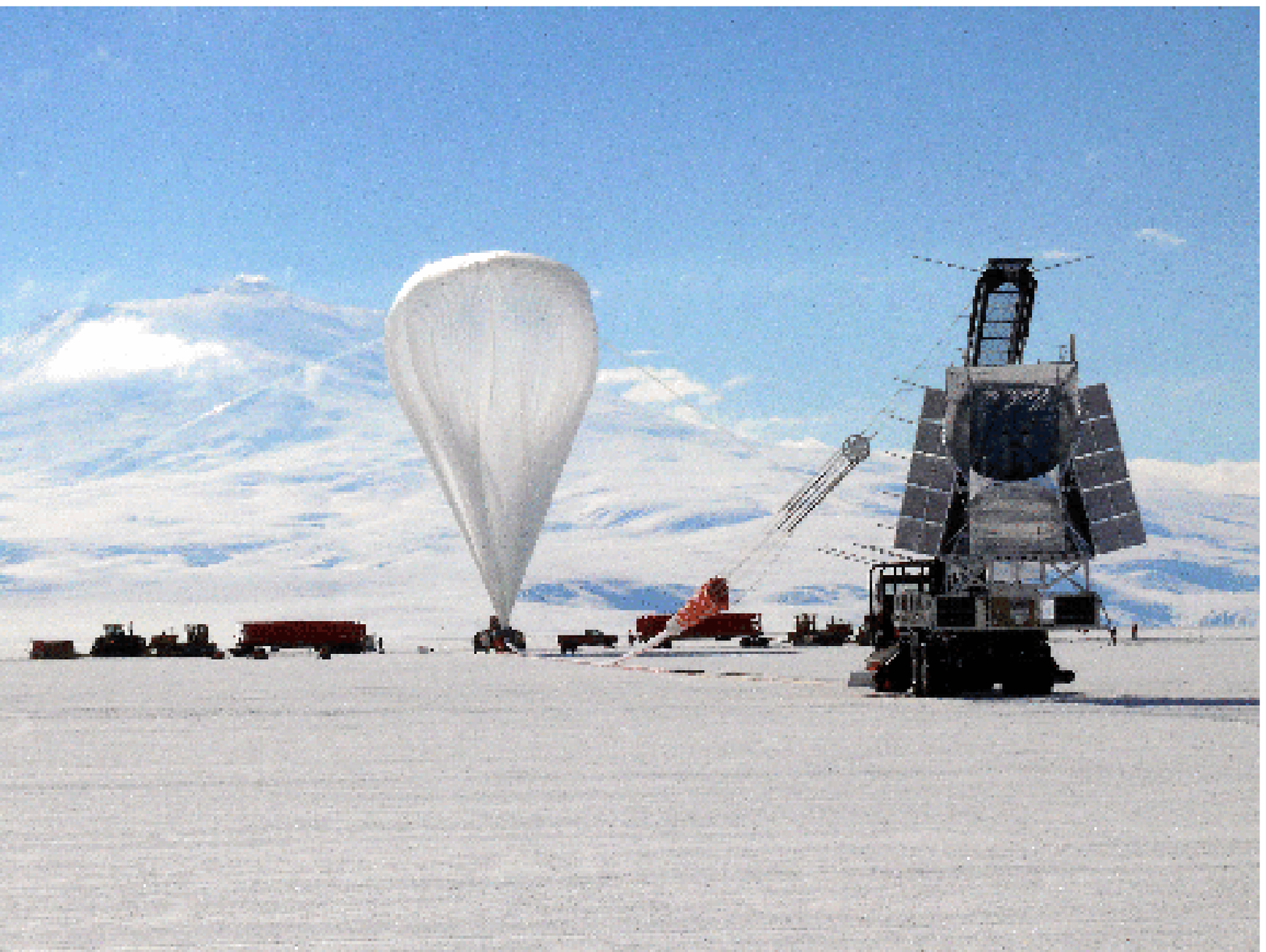}
\end{center}
\caption{Launch of the 2\,m BLAST sub-millimetre telescope from the Long 
Duration Balloon Facility (LDBF) on Williams air field at the US coastal
station of McMurdo in December 2006. The 3,800\,m volcano Mount Erebus
is in the background. Credit: Mark Halpern.}
\label{fig:mcmurdo}
\end{figure*}

The South Pole, at 2,835\,m elevation, lies on the flank of the
Antarctic plateau.  While there are better places in Antarctica for
many kinds of observation, the logistics of ready access have made the
Pole the place where most of the astronomy has so far been conducted
in Antarctica, at the US Amundsen-Scott South Pole station. Named
after the pioneers who first reached the Pole in 1911-12, the station
was established in 1957 during the International Geophysical Year
(IGY). It recently completed a major refurbishment, and supports
around 60 personnel over-wintering and $\sim 250$ in the summer
months.  Access is via ski-equipped LC130 aircraft from McMurdo
station on the coast, with daily flights over the summer period from
November to February.

Concordia station at Dome C, built by France and Italy, was opened for
winter operation in 2005 following a decade-long construction phase.
At 3,268\,m Dome C is one of the high points of the Plateau, where the
conditions are at their most stable. Extensive site quantification has
taken place over the past decade, starting from before the station
opened for winter operation. Plans for a range of facilities to
exploit the conditions there have been developed by several European
nations and by Australia.  Dome C is accessed, for cargo, by three
overland tractor traverses per season.  Each traverse lasts about two
weeks and originates from the coastal station of Dumont d'Urville
(France).  For transportation of people, Twin Otter aircraft are used,
generally starting from either Dumont d'Urville or from Mario
Zucchelli station (Italy).

The highest location on the plateau is the 4,083\,m Dome A, first
visited by humans in just 2005 by a Chinese team on a tractor traverse
from Zhongshan station.  In 2009 China began the construction of
Kunlun station at the site.  Already several astronomical experiments
have been conducted here, making use of an Australian robotic observatory.
Ambitious plans are being formulated for IR and THz telescopes within the
next decade.  Currently one traverse is conducted each summer season,
carrying all people and $\sim 570$ tonnes of equipment.  No
winter-overs have yet been conducted.

There are other high plateau sites where astronomy could be conducted.
These include the Russian station of Vostok (3,488\,m) and the
Japanese station at Dome F (3,810\,m), both of which have stations
that have been used for wintering.  Site testing is being initiated at
Dome F\@.  Another ice dome, Dome B (3,809\,m), lies on the SE end of
the ridge from Dome A, before the plateau drops towards Vostok. Ridge
A (4,053\,m), lying at the end of a ridge running SW from Dome A, has
also been proposed as a site where conditions may even be superior to
Dome A for some applications.

Not all Antarctic astronomy needs to be done on the plateau, however,
in particular cosmic-ray observations.  Facilities for such
measurements exist at the coastal stations of Mawson (Australia) and
McMurdo (USA). Neutron monitors also operate at several coastal
locations.  In addition, McMurdo houses a long-duration balloon
facility, taking advantage of the high-altitude circumpolar winds
which permit balloons to remain aloft for periods of 10-30 days at
times during summer, circling the continent as they do so.

A map showing these plateau sites, together with some of the coastal
sites associated with their logistic support, is shown in
Fig.~\ref{fig:map}. Images of the four plateau stations, together with
the McMurdo Long Duration Balloon Facility, are shown in
Figs.~\ref{fig:southpole}--\ref{fig:mcmurdo}. Table~\ref{table:sites}
summarises these sites and lists the astronomical applications
currently of interest at each. Access to the continent is by both air
and sea, and from there to the plateau by ski-equipped planes and/or
overland traverse, depending on the logistic capabilities of the
participating nations.

\section{Site Conditions for Astronomical Observations from the Antarctica Plateau}
\begin{table}
  \caption{Typical gains achievable from an ice Dome of the Antarctic plateau 
    compared to excellent temperate latitude sites (adapted from Storey,
    2009), together with a synopsis of the applications that are facilitated.}
\label{table:gains}       
\begin{tabular}{lll}
\hline\noalign{\smallskip}
Parameter & Antarctic & Astronomical Consequence \\ & Advantage & \\
\noalign{\smallskip}\hline\noalign{\smallskip}
Atmospheric Seeing & $2-3 \times$ better & Better spatial resolution \\
(above 10-40m Boundary Layer) && Better point source sensitivity\\
Isoplanatic Angle & $2-3 \times$ larger & Better Adaptive Optics sky coverage \\
& \\
Coherence Time & $2.5 \times$ longer & Increased sensitivity for \\
&& Adaptive Optics \& Interferometry \\
Scintillation Noise & $3-4 \times$ less & High dynamic range imaging\\
&& More precise photometry \\
IR sky background & $20-100 \times$ less & Increased infrared sensitivity \\
&& Improved photometric stability \\
Aerosols & up to $50 \times$ lower & Better infrared windows \\
&& Improved sky stability \\
Water Vapour & $3-5 \times$ lower & New windows from IR to sub-mm \\
&& \\
Ice & Vast quantities! & Neutrino detectors \\
&& Paleoclimate \\
\noalign{\smallskip}\hline
\end{tabular}
\end{table}

Two primary factors have driven much of the current interest in
Antarctic astronomy -- the coldest and driest conditions on Earth that
are found in the air above the Antarctic plateau.  The cold reduces
background fluxes in the infrared. Over infrared to sub-millimetre
wavelengths the dry air results in many windows in the atmosphere
opening up to observation.  A host of secondary factors provide
additional reasons for astronomers to conduct observations from
Antarctica (see also Tables~\ref{table:gains} \&
\ref{table:techniques}).  These include the stability of the
atmosphere and its thin surface boundary layer, low levels of
pollution and dust aerosols in the air, high cloud-free fractions, the
ability to conduct continuous or long-duration monitoring, increased
low-energy cosmic-ray fluxes arising from the proximity to the
magnetic pole, low levels of seismic activity and the vast quantities
of pure ice available as an absorber of particles.  Some of these
secondary factors have proven to be particularly potent.  For
instance, the ice is being used to construct a neutrino detector at
the South Pole with a cubic kilometre of collecting volume (IceCube).
The confinement of most of the turbulence in the air to a thin layer
just above the ice, over the summits of the plateau, creates
conditions that are particularly favourable for wavefront correction
of light, as well as providing telescopes with extraordinary good
seeing if they were to be raised beyond the boundary layer.

\begin{table}
\caption{Astronomical Techniques for Antarctica}
\label{table:techniques}       
\begin{tabular}{ll}
\hline\noalign{\smallskip}
Technique & Advantage \\
\noalign{\smallskip}\hline
Optical & Imaging quality ($\epsilon_0$); exceptional seeing above shallow surface boundary layer \\
Infrared & Low sky \& telescope background \\
& Sky flux stability \\
& Image quality \\
THz & Opens the window \\
Sub-millimetre & Improved windows \\
& Sky stability \\
Millimetre (CMBR) & Sky stability \\
Time-series & High duty-cycle measurements \\
& Stability \\
& Long, uninterrupted periods of darkness \\
Precision Photometry & Low scintillation noise \\
\{IR Interferometry +  & \{Improved values for $\tau_0$, $\theta_0$ \& $r_0$ and Coherence Volume 
($\theta^2 \tau_0 / \epsilon^2$)\\
\{Adaptive Optics & \{+ Temperature-stability for delay lines \\
Cosmic Rays & Low energy threshold (proximity to geomagnetic pole) \\
Neutrinos & Pure ice as an absorber \\
Balloons & Long duration flights in constant environment \\
Solar & Excellent daytime seeing \\
\noalign{\smallskip}\hline
\end{tabular}

This table summarises astronomical techniques and fields where
Antarctica can provide special advantages over other earth-based 
observatory locations, and
the nature of that advantage.
\end{table}

\subsection{Infrared Sky Background: the coldest locations on Earth}
\label{sec:ir}
\begin{figure*}
\begin{center}
\includegraphics[width=1.0\textwidth]{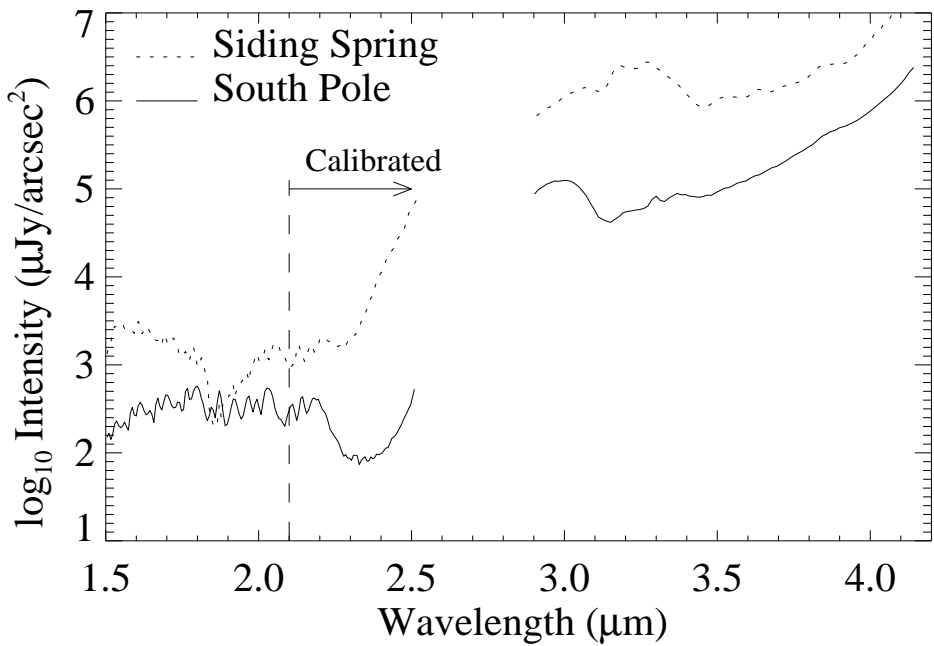}
\end{center}
\caption{The near-IR sky emission, in $\rm \mu Jy \, arcsec^{-2}$,
  above the South Pole in winter, from 1.5-4$\mu$m, as measured by the
  IRPS instrument (Phillips et al.\ 1999), and compared to a
  similar spectrum obtained from Siding Spring Observatory in
  Australia.  The sharp drop in flux between 2.27--2.45$\mu$m is the
  so-called `cosmological window'.  From 2.2$\mu$m longwards the sky
  is between 20 and 100 times darker than the temperate latitude site.
  Note that the data only have a nominal calibration factor applied
  shortward of 2.1$\mu$m.  }
\label{fig:irps}
\end{figure*}

\begin{figure*}
\begin{center}
\includegraphics[width=1.0\textwidth]{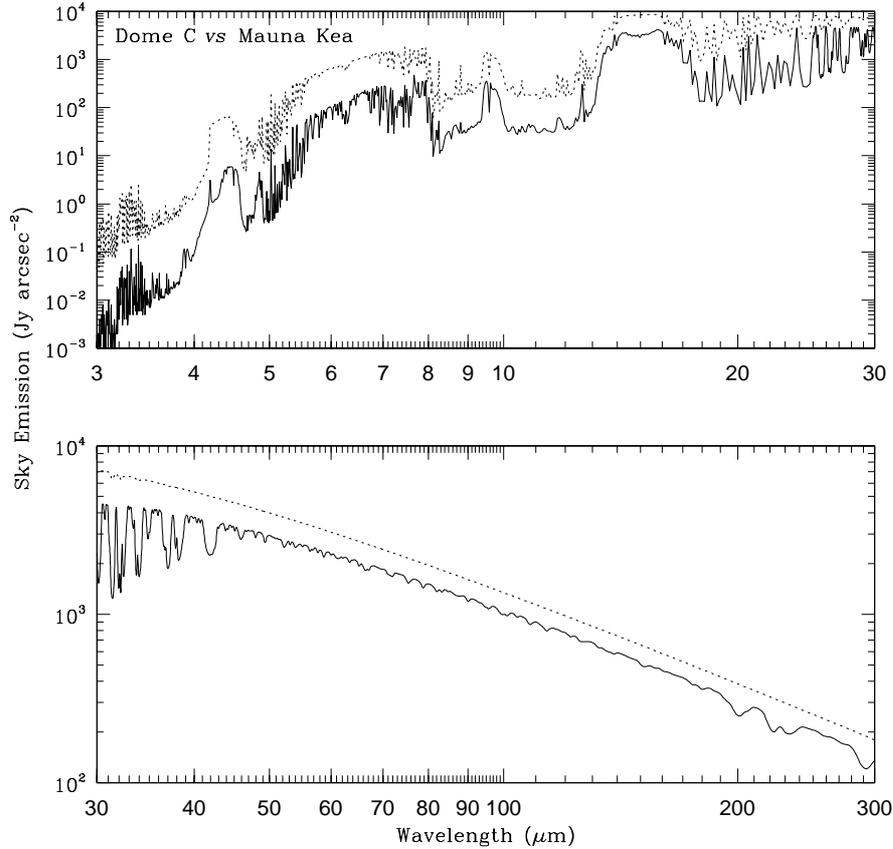}
\end{center}
\caption{Model calculations of the infrared sky emission, in
  Jy\,arcsec$^{-2}$, above Dome C in Antarctica (solid lines;
  250$\mu$m ppt H$_2$O), compared to Mauna Kea in Hawaii (dotted
  lines; 1\,mm ppt H$_2$O), adapted with permission from Lawrence
  (2004).  The upper plot shows the emission across the mid-IR (3 to
  30$\mu$m), and the lower plot the emission across the far-IR to
  sub-mm (30 to 300$\mu$m).  In the observing bands typical background
  reductions are between one and two orders of magnitude at the
  Antarctic site.}
\label{fig:irflux}
\end{figure*}

The temperature drop from freezing point to $-60^{\circ}$C, about the
change going from Mauna Kea in Hawaii to the South Pole in winter,
would lead to a fall in the sky background of 200 times at 2.4$\mu$m,
if the sky were thermally emitting as a blackbody.  The realisation
that this drop could open a new window for deep cosmological studies
drove initial efforts to quantify the infrared sky emission in
Antarctica (Harper 1990, Burton, Allen \& McGregor 1993). Subsequent
measurements found the drop to be a factor of $\sim 50$ at the South
Pole, being limited by residual high-altitude airglow emission (Nguyen
et al.\ 1996, Ashley et al.\ 1996; see Fig.~\ref{fig:irps}).  However,
from 3-30$\mu$m the sky background was also found to be 10--20 times
darker than at the best temperate latitude sites (see
Fig.~\ref{fig:irflux}; Smith \& Harper 1998, Phillips et al.\ 1999,
Chamberlain et al.\ 2000, Lawrence 2004).  Such gains at mid-IR
wavelengths were unexpected, for the drop in the blackbody flux near
its peak emission wavelength is only a factor of 2--3 for the fall in
temperature.  It turned out that in the mid-IR the emission from
aerosols also dropped significantly, with only ice crystals
contributing to the emission, and not the dust found at temperate
sites.  Longwards of 20$\mu$m the drop in background is less, and
arises principally from the reduced water vapour in the air, lowering
the emissivity of the atmospheric windows.

Only summer (i.e.\ day time) thermal IR measurements have so far been
obtained from a high plateau site (Dome C; Walden et al.\ 2005).
However the sky fluxes from 3--20$\mu$m were found to be comparable
with those measured at the South Pole in winter, as well as being
exceptionally stable. Table~\ref{table:skyfluxes} summarises the sky
backgrounds at Dome C inferred from all these measurements and
modelling, as listed in Burton et al.\ (2005).

\begin{table}
\begin{center}
\caption{Comparison of Sky Fluxes at Dome C, Antarctica to Mauna Kea, Hawaii}
\label{table:skyfluxes}
\begin{tabular}{cccc}
\hline\noalign{\smallskip}
Band & Wavelength & Mauna Kea & Dome C \\
& $\mu$m & \multicolumn{2}{c}{Jy arcsec$^{-2}$} \\
\noalign{\smallskip}\hline
V & 0.55 & 6 (-6) & 6 (-6) \\
R & 0.65 & 1 (-5) & 1 (-5) \\
I & 0.80 & 2 (-5) & 2 (-5) \\
J & 1.21 & 9 (-4) & 5 (-4) \\
H & 1.65 & 3 (-3) & 1 (-3) \\
K$'$ & 2.16 & 2 (-3) & $\ldots$ \\
K$_{dark}$ & 2.30 & $\ldots$ & 1 (-4) \\
L & 3.76 & 2 (+0) & 2 (-1) \\
M & 4.66 & 4 (+1) & 5 (-1) \\
N & 11.5 & 2 (+2) & 2 (+1) \\
Q & 20 & 3 (+3) & 5 (+2) \\
\noalign{\smallskip}\hline
\end{tabular}
\end{center}
These fluxes are derived from a combination of measurements and modelling, as
listed in Burton et al.\ 2005.  For the K--band we compare K$'$ to
K$_{dark}$ as these are the parts of the window where the deepest
observations may be made at Mauna Kea and Dome C, respectively. Note 
that $6 (-6) \equiv 6 \times 10^{-6}$ etc.
\end{table}

\subsection{Atmospheric Transparency: the driest locations on Earth}
\label{sec:transparancy}
The amount of precipitable water vapour in the atmosphere determines
the quality of a site from longwards of 20$\mu$m to the millimetre regime,
as well as having a substantial bearing on performance from
3-20$\mu$m.  At many `good' observing sites, such as Kitt Peak in the
USA and Siding Spring in Australia, typically 1\,cm of water is found
between the telescope and space.  At excellent sites like Mauna Kea
this falls to 1\,mm on the best of days.  On the 5,000m elevation
Chajnantor plateau in Chile (site of ALMA) the level occasionally
reaches as low as 300$\mu$m of precipitable water vapour. At the South
Pole (Chamberlin et al.\ 1997, Lane 1998, Peterson et al.\ 2003) and
Dome C (Valenziano \& dall'Oglio 1999, Calisse et al.\ 2004, Tomasi et
al.\ 2008) such a level is maintained for most of the year, but can
fall further still.

Measurements of water vapour content over the Antarctic plateau were
in fact first carried out at the Russia's Vostok station in 1972
(Burova 1986, Bromwich 1988) and from the South Pole in 1975 (Smythe
\& Jackson 1977).  From these and subsequent data a year round average
was then determined for Pole of 450$\mu$m ppt H$_2$O and 350$\mu$m for
Vostok, with levels occasionally dropping below 100$\mu$m ppt H$_2$O
in winter.  As then studied by Townes \& Melnick (1990), with such
levels of atmospheric water vapour the sub-mm bands from 350$\mu$m to
1\,mm are open virtually continuously.  New bands open at times at
200$\mu$m -- in the THz spectral regime.  Similarly, observations from
20-30$\mu$m could be routinely made.  From the highest plateau site,
Dome A, windows are conjectured to appear across the 30 to 300$\mu$m
range that are opaque from any other ground-based location.

\begin{figure*}
\begin{center}
\includegraphics[width=1.0\textwidth]{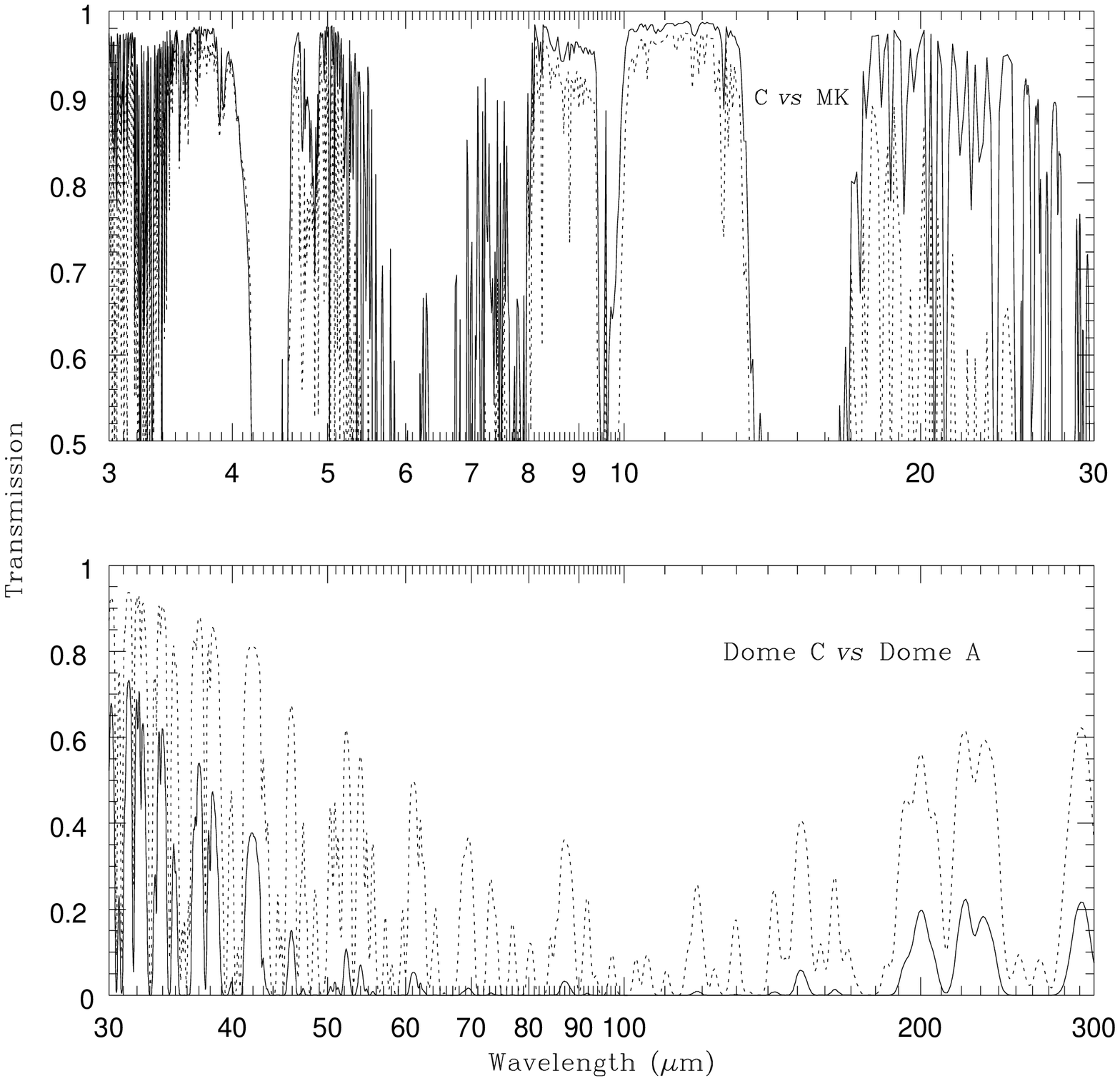}
\end{center}
\caption{Model calculations of the atmospheric transmission across the
  mid-IR (3 to 30$\mu$m; upper plot), and the far-IR to sub-mm bands
  (30 to 300$\mu$m; lower plot), adapted from Lawrence (2004).  The
  mid-IR plot compares the transmission between typical conditions at
  Dome C (250$\mu$m ppt of H$_2$O; solid line) and excellent
  conditions at Mauna Kea (1mm ppt; dotted line), whereas in the
  far-IR/sub-mm plot the comparison is between average conditions at
  Dome C (solid) and the best conditions then predicted for Dome A
  (30$\mu$m; dotted), since Mauna Kea is effectively opaque in this
  wavelength range. Note that 25$\mu$m ppt H$_2$O has subsequently
  been determined as the best value at Dome A in the analysis
  conducted by Yang et al.\ 2010.}
\label{fig:transmission}
\end{figure*}

The first measurements from Dome A of water vapour content have
recently been made, supporting these assertions.  They used a tipping
radiometer operating at 450$\mu$m (661\,GHz) on the PLATO laboratory
(Saunders et al.\ 2009), further correlated with satellite radiometry
from above (the NOAA-18 satellite observing in the 183\,GHz H$_2$O
line). A median value of 140$\mu$m ppt H$_2$O was found, a 25\%
quartile of just $100 \mu$m and the ppt H$_2$O level fell to 25$\mu$m
at its lowest, comparable to that experienced by an aircraft flying in
the stratosphere ({\em e.g.}\, SOFIA).  These conditions dramatically
open up atmospheric windows across the entire infrared and
sub-millimetre regimes (see Fig.~\ref{fig:transmission}; Hidas et al.\
2000, Lawrence 2004).  For instance, for 100$\mu$m of ppt H$_2$O the
atmospheric transmission is 28\% at the 205$\mu$m wavelength (1.45
THz) of the important [NII] line.

Initial summer time measurements of the atmospheric transparency have also
been obtained from Dome F (Ishii et al.\ 2010), using a tipping
radiometer operating at 220\,GHz.  An extremely low mean opacity of
0.045 was determined (comparable to South Pole and Chajnantor winter
median values), and varying little over the 1 month period of the
measurements. The authors model this data with 0.6\,mm of precipitable
water vapour.

There may be even better sites for astronomy on the plateau.  A ridge
runs south and west from Dome A and $\sim 400$\,km SW is a location
called ``Ridge A''.  Examining the NOAA--18 183\, GHz satellite
measurements for this location, Yang et al.\ (2010) deduce even lower
25\% and 50\% quartiles for the water vapour than for Dome A, of
80$\mu$m and 120$\mu$m, respectively.  The 158$\mu$m window (1.9\,THz
and dominant [CII] line) would even open at times, with a transmission
of more than 11\% for 10\% of the time.  Such remarkable conjectures
call for in situ measurements of the water vapour content to be
carried out at Ridge A to examine their veracity.

In the near--IR, from 1--2.5$\mu$m, models by Siebenmorgen (2010) show
that the average transmission at Dome C exceeds 40\% at all
wavelengths, when measured with a resolution R=250 (there are, of
course, narrow spectral regions which remain saturated).  This results
in a widening of the J, H \& K bands from 1-2.5$\mu$m, and of the L \&
M bands at 3.6 \& 5.0$\mu$m.

\subsection{Atmospheric Stability and Seeing: the thin surface boundary layer}
\label{sec:seeing}
\paragraph{Stability}
Less appreciated, but equally important for the conduct of many kinds
of observation, is the stability provided by the atmosphere,
particularly for measurements that can mitigate the effects of a thin,
but turbulent, surface boundary layer.  This provides for significant
gains, not just in imaging quality, but also in the photometric
precision that can be obtained for a measurement and for long
time-series observations. In turn, these gains have implications not
only for observations using conventional single-mirror telescopes but
also for solar telescopes and for interferometers.

At the South Pole there is no diurnal cycle and so no corresponding
daily temperature changes -- simply the 6-month cycle of night and
day. Any source beyond the Solar System that can be observed can also
be seen continuously, and at constant zenith angle.  This greatly
facilitates experiments which require continuous monitoring or high
duty cycles.  It is also favourable for following particle cascades in
air showers as a source is always seen through the same depth of
atmosphere.  In the IR, beyond 3$\mu$m, it is perfectly feasible to
observe all year round, the temperature being cold enough to
facilitate day time observations.  Away from the Pole there is, of
course, a latitude-dependent seasonal diurnal cycle, and the length of
the period of continuous darkness is reduced, but many of the
advantages remain.

Not only is the strength of the sky background emission reduced in
Antarctica, its level is more stable at IR and sub-mm wavelengths than
at temperate sites (Smith \& Harper 1998, Peterson et al.\ 2003).
Indeed, it is fluctuations in the level of the background (both
spatially and temporally) that can provide the limiting factor for
photometric measurements in these bands.  At 350$\mu$m (860\,GHz),
while Peterson et al.\ found that the median sky opacity at South Pole
and Chajnantor (Chile) to be similar (1.20 {\it vs.} 1.39,
respectively), the fluctuations in the opacity level were found to be
$\sim 3$ times lower at Pole during the best observing conditions,
when the opacity is below unity -- i.e.\ when observations at these
wavelengths could be made.

For CMBR measurements, made in the microwave bands, the South Pole has
proved to be a superb site, as attested by the wealth of science that
has been achieved (see \S\ref{sec:cmbr}). For instance, at 220\,GHz
the zenith opacity is just $\sim 0.03$ and the median brightness power
fluctuations are an order of magnitude lower than at any established
temperate sites (at $\rm \sim 38\,mK^2 \, rad^{-5/3}$ in
Rayleigh-Jeans temperature units; Bussmann, Holzapfel \& Kuo, 2005).

\paragraph{The Turbulent Surface Boundary Layer}
A strong temperature inversion exists above the ice during the most
stable of conditions in winter, produced by radiative cooling from the
ice surface in the absence of sun light.  The temperature can rise by
up to $\sim20^{\circ}$C in heights of a few tens of metres.  Virtually
all the turbulence responsible for seeing at optical and infrared
wavelengths, the result of micro-thermal temperature fluctuations
causing refractive index changes, is confined to this boundary layer.
Dealing with the boundary layer is one of the most challenging
problems facing designers of optical and infrared telescopes in
Antarctica (see \S\ref{sec:challenges}), but if its mitigating effects
can be overcome then the Antarctic plateau offers sharper imaging,
more precise photometry and superior astrometric precision than
measurements made from any other place on Earth. Of course, such a
statement applies to any ground-based telescope, but the narrow
boundary layer above the Antarctic plateau suggests that this might be
readily accomplished there, for instance by building a tower that
extends above it.

The free air seeing in the visible, above the surface boundary layer,
is typically a factor of two better than the $\sim 0.7''$ seeing above
the best temperate sites.  The proximity of the turbulent boundary
layer to the telescope also results in reduced values for the
scintillation noise, since the `lever arm', from the turbulent
atmospheric cells that de-focus the light to the telescope, is much
reduced.  Since scintillation provides the limiting factor to the
error in a flux measurement, this implies that improved photometric
precision is also possible.  

Two further quantities are also very important in determining site
quality; the coherence time and the isoplanatic angle of the turbulent
cells.  Since the cells are located close to the surface over the
Antarctic plateau, and pass across the telescope field of view
relatively slowly, the above two parameters are larger than found at
temperate sites, where the cells are at high altitude and move
rapidly.  Longer coherence times and larger isoplanatic angles feeds
into improved performance for adaptive optics systems (e.g.\ larger
bandwidth, fainter reference stars).  For an interferometer, it also
results in improved signal to noise and improvement in the achievable
astrometric precision (e.g.\ Lloyd et al.\ 2002, Storey 2004).

Considerable effort has been put into quantifying the relevant
parameters -- the free air seeing ($\epsilon_0$), isoplanatic angle
($\theta_0$), Fried parameter ($r_0$), coherence time ($\tau_0$), the
boundary layer turbulence ($C_N^2$), its height ($h$) and outer scale
length ($L_0$) -- at the South Pole, Dome C and, most recently, Dome
A, so as to determine the quality of these sites for prospective
telescopes.  While this qualification of site properties remains
incomplete, substantial information is now available, or can be
inferred through modelling, so that a quantitative picture of
observing conditions over the Antarctic plateau can be deduced.  It is
sufficient to allow a detailed comparison with conditions at
mid-latitude sites.  We discuss some of this work below.

\subsubsection{Comparison of Seeing Conditions at Plateau Sites}
\label{sec:siteseeing}
\paragraph{South Pole}
At the South Pole a steady breeze blows most of the time (the
katabatic wind falling from the summit of the plateau, Dome A). This
causes substantial seeing at the ice surface level, of order $1.5''$
in the visual (Loewenstein et al.\ 1998).  It is produced almost
entirely in the lowest 200-300\,m above the ice (Marks et al.\ 1996,
1999, Travouillon et al.\ 2003a).  On the other hand, there is no jet
stream to produce the high-altitude seeing that temperate sites
experience.  As first pointed out by Gillingham (1989, 1993), the free
air seeing above the surface boundary layer is much smaller than the
surface seeing at temperate sites, a concept termed by Gillingham
``super-seeing''.

The depth of the boundary layer at the South Pole, while shallow, is
still too high to consider raising a telescope beyond it. On the
summits of the plateau, however, the wind speed is much reduced,
leading to a significantly thinner surface boundary layer (Marks
2002, 2005) than at the Pole.

\begin{figure*}
\begin{center}
\includegraphics[width=1.0\textwidth]{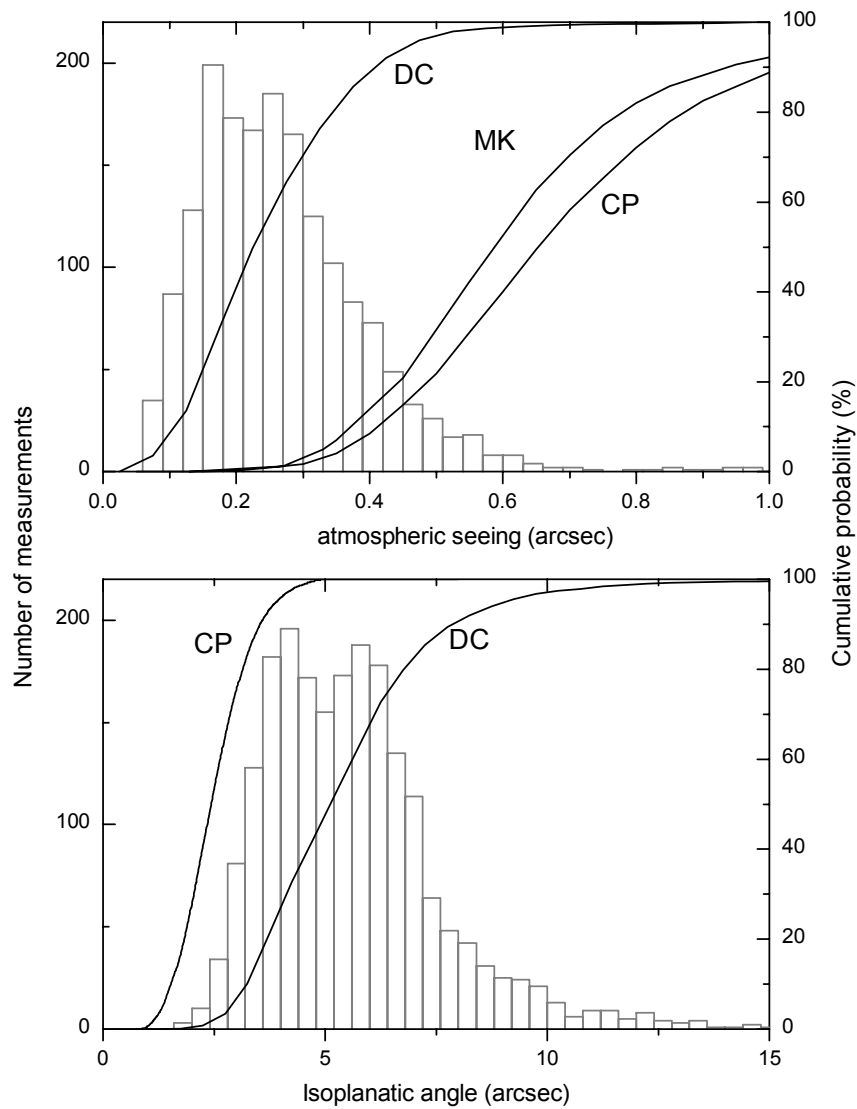}
\end{center}
\caption{Histograms and cumulative distributions for the atmospheric
  seeing (top plot) and isoplanatic angle (bottom plot) for Dome C,
  above $\sim 30$\,m (a typical boundary layer height), as measured by
  Lawrence et al.\ (2004) using a combination of MASS and SODAR
  instruments (see text in \S\ref{sec:siteseeing}). The cumulative
  distributions also compare Dome C (DC) to Mauna Kea (MK) and Cerro
  Paranal (CP) (including the contribution from the surface boundary
  layer for these latter two sites).  The superb seeing attainable
  close to the surface at Dome C (median value $0.27''$ in V-band) is
  evident, with the values even falling being below $0.1''$ at times.}
\label{fig:mass}
\end{figure*}

\paragraph{Dome C}
Two decades of measurements of wind speeds at Dome C, made with an
automated weather station (AWS), showed them to be exceptionally low,
averaging $\rm \sim 3\,m\,s^{-1}$ at the ice surface, with maximum
gusts reaching only $\rm \sim 10\,m\,s^{-1}$ (Aristidi et al.\ 2005b).
Such data gave promising intimations of a stable atmosphere that would
provide superb seeing.  When such seeing measurements were first made
(with a robotic observatory at Dome C (the AASTINO; Lawrence et al.\
2004; see Fig.~\ref{fig:mass}) using both an acoustic radar (SODAR),
sampling the boundary layer, and a multi-aperture scintillation sensor
(MASS), sampling the whole atmosphere), a median value was obtained
for the visual seeing of just $0.27''$, above a boundary layer whose
height was found to be only $\sim 30$\,m thick.  Subsequent
measurements from Dome C by Agabi et al.\ (2006), after the winter
opening of the station, making use of direct measurements of the
turbulence using microthermal sensors flown on balloons, combined with
total seeing measurements using differential image motion monitors
(DIMMs), obtained similar results within the errors (median seeing
$0.36''$, boundary layer height 36\,m), so confirming that the
turbulence is primarily confined to the boundary layer.  These results
indicated that the visual seeing at Dome C, above this boundary layer,
is typically about a half that found at the best temperate sites.  For
$\sim 10\%$ of the time Lawrence et al.\ found the seeing, above the
narrow surface boundary layer, to be $< 0.1''$, the lowest values ever
recorded.  The determination of the isoplanatic angle and coherence
time, as measured by this combination of instruments, was also in
reasonable agreement ($5.7''$ and 7.9\,ms for Lawrence et al.; $4.7''$
and 8.6\,ms for Agabi et al.; both above 30\,m), and indicate that
Dome C is an excellent site for both wide-field, high spatial
resolution imaging and for infrared interferometry.  Furthermore,
measurements of the scintillation noise (Kenyon et al.\ 2006) show it
to be the lowest so far measured on the Earth, making it an excellent
site for precision photometric measurements as well.  As a caution to
the reader, however, it should be remarked that the above results have
been deduced from relatively limited measurements.  It would be
desirable to improve the statistics of these data sets to fully
characterise the seeing conditions at Dome C\@.

\begin{figure*}
\begin{center}
\includegraphics[width=1.0\textwidth]{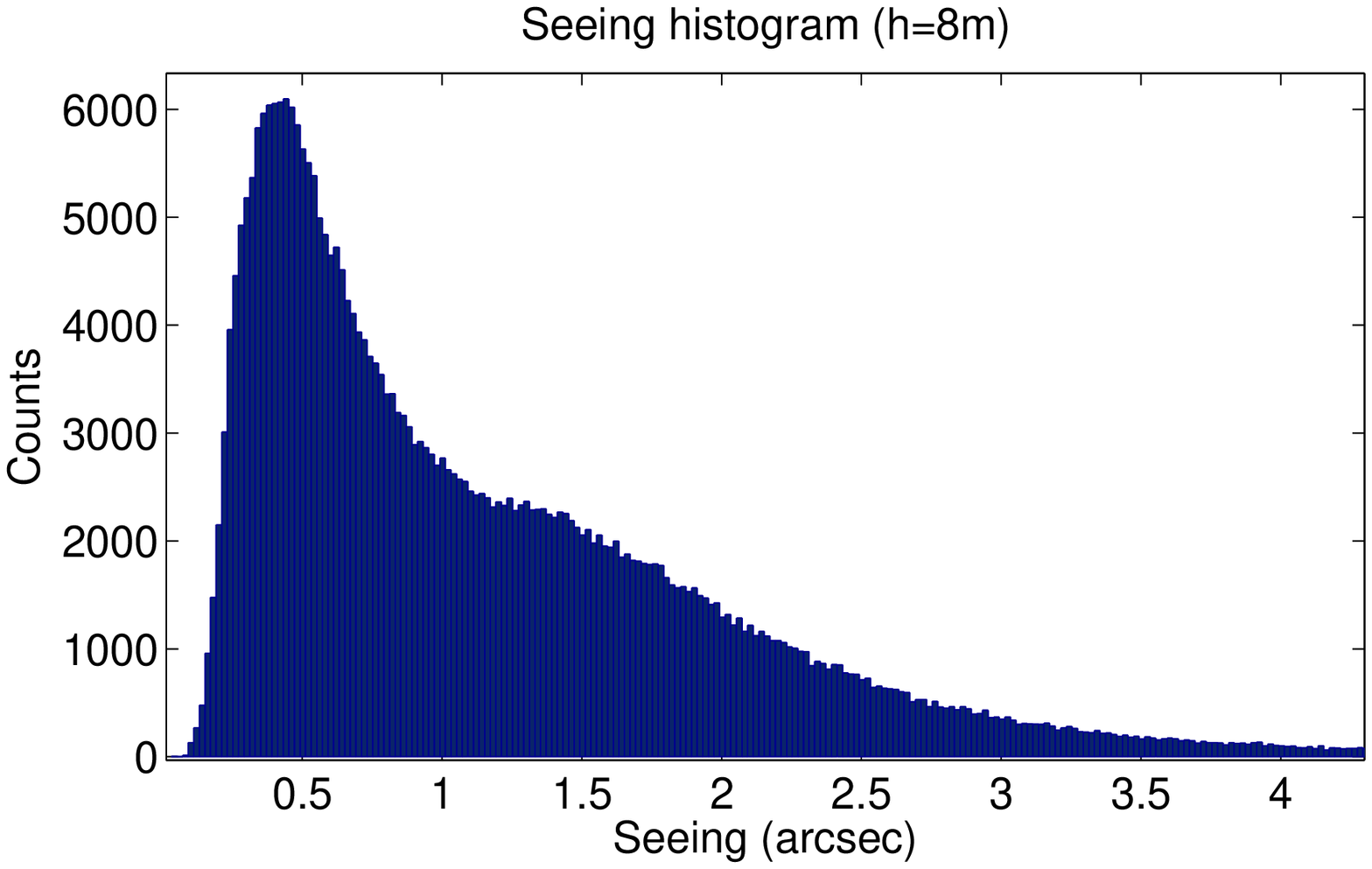}
\includegraphics[width=1.0\textwidth]{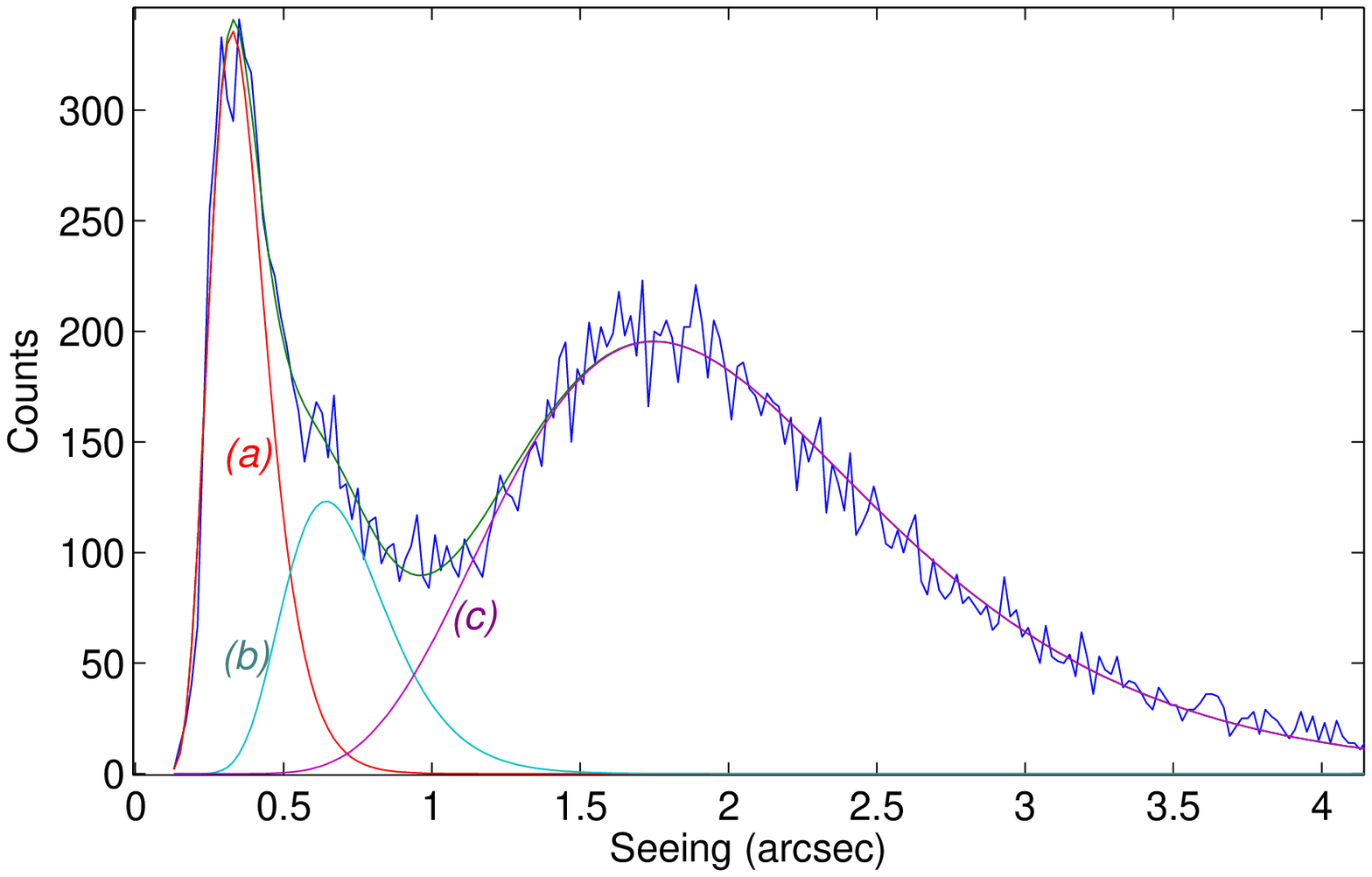}
\end{center}
\caption{Histograms of seeing measured with a DIMM at Dome C from
  Aristidi et al.\ (2009).  The shaded histogram is the data from
  an instrument mounted 8\,m above the ice, and includes the authors'
  full 3.5 year data set (summer \& winter).  The bimodal distribution
  reflects periods of excellent seeing at this height, when the
  boundary layer drops below 8\,m, together with the more normal
  moderate seeing when the instrument lies within the boundary
  layer.  The bottom curves show the data from the 2006 winter, with
  the three log-normal fits reflecting times when the instrument is
  above the boundary layer (a), within it (c) and a few intermediate
  cases (b).}
\label{fig:bimodal}
\end{figure*}

The statistics of 3.5 years of DIMM measurements from Dome C, placed
on towers at different heights within the surface boundary layer, have
been summarised by Aristidi et al.\ 2009 (also incorporating results
from Trinquet et al.\ 2008 on the vertical distribution of
turbulence).  The boundary layer height was found to be variable, as
well as volatile on short timescales, but has a sharp boundary, whose
median height was inferred to be between 23-27\,m based on its
statistical properties.  When the boundary layer is depressed below
the height of the DIMMs, excellent seeing from the free air is
obtained (median value $0.36''$), as clearly seen in the bimodal
distribution of the seeing shown in Fig.~\ref{fig:bimodal}.  There is
also a clear summer--winter difference in the seeing distributions;
while the bimodal separation in each plot indicates when the seeing
monitor was either lying inside or above the boundary layer (and the
clear separation indicating its sharp boundary). More turbulent energy
resides within the boundary layer in winter, as indicated by the
larger values of seeing measured then.  The seeing data also provide
an estimate of the median value of size of the outer scale for
turbulence (Ziad et al.\ 2008), of $\sim 7$\,m in winter (measured
3.5\,m above the ice). These values are 2--3 times lower than measured
at Mauna Kea and Paranal observatories, and have implication for
improved AO and interferometric performance by reducing fringe
excursion errors (though the extension of these results to a telescope
placed above the boundary layer is not clear).

\begin{figure*}
\begin{center}
\includegraphics[width=0.4\textwidth]{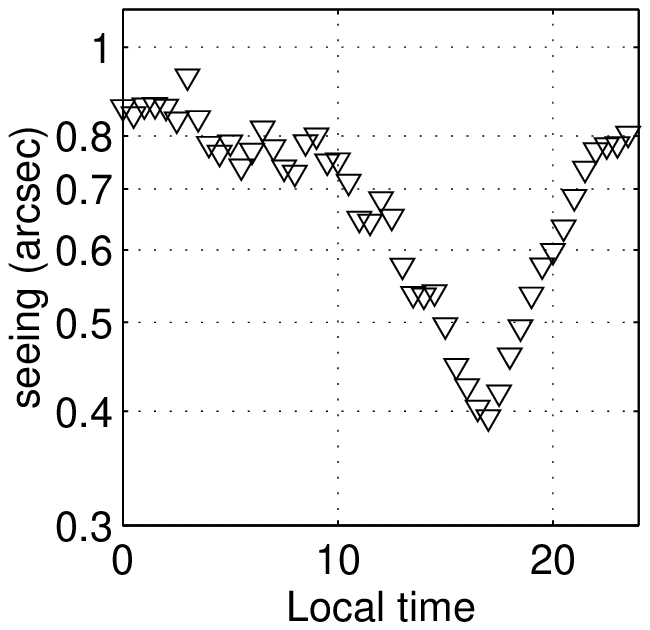}
\includegraphics[width=0.4\textwidth]{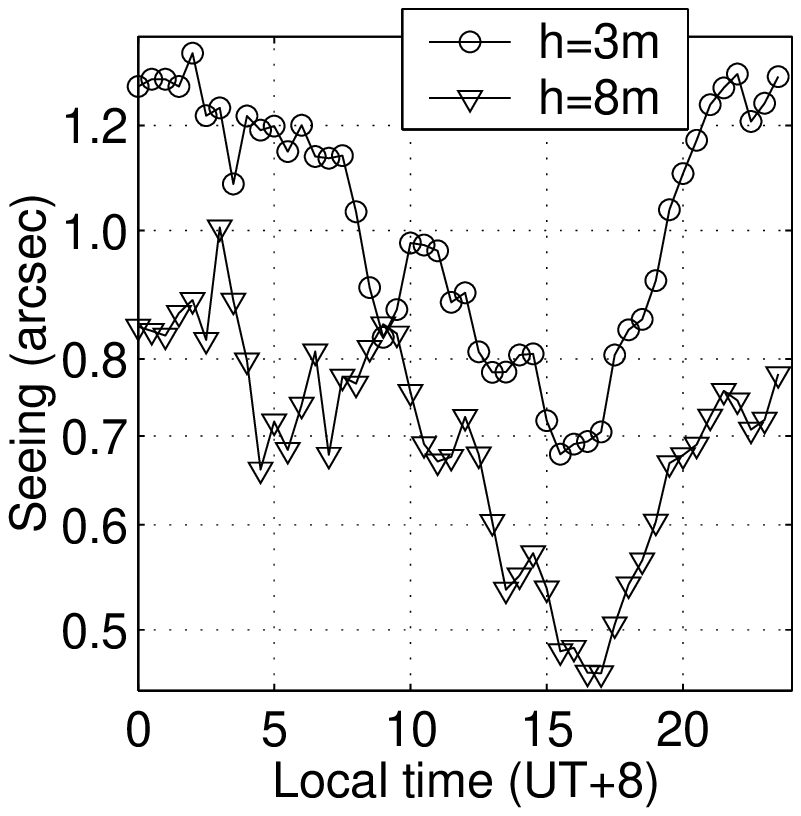}
\end{center}
\caption{Diurnal variation of the daytime seeing at Dome C, measured
  by Aristidi et al.\ (2005a).  The plot to left shows data averaged
  over two summers (2003-04 \& 2004-05) measured from a DIMM placed
  8\,m above the ice.  The plot to right compares data from two DIMMs,
  one at 8\,m and the other at 3\,m above the ice, from the 2004--05
  summer. Note especially the behaviour each afternoon, around 5 p.m.
  local time, with the boundary layer often disappearing.  This is
  caused by the temperature gradient above the ice disappearing for a
  few hours then.  This results in excellent ``on-ice'' seeing of
  around $0.4''$, with implications for obtaining high image quality
  for solar telescopes.}
\label{fig:daytimeseeing}
\end{figure*}

A particularly interesting result (Travouillon 2004, Aristidi et al.\
2005a) relates to the daytime seeing in summer.  Late in the
`afternoon' each day, around 5~p.m. local time, the surface boundary
layer often disappears, leading to a free-air seeing of $\sim 0.4''$
just 0.8\,m above the ice (see Fig.~\ref{fig:daytimeseeing}).  At this
time of the day the temperature gradient above the ice disappears.
The isothermal temperature profile then leads to a great reduction in
the turbulent energy and hence excellent surface seeing.  This has
particular implications for measurements of the Sun, which of
necessity must be made in daytime, providing a period of $\sim
2$~hours each day where excellent imaging quality is directly
obtainable.

The paper by Gredel (2010) provides a concise and clear summary of the
site testing results from Dome C, conducted under the auspices of the
European Union ARENA consortium, including the results described above.

An interesting recent development is the application of numerical
meteorological models to the calculation of turbulence and seeing
conditions over the Antarctic plateau, with eventual application
towards the forecasting of the seeing.  Lascaux et al.\ (2009, 2010)
have compared (low-resolution) general circulation models and (higher
resolution) mesoscale models for the calculation of meteorological
parameters, with the particular application being to the modelling of
$C_N^2$ turbulence profiles obtained from 15 balloon flights from Dome
C in the 2005 winter by Trinquet et al.\ (2008).  Perhaps
unsurprisingly, the higher resolution, computationally more expensive,
models work best.  Lascaux et al.\ find they can model the wind
gradients and temperature profiles with height reasonably well,
including their rapid changes over the extent of the surface boundary
layer.  The boundary layer thickness in these mesoscale models is, in
general, currently calculated to be $\sim 50$\% higher than that
measured.  The turbulence profile ($C_N^2$) within the boundary layer
is reproduced well, but deviates from the data above the boundary
layer.  There is a reasonable correlation between observed and
predicted seeing, although still considerable scatter for individual
data points.  The overall agreement between models and data shows that
this is a promising area of study, as well as emphasising the need for
a much more comprehensive set of measured turbulence profiles to
compare the models with.  The authors intend to extend their
calculations to other sites over the Antarctic plateau, and so provide
quantitative model comparisons for the behaviour of the seeing
conditions at each site over the year.

\begin{figure*}
\begin{center}
\includegraphics[width=1.0\textwidth]{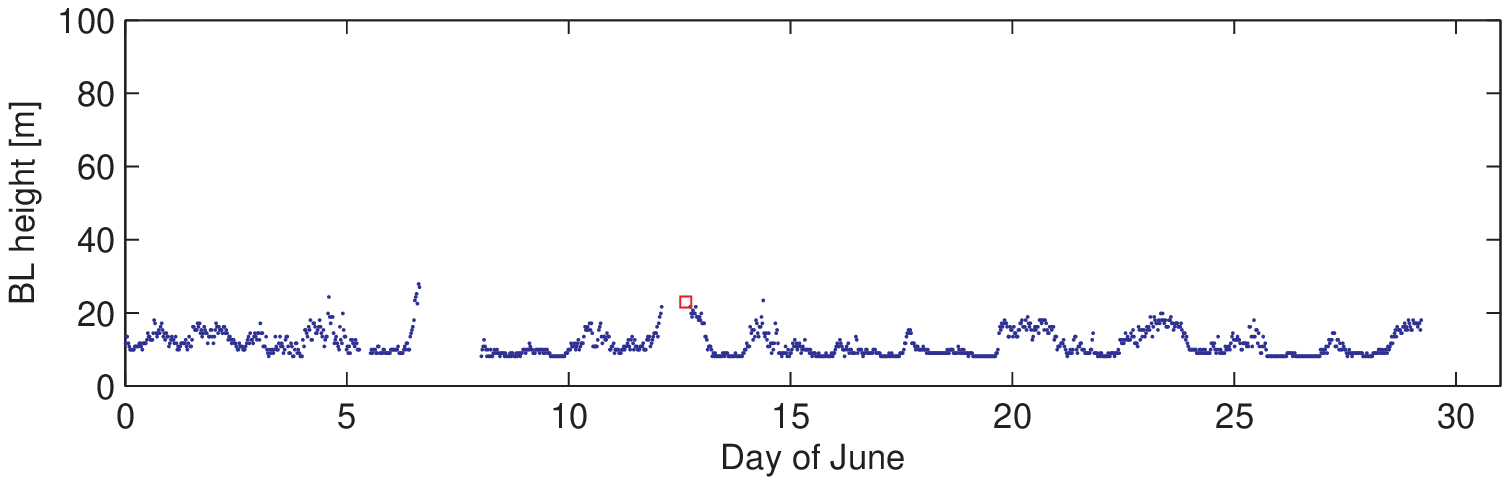}
\includegraphics[width=1.0\textwidth]{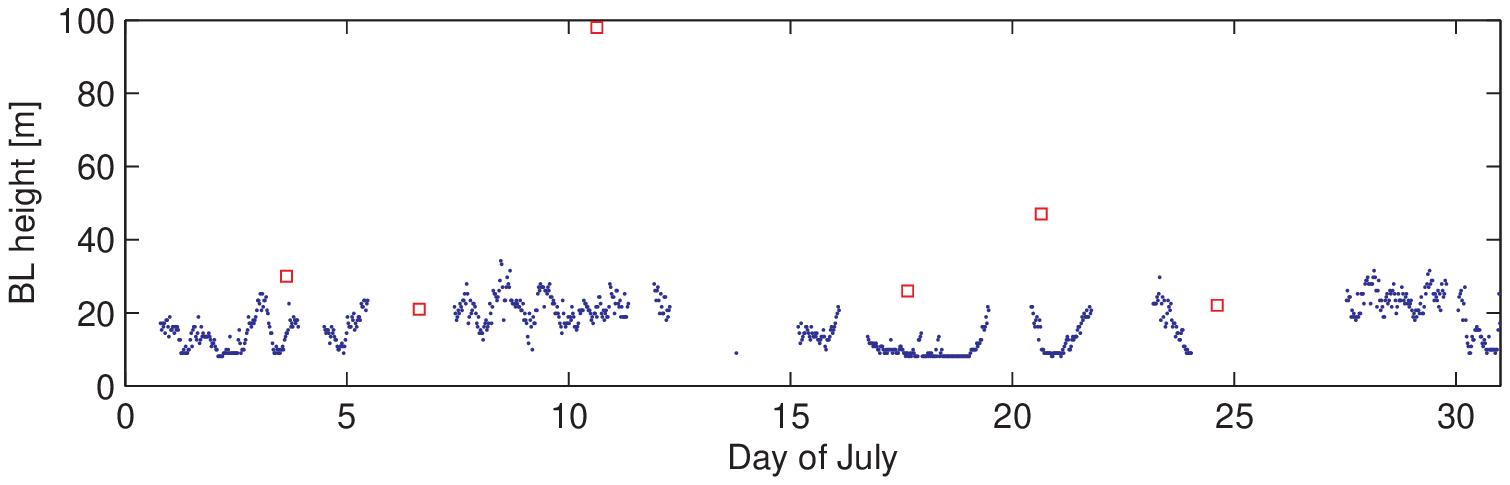}
\includegraphics[width=1.0\textwidth]{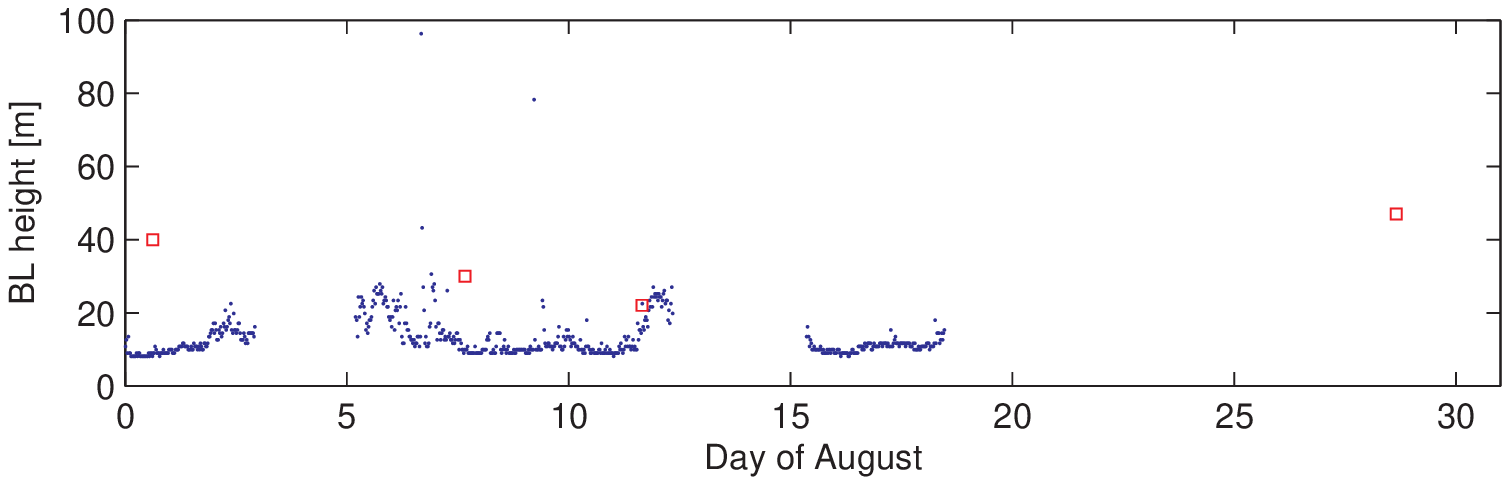}
\end{center}
\caption{Height of the boundary layer at Dome A from June to August
  2009, as measured using the SNODAR instrument (an acoustic radar;
  Bonner et al.\, 2010). Each dot represents the average of 30 minutes
  of data.  Squares show measurements made at Dome C on the same day
  of the year in 2005 by Trinquet et al.\, (2008) from balloon
  measurements.  The SNODAR provided a vertical resolution of 0.9\,m
  and sampled the turbulence in the atmosphere every 10\,s to a height
  of 180\,m above the ice, to determine the temperature structure
  constant, $C_T^2$.  The boundary layer height was defined to be the
  height at which $C_T^2$ reduces to 1\% of its initial value.  The
  25\%, 50\% and 75\% quartiles determined for it by Bonner et
  al.\,(2010), over the Feb--Aug period of their whole data set, are
  9.7\,m, 13.9\,m and 19.7\,m, respectively.}
\label{fig:snodar}
\end{figure*}

\paragraph{Dome A}
Site testing at Dome A, the highest point on the Antarctic plateau,
only began in 2008, making use of a series of instruments attached to
the PLATO automated observatory (Yang et al.\ 2009, Lawrence et al.\
2009).  Of necessity, the information obtained so far is limited, but
it does suggest some remarkable properties for the site.  While direct
seeing measurements have yet to be made, the boundary layer has been
probed using a specially designed acoustic radar able to sample on the
1m-scale (SNODAR; see Bonner 2009, 2010 and Fig.~\ref{fig:snodar}),
indicating that it is even thinner than at Dome C\@.  Over the first
part of the winter period of 2009 its median height was found to be
14\,m and its 25\% quartile value to be just $\sim 10$m thick.

\paragraph{Ridge A}
An analysis by Saunders et al.\ (2009) of three of the key parameters
($\epsilon_0$, $\theta_0$ \& $\tau_0$), together with their `coherence
volume' ($\theta_0^2 \tau_0 / \epsilon_0^2$; Lloyd 2004) was
undertaken to determine the efficacy of an interferometer placed at
different sites over the plateau.  The study ranks the South Pole as
the best site of any station, followed by Dome A and then Dome C (an
order of magnitude worse). However, Ridge A was found to be the best
location of all.  Ridge A is closer to the null point for origin of
the katabatic winds than Dome A\@.  It also has lower temperatures and
less cloud cover.

While the estimates made in this study are, of necessity, preliminary
and based on extrapolations of current data sets combined with
satellite telemetry, they do highlight the need to obtain extensive
site data over the plateau before choosing a site that may be the best
for a particular application.  There is no single best site on the
Antarctic plateau for all astronomical applications.

\subsection{Other Site Factors}
\label{sec:other}
\paragraph{Clear Skies}
While the South Pole experiences clouds about half the time, Dome C appears to
be a much clearer site. A variety of measurements made from automated
cameras and by direct observations indicate photometric skies occur an
exceptional $\sim 75-90$\% of the time (e.g.\ Ashley et al.\
2005; Mosser \& Aristidi, 2007, Moore et al.\ 2008a). Satellite
measurements indicate that Dome A and F should have similar cloud
characteristics to Dome C, being clear $\sim 75$\% of the time in
winter (Saunders et al.\ 2009).

Direct measurements of the clear sky fraction at Dome C, as well as of
the effective duty cycle, were also obtained in astronomical
experiments during the 2007 \& 2008 winters as part of the small-IRAIT
(Strassmeier et al.\ 2008) and ASTEP projects (Crouzet et
al.\ 2010), which we discuss below.

Small-IRAIT used an optical CCD camera mounted on a 25\,cm telescope
to follow photometric variations simultaneously of two bright variable
stars, continuously over a 10 day period in July 2007.  The weather
was stable and only $\sim 6$ hours were lost over this time period,
providing a duty cycle of 98\%.  Furthermore, a 3\,mmag photometric
precision in V--band (and 4.2 mmag in R--band) was obtained for a
2.4\,hour subset of the data, 3--4 times better than has been achieved
with an equivalent telescope in Arizona, attesting to the improved
photometric precision resulting from reduced scintillation noise.

ASTEP used a CCD camera on a fixed 10\,cm refractor pointed towards
the celestial South Pole.  Excellent sky conditions were recorded
$\sim 60$\% of the time during the 4 months of operation (June --
September, 2008).  A duty cycle of 50\% for photometric measurements
in the optical was obtained (taking into account cloud, moonlight,
scattered sunlight etc.), a higher value than has been obtained at
temperate sites.

Sky brightness and transparency measurements have also been made at
Dome A in the winter of 2008 by the CSTAR (the Chinese Small Telescope
ARray) instrument (Zou et al.\ 2010), controlled from the PLATO
laboratory. This comprises four fixed 14.5\,cm telescopes, pointed at
the south celestial pole, each with a $4.5^{\circ}$ field and a
different optical filter. The primary purpose of this experiment was
to search for variability in stellar flux, for which a catalogue of
10,000 stars and 30,000 images has been produced (Zhou et al.\ 2010).
Clear skies were found in 74\% of the images.
 
\paragraph{Aircraft Contrails}
Aircraft contrails add to high-altitude turbulence and increase
optical sky brightness via scattering of moonlight. They are an
increasing problem at temperate sites as they are often overflown by
air traffic, whose frequency continues to increase.  Contrails are,
however, non-existent over the Antarctic plateau during the winter
months.

\paragraph{Low Aerosols and Pollution}
Antarctica has the cleanest air on the Earth, in particular in the
amount of aerosols in the atmosphere (there is no dust). For instance,
as discussed in Kenyon \& Storey (2006), aerosol levels are $\sim 50$
times lower than found at the atmospheric observatory on Mauna Loa,
Hawaii (Bodhaine, 1995).  This results in reduced emissivity at
infrared wavelengths (and so lower sky emission) and reduced levels of
scattering at optical wavelengths.  It is, of course, important to
avoid local degradation of these conditions at an observatory, for
instance as might be caused by emissions from diesel generators.  This
might be achieved by the use of clean energy sources (see also
\S~\ref{sec:challenges}).

There is also minimal light pollution to consider during the winter
months, as there are no settlements nearby other than the scientific
station the observatory is based at, and this will remain the case for
the foreseeable future.

\paragraph{Seismically Quiet}
Antarctica is the quietest continent seismically, quite different to
the geologically active regions where temperate observatories are
often located (e.g.\ in Hawaii or Chile).  While not normally a
critical part of the assessment in designing telescopes, consideration
of the effects of seismic disturbances have implications for the
design of the largest telescopes, such as the ELTs, or interferometers
where path lengths must be maintained to sub-micron precision.  While
the Antarctic ice sheet itself is moving, at about 10\,m per year,
this is a bulk motion that is straightforward to measure and to
compensate for.

\subsection{Disadvantages of Antarctica for Astronomy}
There are, of course, some disadvantages relating to the site
conditions on the Antarctica plateau.  The cold and high altitude pose
difficulties for humans, but have relatively simple engineering
solutions for the operation of facilities. Other facets of the sites
are less easily overcome. Here we summarise the principal issues, and
discuss their effects further in \S\ref{sec:conduct}.

\label{sec:disadvantages}
\paragraph{Humidity and Super-Saturation}
While the extreme cold means that the absolute humidity level of the
air is low (up to three orders of magnitude lower than at temperate
sites), the relative humidity in Antarctica is always high.
Calculations assuming 100\% humidity in fact provide reasonable
approximations for the atmospheric transmission, without needing a
detailed knowledge of its actual level and distribution through the
atmosphere.  Close to the ice surface, within the boundary layer, the
air can in fact be super-saturated.  This has the adverse consequence
of causing ice growth on any exposed surfaces, unless they are heated
slightly above the ambient temperature, flushed with dry air, or
placed within an enclosure in a thermally controlled environment (see also
\S~\ref{sec:challenges}).

\paragraph{Sky Coverage}
The further a telescope is sited from the Equator the less sky it has
access to.  In the extreme case of the Poles only half the celestial
sphere is visible, though of course it is visible the whole time.  An
even smaller portion of sky is accessible above a telescope elevation
of $30^{\circ}$, the usual elevation limit considered for most
observations. At Dome C (75S) only 37\% of the sky can be seen above
$30^{\circ}$ compared with, for instance, 81\% from Mauna Kea (see
Kenyon \& Storey, 2006).  Those objects that can be seen, however, may
be observed for longer, and suffer less diurnal change in elevation
than they do when seen from temperate sites, providing gains in
precision for long time-series measurements.

\paragraph{Twilight}
The Sun also spends long periods of time just below the horizon at
high latitude sites, meaning that the fraction of time that is
``astronomically dark'' (i.e.\ when the Sun is more than $18^{\circ}$
below the horizon) is smaller than at temperate latitudes.  This is
only relevant at optical wavelengths, as at infrared wavelengths the
sky is bright (even in Antarctica).  However, the reduced aerosol
content reduces the scattering, so that the sky is dark in Antarctica
with the Sun closer to the horizon than is the case at temperate
sites.  In the short infrared wavebands (1.6 \& 2.4$\mu$m), Phillips
et al.\ (1999) found the effect of scattered sunlight disappeared when
the Sun is more than $10^{\circ}$ below the horizon at the South
Pole. Moore et al.\ (2008a) and Crouzet et al.\ (2010) found that at
Dome C the sky is optically dark when the Sun is more than
$13^{\circ}$ below the horizon (as did Zou et al.\ 2010 at Dome A).
Combining this limit with the fraction of clear skies and with the
phase of the Moon, Kenyon \& Storey (2006) find that the number of
hours of optically dark time available at Dome C per year is at least
as many as at Mauna Kea.

\paragraph{Auroral Activity and Optical Sky Brightness}
Auroral activity in Antarctica is frequent, but its intensity varies
over the continent, dependent on whether the auroral circle is
visible.  The South Pole and Dome F lie nearly under the auroral
circle, so suffer frequently, whereas Domes A, B \& C lie just
$6^{\circ}$ from the geomagnetic South Pole, so that aurorae there
generally lie below the horizon.  Auroral emission is concentrated
into a few spectral lines, particularly atomic oxygen and bands from
molecular nitrogen and oxygen. Aurorae will therefore only impact on a
small subset of observations that might be undertaken from Antarctica.
Saunders et al.\ (2009) re-analyse the work of Dempsey \& Storey
(2006)\footnote{This paper contains an error in the position used for
  the Geomagnetic Pole, implying that Dome C is closer to the
  Geomagnetic Pole than Dome A\@.  The error is rectified in Saunders
  et al.\, (2009).} on auroral emission, extending it over the
principal sites of interest on the plateau.  They find the auroral sky
contribution ($\rm \sim 23\, mags\, arcsec^{-2}$ at B \& V) at the
best plateau sites (Domes A, B \& C), will make the sky there about a
factor 2 brighter than at the best temperate sites at B band (and
20--30\% brighter at V). The South Pole and Dome F, on the other hand,
are another 1-2 magnitudes worse.  By way of a direct comparison of
their impact on an experiment, ASTEP, operating in the optical over
the 2008 winter at Dome C (Crouzet et al.\ 2010), found negligible
influence from aurorae on their data set.  The CSTAR experiment at
Dome A (Zou et al.\ 2010) determined the typical sky brightness in the
Gunn $i$ band, in dark sky conditions in mid-winter, to be 20.2-20.5
mags arcsec$^{-2}$. Converting this to the Mould I band (subtracting 0.75
magnitudes; see Windhorst et al.\ 1991), this is comparable to darkest
skies measured at dark temperate sites such as Siding Spring
Observatory in Australia (19.3 mags arcsec$^{-2}$) and CTIO in Chile
(19.9 mags arcsec$^{-2}$).

\section{A Brief History of Antarctic Astronomy}
\label{sec:history}
Captain Cook, and the crew of the ships Resolution and Adventure, were
the first people to cross the Antarctic Circle in 1772, carrying with
them a collection of telescopes, quadrants and chronometers, together
with two astronomers, William Wales and William Bayly (Wales \& Bayly
1777).  However the prime purpose of these instruments was for
navigation rather than astronomy.  Douglas Mawson's Australasian
Antarctic Expedition of 1911-1914 can be credited with first advancing
the science of astronomy in Antarctica, through the discovery of the
Adelie Land Meteorite.  This was found on December 5, 1912, during one
of the many sledging expeditions undertaken to chart the coastal
fringes around Commonwealth Bay in Adelie Land (Bayly \& Stillwell,
1923).  It was not until 1969, however, before Japanese scientists
found a number of meteorites of different types in close proximity and
realised the favourable conditions that Antarctica provides for their
collection -- the meteorites are transported from where they fall to
ablation zones in blue ice fields, where they can simply be picked up
off the snow (see Nagata 1975). Over 30,000 meteorites have now been
found in Antarctica, more than from the rest of the world put
together.  Meteorite research is currently of particular interest due
to the discovery of several Martian meteorites in Antarctica, and the
possibility they may contain signatures of past biotic structures from
that planet (e.g.\ McKay et al.\ 1996).

Admiral Peary, who had led the first successful expedition to the
North Pole in 1909, wrote to the Director of Yerkes Observatory, EB
Frost, in 1912\footnote{Exchange of letters between RE Peary \& EB
  Frost, held at Yerkes Observatory, Wisconsin, USA and reproduced in
  the newsletter of the Center for Astrophysical Research in
  Antarctica (CARA), Yerkes Observatory (\#5, January 1994).}
suggesting that Antarctica might provide a suitable place to consider
continuous observations of astronomical sources on account of the long
winter night.  Frost replied largely in the negative, and the subject
was not pursued further.  The first measurements to be made for
astronomical research were to study cosmic rays.  Cosmic-ray detectors
were installed at the Australian base of Mawson in 1955 (Parsons 1957)
and at the US base of McMurdo during the International Geophysical
Year of 1957 (Pomerantz, Agarwal \& Pontis 1958), the year when
Antarctic science began in earnest.  It was not until 1979 that the
first optical research programs were conducted in Antarctica, when 120
hours of continuous observation of the Sun was made using an 8\,cm
diameter heliostat at the South Pole (Grec et al.\ 1980).  A 45\,cm
sub-mm telescope was then re-deployed from the Canada-France-Hawaii
3.6\,m Telescope and used for site testing at the Pole in 1984
(Pomerantz 1985), and by the end of that decade several experiments
had been trialled for measurement of CMBR anisotropies.  A $2''$
periscope-style optical telescope, SPOT (the South Pole Optical
Telescope), was also operated at the Pole from 1984--1988, managing to
obtain a continuous light curve of 1 weeks duration of the Wolf-Rayet
star $\gamma^2$\,Velorum (Taylor 1990).

Antarctic astronomy began to flourish in earnest from 1991, following
the formation in the USA of CARA, the `Center for Astrophysical
Research in Antarctica' at the South Pole, with Doyal Harper, Director
of Yerkes Observatory (a part of the University of Chicago), as its
first Director.  CARA instigated research programs into CMBR
anisotropy, sub-millimetre and infrared astronomy, as well as
initiating a site testing program.  Australia, through the University
of New South Wales, joined this last program in 1994, and began
designing experiments to allow uninhabited high plateau sites to be
evaluated (Storey et al.\ 1996).

Construction of the Martin A Pomerantz Observatory (MAPO) began at the
Pole with CARA, and continues to grow today.  A series of telescopes
have been deployed for CMBR measurements: the 0.75\,m Python (Coble et
al.\ 1999), the 2.1\,m Viper (Peterson et al.\ 2000), the 13-element
DASI interferometer (Leitch et al.\ 2002), the 25\,cm BICEP (Takahashi
et al.\ 2010) and currently the 10\,m South Pole Telescope (SPT;
Carlstrom et al.\ 2010). CMBR experiments were also flown on
high-altitude balloons, launched from the Long-Duration Balloon
Facility at the coastal station of McMurdo (e.g.\ BOOMERanG, de
Bernadis et al.\ 2000; BLAST, Pascale et al.\ 2008).  For the
sub-millimetre, the 1.7\,m off-axis AST/RO telescope was constructed
at the South Pole, and used with a series of increasingly more
sophisticated instruments (Stark et al.\ 2001).  In the near-infrared
the 0.6\,m SPIREX telescope was deployed, equipped with cameras from
1--5$\mu$m (Hereld et al.\ 1990, Fowler et al.\ 1998).

The CMBR facilities, in particular, have produced a wealth of
discoveries (see also \S\ref{sec:cmbr}), including the first
conclusive measurements that the geometry of the Universe is flat (de
Bernardis et al.\ 2000), of the polarization of the E--modes of the
CMBR (Kovac et al.\ 2002) and of the thermal--SZ effect in galaxy
clusters (Staniszewski et al.\ 2009).

Concurrently with these developments, a number of increasingly
sophisticated high-energy experiments were being developed at the
Pole, seeking cosmic sources of gamma rays, cosmic rays and neutrinos
(see also \S\ref{sec:highenergy}). These started with GASP (Morse \&
Gaidos, 1990), a gamma-ray telescope at the South Pole, developed
through the SPASE cosmic ray air shower array (Smith et al.\ 1989) and
then led to the AMANDA neutrino detector (Andr\'{e}s et al.\ 2000). This
was a fore-runner to IceCube, which makes use of a cubic kilometre of
pure ice to search for sources of high-energy neutrino emission
(Ahrens et al.\ 2004).  At a cost of over US\$300M, IceCube is the
single most expensive scientific experiment ever to be conducted in
Antarctica. The USA has significantly upgraded the infrastructure at
the South Pole to cope with the increasingly sophisticated
astronomical experiments being deployed there.

The latest chapter in the Antarctic astronomy is taking place on the
summits of the high plateau, Domes A, C \& F\@.  It began at Concordia
station (France \& Italy) at Dome C, with the first summer time
measurements made in 1996--97 by Valenziano \& dall'Oglio (1999), of
precipitable water vapour in the atmosphere.  The first winter-time
operations at the station took place in 2005 (Fossat 2005), though the
first winter astronomical measurements were carried out in 2003 using
the AASTINO automated observatory (Storey et al.\ 2003, Lawrence,
Ashley \& Storey 2005).  Construction of Kunlun station at Dome A
began in 2009 (China), but astronomical measurements began there in
the 2008 winter season using the PLATO automated observatory (Yang et
al.\ 2009).  At Dome F site testing has been initiated, with summer
time measurements of the millimetre sky transparency and turbulence in
the lowest kilometre of the atmosphere obtained (Ichikawa, 2010).
However, analysis of an ice core drilled in 2001 at Dome F for
paleoclimate research has found spikes of enhanced nitrate ion
concentration.  These may be attributable to atmospheric ionization
caused by gamma rays produced by two historical supernovae (Motizuki
et al.\ 2010).  There is also evidence for the 11-year solar cycle in
the same data.  If these findings are substantiated, there is the
exciting prospect of extending this technique further back in time to
search for pre-historic supernovae in the ice core record.

A full account of the historical development of astronomy in
Antarctica to 2004 can be found in Indermuehle, Burton \& Maddison
(2005).

\section{Astronomical Results from Antarctica}
\label{sec:results}
Despite the relative youth of the field, and the small size of the
science community working in Antarctica, a diverse and extensive range
of astronomy has been conducted from the continent. Here we briefly
overview some of the science results that have been obtained.

\subsection{Helioseismology}
\label{sec:helio}
\begin{figure*}
\begin{center}
\includegraphics[width=1.0\textwidth]{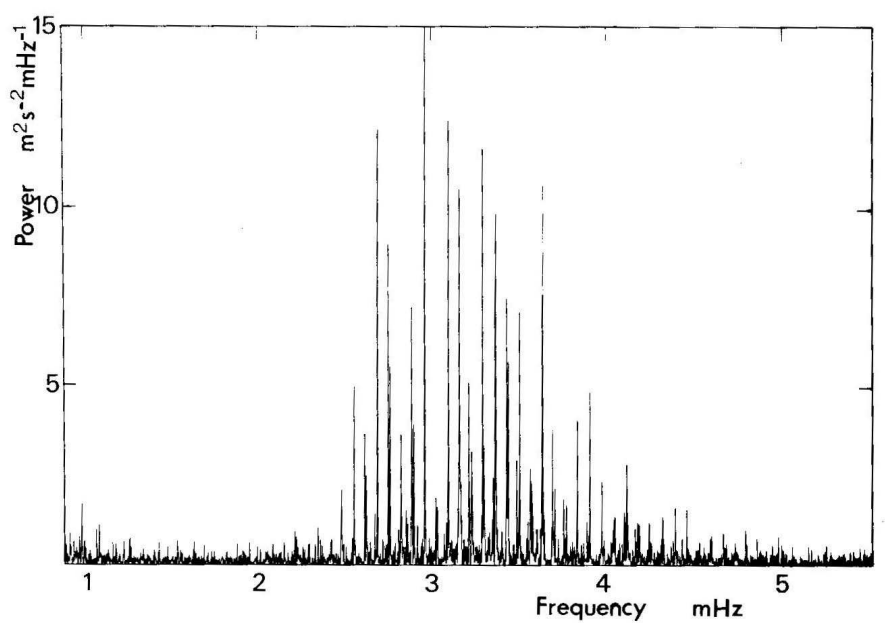}
\end{center}
\caption{A power spectrum obtained from 5 days of continuous
  observations of motions on the surface of the Sun using a sodium
  resonance cell, as made at the South Pole by Grec, Fossat \&
  Pomerantz (1980).  Over 80 harmonics are seen, with periods ranging
  from 3 to 8 minutes (i.e.\ the 5 minute solar oscillation).  The
  many lines separated by 68$\mu$Hz represent the global $p$-modes of
  oscillation of the Sun, with the smallest reliable features, seen
  around 2.4\,mHz, corresponding to oscillations with speeds of $\rm <
  10\, cm\, s^{-1}$ and displacements $< 5$\,m on the Sun's surface. }
\label{fig:helio}
\end{figure*}

The first observations for astronomical research at the South Pole
took place in 1979 in a program to study the interior of the Sun.  A
sodium resonance cell was attached to an 8\,cm telescope. An unbroken
run of measurements over an unprecedented continuous 5 days length was
obtained, of the 5,896\AA\, Na D1 line over the full disk of the Sun,
and reported in a paper to {\it Nature} (Grec, Fossat \& Pomerantz,
1980).  A clear signal of pulsation was seen, dominated by the 5
minute global solar oscillation.  An array detector was added in the
1981-82 summer, allowing features to be resolved as small as $10''$ on
the solar disk.  A full analysis (Grec, Fossat \& Pomerantz, 1983; see
Fig.~\ref{fig:helio}) found $\sim 80$ harmonics of solar eigenmodes,
with periods ranging from 3 to 8 minutes. The latitude-dependent
measurements provided evidence that the structure of the convection
zone in the Sun is different near the equator to that at higher
latitudes.  These pioneering measurements have been an important step
leading to our detailed knowledge today of the temperature,
composition and motions in the Sun's interior, helping develop the
techniques later used in spacecraft (SOHO -- the Solar and
Heliospheric Observatory; Gabriel et al.\ 1995) and with the GONG
telescope network (Global Oscillation Network Group; Kennedy et al.\
1994).

An 80\,cm solar telescope was also flown on two long duration balloon
flights, launched from McMurdo in 1996 \& 2000 -- the Flare Genesis
Experiment (Bernasconi et al.\ 1999).  Imaging of solar flares
and filament eruptions was carried out, although the image quality
obtained, of $0.5''$, was poorer than the diffraction-limited $0.1''$ aimed
for. The balloon gondola and telescope is to be re-used for the STO
THz astronomy flight in 2011 (see \S\ref{sec:thz}).

Interest in using Antarctica as a platform for Solar studies has
recently been revived by the discovery of regular periods of excellent
daytime seeing at Dome C, as well as the high clear sky fractions at
that site (see \S\ref{sec:seeing}; e.g.\ Dam\'{e} et al.\ 2010).  The
scientific focus for future Antarctic solar observations is now on
high spatial resolution imaging of the corona--chromosphere interface,
including direct measurements of magnetic fields.

\subsection{Cosmic Microwave Background Anisotropies}
\label{sec:cmbr}
Measurements of CMBR anisotropies have provided the most prominent
science to emerge from astronomy in Antarctica to date. This has been
facilitated by the extreme stability of the atmospheric microwave
emission over the Antarctic plateau, making it the pre-eminent
earth-based site for CMBR measurements. There were a series of
experiments, starting in the mid--80's (see Indermuehle et al.\
2005 for the early history), but it was not until the 1.4\,m ``White
Dish'' experiment (Tucker et al.\ 1993) that new science results
in the field emerged, placing tighter constraints on CMBR anisotropies
at high angular scales than had been determined by the COBE satellite.

\begin{figure*}
\begin{center}
\includegraphics[width=1.0\textwidth]{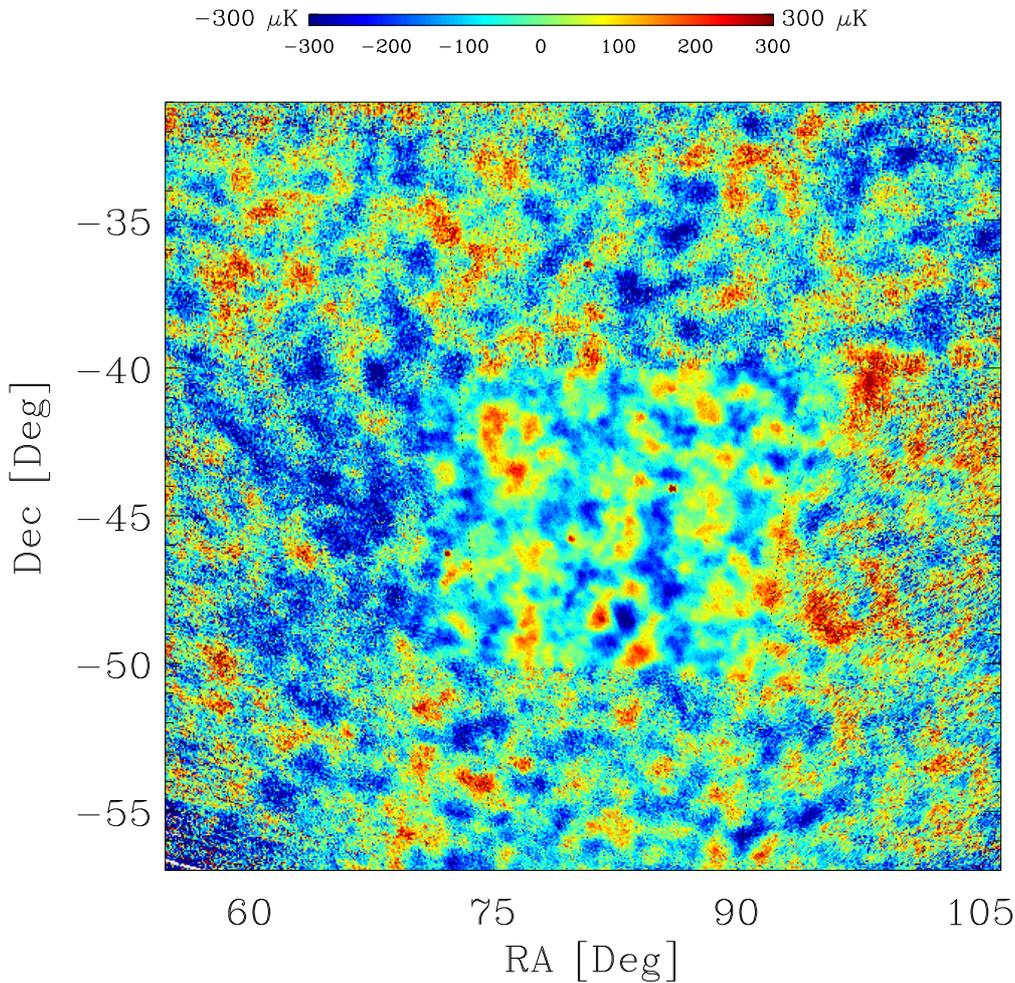}
\end{center}
\caption{Map of the CMBR fluctuations measured at 145\,GHz by the
  BOOMERanG long duration balloon experiment flown from McMurdo (de
  Bernardis et al.\ 2000; this image as reported with further data
  from Masi et al.\ 2006).  The colour scale marks fluctuations
  between $\rm \pm 300 \mu K$ about the CMBR mean temperature of
  2.725\,K.  The pixel scale of the map is $3.4'$. This data provided
  the most convincing evidence then available for the Euclidean nature
  of the Universe; i.e.\ that the geometry is flat.}
\label{fig:boomerang}
\end{figure*}

At McMurdo, the BOOMERanG experiment (Balloon Observations Of
Millimetre Extragalactic Radiation ANd Geomagnetics) carried a 1.2\,m
microwave telescope to an altitude of 38\,km on a long-duration
balloon flight in 1998.  The experiment combined broad frequency
coverage with the high sensitivity obtained from the 10 days duration
of the flight through exceptionally clear atmospheric windows.  The
data provided the best evidence at the time that the geometry of the
Universe was indeed flat (i.e.\ Euclidian; de Bernardis et
  al.\ 2000 -- see Fig.~\ref{fig:boomerang}).  A second flight in 2003
then obtained polarization maps (Masi et al.\  2003) of the
E-mode of the CMBR anisotropy fluctuations.

\begin{figure*}
\begin{center}
\includegraphics[width=1.0\textwidth]{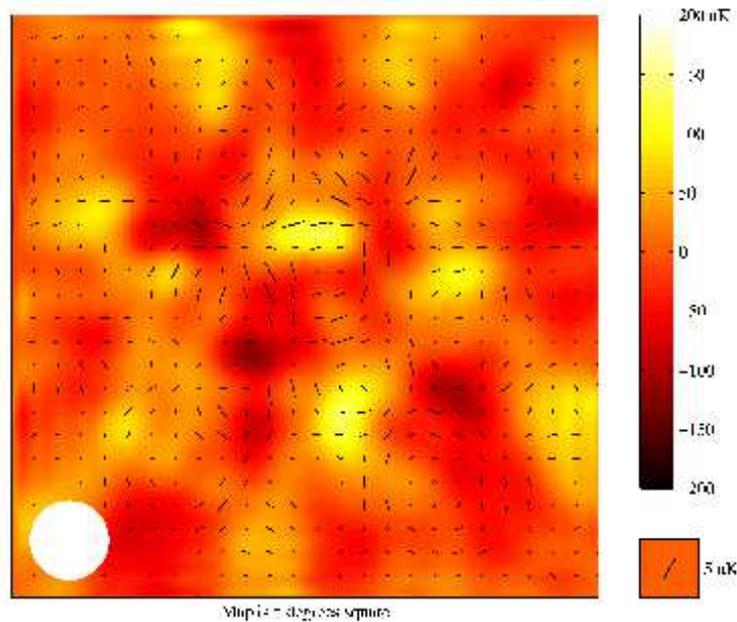}
\end{center}
\caption{The first measurement of polarization made of the CMBR,
  obtained by the DASI experiment at the South Pole (Kovac et
    al.\ 2002). The figure shows the E--mode polarization vectors
  overlaid on an image showing the CMBR fluctuations over the $5^{\circ}$
  field of view. The beam size is shown by the white circle.}
\label{fig:dasi}
\end{figure*}

DASI (Degree Angular Scale Interferometer) was a 13-element
interferometric array built at the South Pole designed to extend the
angular coverage of the CMBR to values in the range $140 < l < 900$.
It made the first detection of polarization in the CMBR (Kovac et al.\
2002 -- see Fig.~\ref{fig:dasi}), obtaining a $5 \sigma$ detection of
the ``E--mode'' of the CMBR polarization.  The QUaD experiment, which
combines a 31-element bolometer array with DASI, has now extended this
imaging of the E--mode polarization to over $\sim 800$ square degrees, at
an angular resolution of 5 arcminutes (Castro et al.\ 2009;
Culverhouse et al.\ 2010).

With the 16-pixel ACBAR array receiver on the 2.1\,m Viper telescope
at the South Pole, Kuo et al.\ (2004) obtained the highest
signal-to-noise measurements then made of CMBR anisotropy at high
angular scales ($l = 100-3,000$).  Combined with data from other CMBR
experiments (in particular the WMAP satellite), this led to the best
estimates then available of important cosmological parameters such as
the Hubble constant, the age of the Universe, and the contributions of
matter \& dark energy to the overall composition of the Universe (see
Spergel et al.\ 2003).

BICEP is a 25\,cm telescope currently operating at the South Pole.
Using a 49-element bolometer array, it is designed to probe the
polarization of the CMBR on degree angular scales, to search for the
signature of ``B--mode'' gravitational waves. These are produced
during the epoch of inflation.  After 2 years of operation BICEP has
probed for B--modes an order of magnitude deeper than any previous
experiment (though they are yet to be detected; see Chiang et al.\
2010), as well as making the first detection of the first peak in the
angular power spectrum of the E--mode polarization.

\begin{figure*}
\begin{center}
\includegraphics[width=1.0\textwidth]{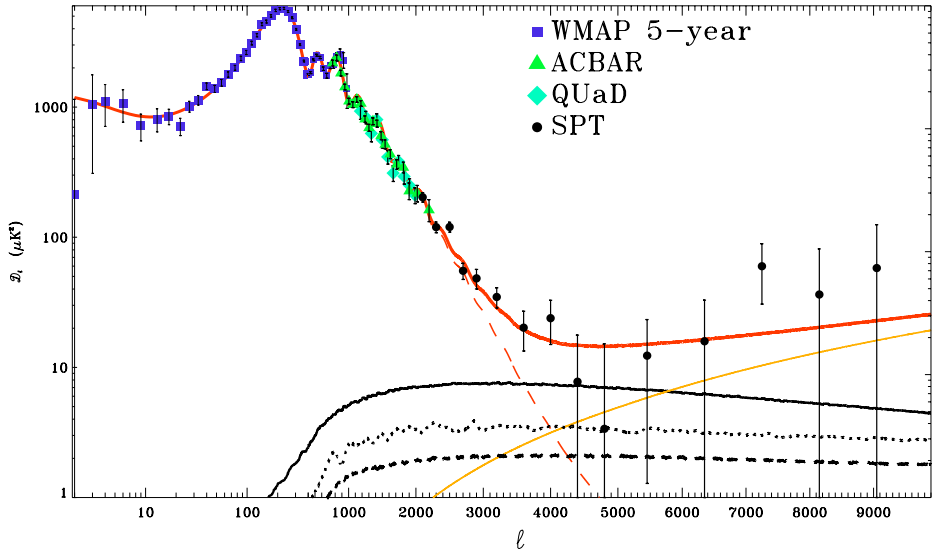}
\end{center}
\caption{Angular power spectrum of the temperature anisotropies in the
  CMBR, measured out to an angular scale of $l = 9,500$, as shown in
  Lueker et al.\ (2010).  The power spectrum incorporates results from
  the WMAP satellite at small values of $l$, from the ACBAR \& QUaD
  experiments at the South Pole on intermediate scales, and from the
  SPT results on the largest scales.  For small values of $l$ the data
  is consistent with predictions from $\Lambda$CDM cosmology
  (red-dashed line), but at high $l$ there is a clear deviation from
  this model.  A combination of contributions from the thermal
  SZ-effect (solid black line), the kinematic SZ--effect (dashed black
  line; note the dotted black line also represent the kSZ-effect, from
  `patchy' re-ionization models) and from point sources (orange line)
  gives the red curve that provides the best fit to the data.  This
  last component is presumed to arise from dusty, star-forming
  galaxies. }
\label{fig:cmbrpower}
\end{figure*}

The largest telescope to be so far operated in Antarctica is the 10\,m
off-axis South Pole Telescope (SPT), employing a bolometer camera with
nearly 1,000 pixels (Carlstrom et al.\ 2010) and working at three
frequencies (90, 150 \& 220\,GHz).  Its key projects are to detect
clusters of galaxies via the Sunyaev-Zeldovich (SZ) effect and to
measure the high-$l$ angular power spectrum of the CMBR\@, over a
$\sim1,000$ square degree area of sky. The first results from SPT
have been published.  Images of the thermal-SZ effect in several
clusters are shown in Staniszewski et al.\ (2009) and Plagge et al.\
(2010).  The power spectrum of the temperature anisotropies, extended
out to an angular scale of $l = 9,500$ is shown in
Fig.~\ref{fig:cmbrpower} from Lueker et al.\ (2010).  This power
spectrum incorporates the results from the WMAP satellite at small
values of $l$, from the ACBAR \& QUaD experiments at the South Pole on
intermediate scales, as well as the SPT results on the largest scales.
While for small values of $l$ the spectrum is consistent with that
predicted by $\Lambda$CDM cosmology, at high $l$ there is clear
evidence for a contribution from point sources, presumed to be dusty,
star-forming galaxies, now seen for the first time in the angular
power spectrum.

\subsection{Sub-millimetre Astronomy}
\label{sec:submm}
AST/RO, the Antarctic Submillimeter Telescope and Remote Observatory,
was a 1.7\,m telescope deployed at the South Pole, in almost constant
use from 1995--2005 (Stark et al.\ 2001).  It is the most productive
astronomical facility to have operated in Antarctica, if judged by the
number of papers produced, with over 50 publications.  This is despite
its small size, for it was the only telescope in the world then able
to access the 350$\mu$m window on a regular basis. It was a general
purpose facility, able to be used with a range of instruments between
0.2--2\,mm, for both astronomical and aeronomy purposes. AST/RO's
success was largely attributable to its off-axis design, with the
instrumentation placed on an optical table at a warm coud\'{e} focus
where it could be readily worked on in comfort by the instrument
teams.  A variety of bolometer and heterodyne systems were used over
the lifetime of the facility.

\begin{figure*}
\begin{center}
\includegraphics[width=1.0\textwidth]{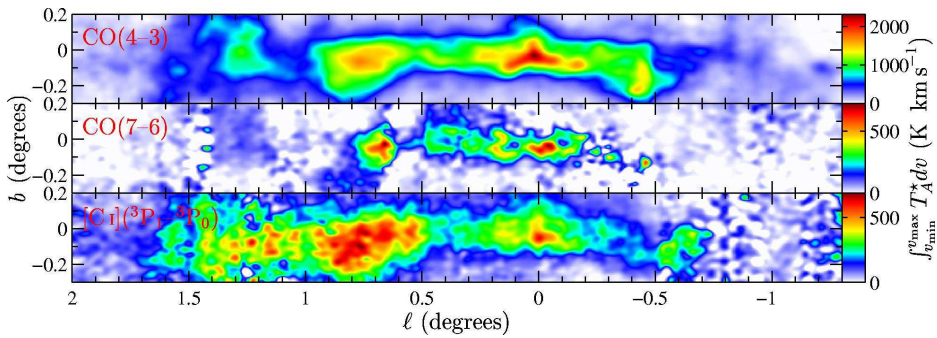}
\end{center}
\caption{Maps obtained in the sub-millimetre by the 1.7\,m AST/RO
  telescope at the South Pole showing the `Central Molecular Zone' of
  our Galaxy, a $3^{\circ} \times 0.5^{\circ}$ region across the
  galactic centre containing $\sim 10$\% of the Galaxy's molecular gas
  (Martin et al.\ 2004).  The three maps show the distribution of CO
  J=4--3, CO J=7--6 and [CI] $\rm ^3P_1-^3P_0$ line emission, at 372,
  651 \& 370$\mu$m respectively.  They have $\sim 1'$ spatial
  resolution.}
\label{fig:cmz}
\end{figure*}

AST/RO was principally used to measure emission lines of atomic carbon
(1--0 and 2--1 lines) and carbon monoxide (CO J=4--3 and J=7--6) emitted
from molecular clouds in the Milky Way and the Large Magellanic
Clouds.  These are among the strongest cooling lines from the dense
interstellar medium.  Results included the first detection of [CI]
emission from the Magellanic Clouds (Stark et al.\ 1997), the first
large-scale maps of [CI] and warm CO from galactic star forming
complexes such as Carina (Zhang et al.\ 2001), NGC 6334 (Kim et al.\
2006) and Rho Ophiuchi (Kulesa et al.\ 2005).  The most ambitious
project was to map the [CI] and warm CO emission from the Central
Molecular Zone (the inner $3^{\circ}$ of our Galaxy; Martin et al.\
2004; see Fig.~\ref{fig:cmz}).  The first ground-based spectrum of the
[NII] 205$\mu$m line was also obtained, emitted from the Carina Nebula
(Oberst et al.\ 2006).  This is in the THz spectral region, and shows
that, while sites like Dome A are being considered for future THz
telescopes, even at the South Pole this window opens at times.

Sub-millimetre polarization measurements were also made using the
SPARO polarimeter on the 2.1\,m Viper Telescope, of the dichroic
emission at 450$\mu$m from aligned, cold dust grains.  This included
the detection of a large-scale toroidal field running through the
Galactic centre (Novak et al.\ 2003), and of magnetic fields in four
other giant molecular clouds in the Galaxy (Li et al.\ 2006).  A
statistically significant correlation was found between the magnetic
field direction and the orientation of the Galactic plane, suggesting
that preservation of the magnetic field direction occurs during the
gravitational collapse to produce stars that follows from molecular
cloud formation.

\begin{figure*}
\begin{center}
\includegraphics[angle=11,width=1.0\textwidth]{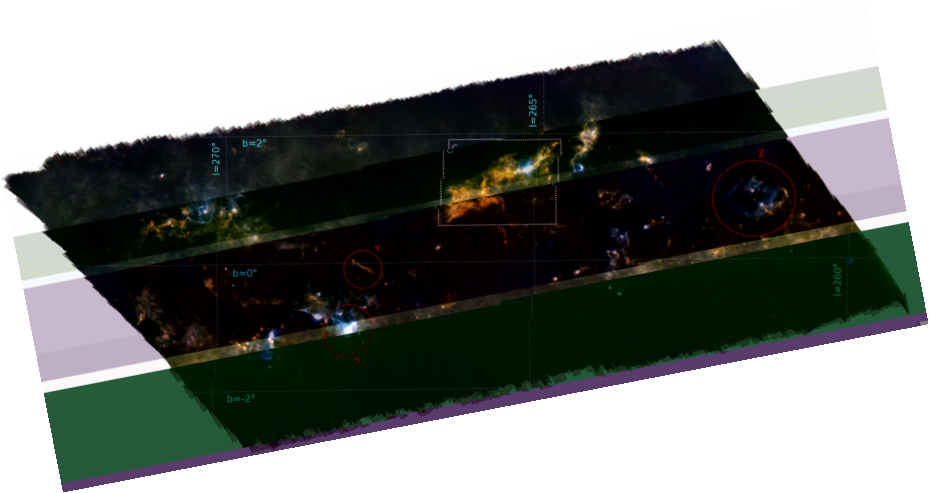}
\end{center}
\caption{A sub-millimetre image of Vela Molecular Ridge obtained by
  the balloon-borne BLAST telescope (Netterfield et al.\ 2009),
  launched from the McMurdo Long Duration Balloon Facility.  The image
  shows the complete range of dust cores within the molecular cloud
  complex, from starless cores to those actively forming stars.  The
  three colours represent emission at 250$\mu$m (blue), 350$\mu$m
  (green) and 500$\mu$m (red).  Blue regions are therefore the warmest
  gas, typically $\ge 25$\,K and red regions the coldest gas ($\le
  13$\,K). The image covers 50 square degrees and the
  horizontal lines lie along the Galactic plane, each
  separated by $b=2^{\circ}$. The angular resolution is $1'$.
}
\label{fig:vela}
\end{figure*}

Sub-millimetre astronomy has also been undertaken in Antarctica on a
long duration balloon flight launched from McMurdo.  BLAST, the
Balloon-borne Large Aperture Submillimeter Telescope, used a 2\,m
telescope and bolometer array to survey many square degrees of sky at
250, 350 \& 500$\mu$m (Pascale et al.\ 2008).  Following a flight from
Sweden in 2005, the dramatic BLAST Antarctic circumpolar flight lasted
11 days in December 2006, as has been spectacularly documented in the
movie of the same name.  Science programs examined galaxy evolution
and the cosmic infrared background, as well as cold dust cores
associated with earliest stages of star formation in molecular clouds
in our Galaxy. Through measurement of over 500 galaxies from a 9
square degree region, which included the GOODS-S deep field, the
sub-millimetre flux-source count relation was determined.  From this
the cosmic far-infrared background was inferred to be dominated by the
emission from individual galaxies, 70\% of the flux coming from known
galaxies at $z \ge 1.2$ (Devlin et al.\ 2009).  In our Galaxy, $\rm
\sim 50$ square degrees of the Vela Molecular Ridge, 700\,pc distant,
was mapped to provide a complete and unbiased sample of the dust cores
within it, and covering all evolutionary stages in the route to star
formation (Netterfield et al.\ 2009, Olmi et al.\ 2009; see
Fig.~\ref{fig:vela}).  Temperatures, luminosities and masses were
determined for over 1,000 cores, so providing the core mass function.
Around 2\% of the mass was found to reside in cores colder than
14\,K\@, and from this the long inferred lifetimes imply the necessity
for non-thermal support mechanisms (against gravitational collapse) to
operate at the very earliest stages of star formation.  A polarization
module is to be added to BLAST for a future flight (Marsden et al.\
2008).  This will enable the relationship between the magnetic field
geometry, and its strength in the cloud cores relative to that in the
large scale cloud, to be determined.

\subsection{Infrared Astronomy}
\label{sec:irastro}
The 60\,cm SPIREX Telescope (South Pole InfraRed EXplorer) operated at
the South Pole from 1994 to 1999, the first four years with the GRIM
1-2.5$\mu$m HgCdTe array camera (Hereld 1994), the last two with the
Abu 2.4-5$\mu$m InSb array (Fowler et al.\ 1998). SPIREX was installed
just before Comet Shoemaker-Levy 9 struck Jupiter, and was the only
telescope in the world with the opportunity of viewing all the
impacts.  Observation of some of these was affected by poor weather.
Nevertheless, SPIREX recorded 16 of the 21 events over the week of the
encounter, in July 1994 (Severson 2000), more than any other telescope
achieved.

\begin{figure*}
\begin{center}
\includegraphics[width=1.0\textwidth]{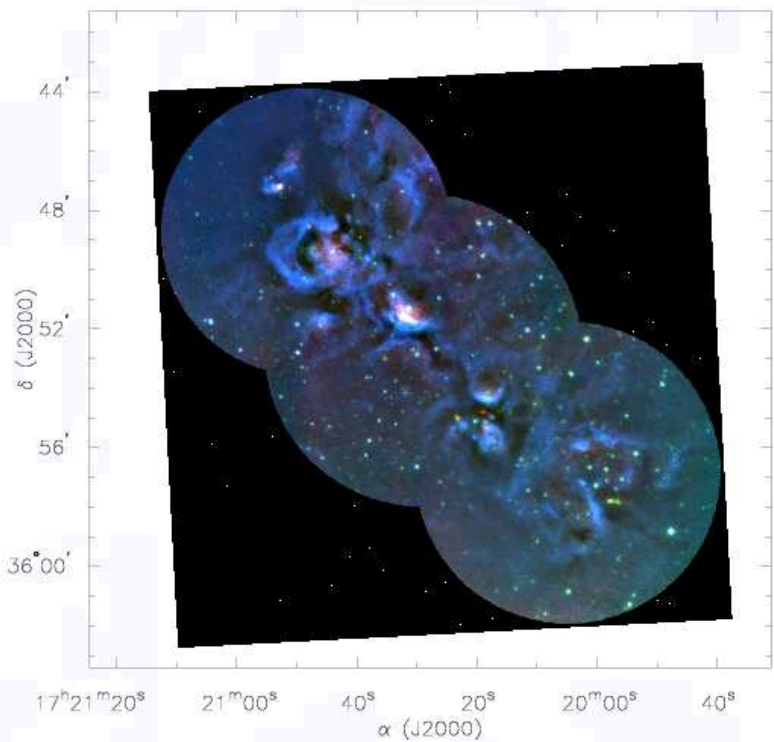}
\end{center}
\caption{An image showing the polyaromatic hydrocarbon (PAH) emission
  at 3.3$\mu$m that envelopes the NGC 6334 star forming complex, as
  imaged by the infrared SPIREX telescope at the South Pole (Burton et
  al.\ 2000). Dark dust lanes (black) and deeply embedded massive
  protostars (red) are evident, surrounded by shells of PAHs (blue),
  excited by the far--UV radiation from the most massive young stars,
  each separated by $\sim 1$~pc along the dust ridge. The image covers
  a $30'$ region, with a spatial resolution of $1.5''$.  The three
  colours represent 3.3$\mu$m PAH (blue), 3.5$\mu$m L--band continuum
  (green) and a 1\% narrow band filter at the wavelength of the
  4.05$\mu$m Br$\alpha$ line. }
\label{fig:6334}
\end{figure*}

SPIREX's principal achievements were in studying the extensive shells
of polycyclic aromatic hydrocarbons, emitting at 3.3$\mu$m, that are
illuminated by far--UV radiation from massive young stars, and in
detecting IR excesses from hot dust in disks around pre-main sequence
stars. PAHs emission was imaged around the galactic star forming
complexes of NGC~6334 (Burton et al.\ 2000; see
Fig~\ref{fig:6334}) and Carina (Brooks et al.\ 2000, Rathborne
et al.\ 2002), revealing the workings of the galactic ecology in
these objects. Thermal-IR excesses, determined from their
[2.2$\mu$m]$-$[3.5$\mu$m] colour, were measured in the low mass star
forming regions of Chamaeleon I (Kenyon \& Gomez 2001) and $\eta$
Chaemaeleontis (Lyo et al.\ 2003), and in the high mass star
forming regions of 30 Doradus in the LMC (Maercker \& Burton 2005) and
RCW57/NGC3576 (Maercker, Burton \& Wright 2006).  A high disk
fraction was found in all cases, including, interestingly, for massive
as well as for low mass stars.

SPIREX demonstrated that the projected sensitivities for the IR, based
on the measured sky background drop, could indeed be achieved. A
summary of all the science results obtained with SPIREX is given by
Rathborne \& Burton (2005). At the time of the observations (1998),
the measurements of 30 Doradus made with SPIREX were the deepest
imaging that had been obtained at 3.5$\mu$m, achieving a sensitivity
of 14.5\,mags.\ in 9 hours of on-source integration.  However, the
small size of the facility (especially in comparison to 8\,m-class
facilities now available), combined with the depth of the surface
boundary layer at the South Pole (necessitating the use of wavefront
correction systems to obtain the free-air seeing), has stalled the
development of IR astronomy in Antarctica since SPIREX was
de-commissioned.

With the opening of Concordia Station at Dome C and the construction
started of Kunlun station at Dome A, both with their thinner surface
boundary layers than South Pole, infrared astronomy may now move
forward again in Antarctica.  An 80\,cm mid-IR telescope, IRAIT (the
International Robotic Antarctic Infrared Telescope), has been built by
the University of Perugia (Busso et al.\ 2010).  It has been equipped
with the 2--28$\mu$m AMICA imager (Dolci et al.\ 2010) and is
currently being installed at Dome C\@.

\subsection{Optical Astronomy}
\label{sec:optical}
A limited number of optical astronomy experiments have been conducted
in Antarctica, generally as part of site testing programs.  As
described in \S\ref{sec:history} the 5\,cm SPOT telescope was used at
the South Pole to obtain 1 continuous week of data monitoring the
light curve of the Wolf-Rayet star $\gamma^2$\,Velorum (Taylor 1990).
In \S\ref{sec:other} the ASTEP (10\,cm telescope; Crouzet et al.\
2010) and sIRAIT (25\,cm telescope; Strassmeier et al.\ 2008)
experiments at Dome C were described, as well as the 14.5\,cm CSTAR
telescope at Dome A (Zou et al.\ 2010).  These latter three
experiments undertook a variety of sky brightness, transparency and
photometric monitoring observations, in the case of CSTAR also
resulting in a catalogue of some 10,000 stars in a field centred on
the South Celestial Pole (Zhou et al.\ 2010).

\subsection{High-Energy Astrophysics}
\label{sec:highenergy}
High-energy astrophysics in Antarctica began with the installation of
two muon telescopes for cosmic ray detection at Mawson Station in
1955.  The Mawson observatory later contributed to the experimental
verification of the spiral nature of the solar magnetic field long
before direct measurements could be made by satellites (McCracken
1962). Cosmic ray research is today pursued at a number of locations
around Antarctica, most notably Mawson, McMurdo and the South Pole
(see Duldig 2002).  Equipment for its conduct includes cosmic-ray
detectors, neutron monitors and muon telescopes.

The GASP gamma ray telescope at the South Pole sought to find cosmic
sources of gamma rays via the Cherenkov light produced by cosmic rays
accelerated by the interaction of gamma rays with nuclei in the
atmosphere (Morse \& Gaidos, 1990).  This search is assisted at the
South Pole by the long polar night and constant source zenith angle.
The experiment was not successful, however, in finding any gamma ray
sources.  It led to the construction of the SPASE air shower array at
the South Pole, increasing the effective collecting area for detecting
such events. SPASE, the South Pole Air Shower Experiment (Smith et
al.\ 1989), also failed to find sources of gamma rays, with the
particle events recorded showing an isotropic distribution across the
sky (van Stekelenborg et al.\ 1993). This was followed by SPASE-2, an
enhanced array (Dickinson et al.\ 2000), which was built on the
ice-surface above the AMANDA neutrino array (see below) so that it
could also work in conjunction with it.  The objective was to measure
the particle compositions in air showers in the TeV range (Ahrens et
al.\ 2004).  Though SPASE-2 again failed to find gamma ray sources,
this detection technique has now matured with the success of the HESS
gamma ray telescope in Namibia, through stereoscopic imaging of the
air showers in the atmosphere (e.g.\ Aharonian et al.\ 2004).

AMANDA, the Antarctic Muon And Neutrino Detector Array (Andr\'{e}s et
al.\ 2000, 2001), was the first experiment at the South Pole to
search for cosmic sources of neutrinos. The technique uses
photomultiplier tubes (PMTs) placed into holes drilled in the ice,
extending from several hundred metres to three kilometres deep.  The
PMTs detect Cherenkov radiation resulting from the exceedingly rare
encounters of neutrinos with ice or rock nuclei.  The PMTs point
downwards to shield the detectors from the vastly greater fluxes from
downward-travelling muons (produced by the interaction of cosmic rays
in the atmosphere).  Hence the array serves to detect
upward-travelling neutrinos that have passed through the Earth,
entering it from the northern hemisphere.  With its extension,
AMANDA-II, over 600 hundred neutrinos were detected, though no
statistically significant cosmic sources of neutrinos were found in
the data set (Ackermann et al.\ 2005).

\begin{figure*}
\begin{center}
\includegraphics[width=1.0\textwidth]{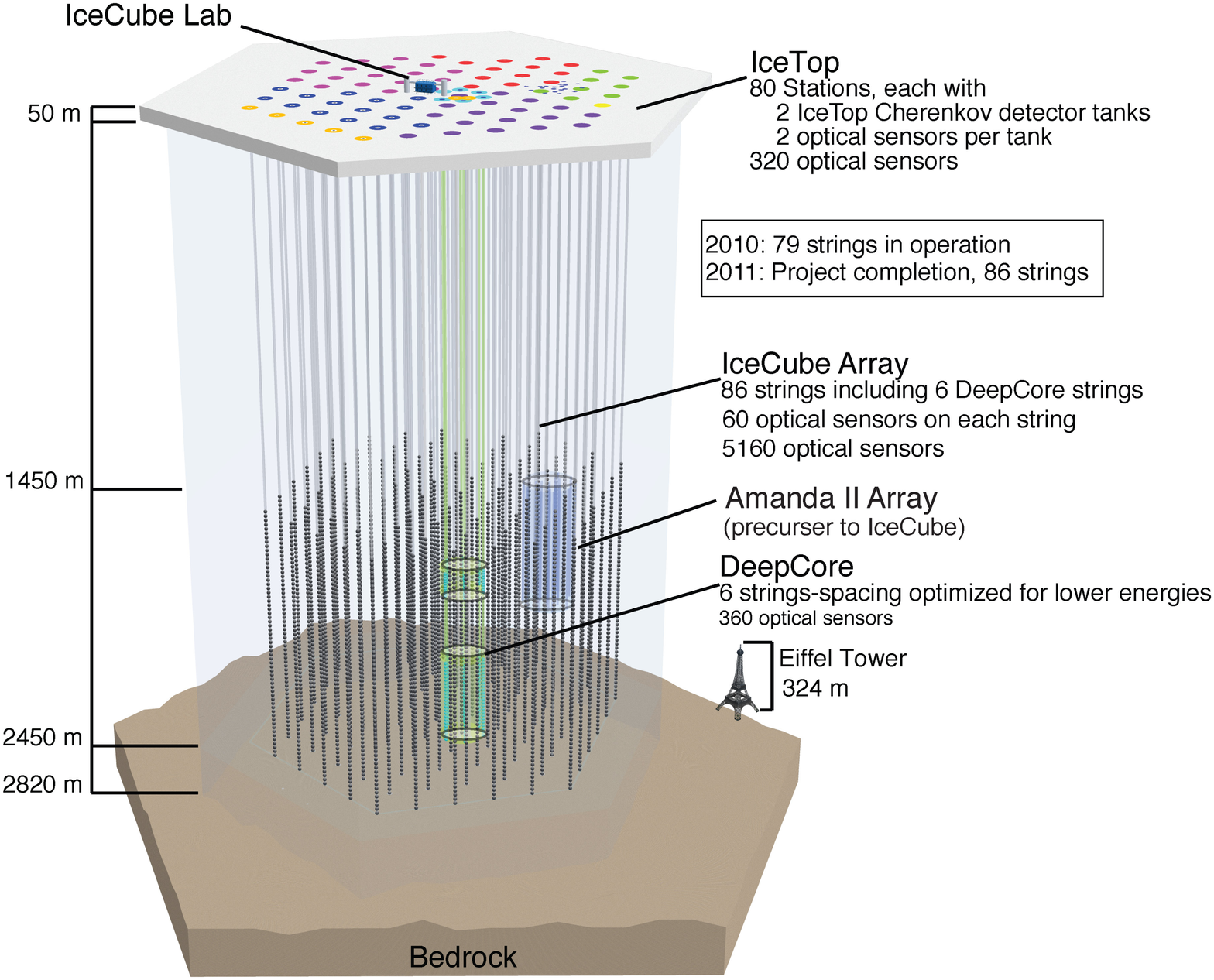}
\end{center}
\vspace*{-3cm}
\caption{A sketch illustrating the construction of the IceCube
  neutrino observatory at the South Pole, together with the Eiffel
  Tower to provide a size comparison.  80 strings, each 1 km long, are
  buried from 1.5-2.5 km deep in the ice (reaching close to the
  bedrock beneath the ice), together with a closer spaced core at the
  centre.  60 digital optical modules are strung along each string. At
  the ice surface is the IceTop experiment, together with the
  observatory laboratory.  Cherenkov radiation from muons generated by
  the rare neutrino interactions with nuclei in the ice creates a cone
  of light that can be detected by the optical modules lying within
  its path, so allowing the direction of the exciting neutrino to be
  inferred.  Credit: IceCube Consortium (icecube.wisc.edu).}
\label{fig:icecube}
\end{figure*}

AMANDA served as the prototype for IceCube, the largest science
experiment yet to take place in Antarctica (Ahrens et al.\ 2004).
Almost complete (see Fig.~\ref{fig:icecube}), IceCube uses a cubic
kilometre of ice as its detector, the volume calculated as necessary
for it to operate as a true telescope, able to image sources of cosmic
neutrinos. When completed 80 strings, each containing 60 digital
optical modules, will have been suspended in the ice. Each string
extends from 1.5 to 2.5\,km beneath the surface, and they are spread
out over a square kilometre, creating a cubic kilometre of collecting
volume.  The array is optimised for detecting high-energy neutrinos,
$\rm E_{\nu} > \sim TeV$, with a closer-spaced core for lower energy
neutrinos. Over the first 6 months of operations $\sim 7,000$
upward-travelling neutrinos were recorded from the first 40 strings
deployed (Filimonov et al.\ 2010). No statistically significant
sources of neutrinos can yet be discerned in this data set.

Another experiment at the South Pole is also seeking to find
ultra-high energy ($\rm E_{\nu} > \sim PeV$) neutrinos -- RICE, the
Radio Ice Cherenkov Experiment -- through measuring coherent pulses of
radio Cherenkov emission produced from neutrino-induced cascades
passing through the transparent ice (the Askaryan effect; Askaryan
1965). The radio receivers used for this have been placed in some of
the AMANDA boreholes.  The status of the experiment, and the neutrino
limits obtained, are discussed further in Kravchenko et al.\ (2008).

The final neutrino experiment underway in Antarctic is ANITA, the
Antarctic Impulsive Transient Antenna (Barwick et al.\ 2003).  This is
flown $\sim 35$\,km above the continent on a long-duration balloon
flight from McMurdo Station, and seeks to measure coherent radio
pulses from EeV ($10^{18}$\,eV) neutrino interactions.  These are
produced by a compact plug of relativistic particles, just a few
centimetres in diameter, but the experiment makes use of the elevation
of the balloon to search for these over a $\sim 1,000$\,km extent of
the Antarctic ice sheet -- an effective collecting area of one million
square kilometres!  No neutrinos were detected from either the first
35 day flight in 2006-2007 (Gorham et al.\ 2009) or from a second,
more sensitive, 31 day flight in 2009 (Gorham et al.\ 2010), but
strong limits have been set for the flux of cosmic neutrinos in the
$10^{18-21}$\,eV range.

As described in \S\ref{sec:history}, an ice core drilled at Dome F has
been used to infer that gamma rays produced by two historical
supernovae also produced ionization events in the Earth's atmosphere
(Motizuki et al.\, 2010).

\section{Conducting Science in Antarctica}
\label{sec:conduct}
\subsection{Life for the Antarctic Scientist} 
\label{sec:life}
In Antarctica logistics determines what it is possible to do.  Without
pre-existing infrastructure and support capability, conducting
frontier science is impossible.  While it may still be possible for an
adventurer to forge new routes across the Trans Antarctic Mountains,
the age of the Antarctic hero is long over.  The exploits of Amundsen,
Scott, Shackleton, Shirase and Mawson may have inspired a fascination
with Antarctica, but they do not provide a model for today's explorer
in the continent.  For the modern explorer the challenges lie
elsewhere, in making sophisticated instrumentation work in conditions
very different to the laboratory back home.  Life in a modern
Antarctic station is certainly no luxury experience, but it is not
arduous either.  One might have to share rooms for sleeping, limit
showers to two minutes, twice a week, and deal with high-altitude
acclimatisation, but these merely serve to make a person slightly
uncomfortable.  Dealing with continuous daylight in summer can be more
of a problem, but this is only a matter of upsetting circadian
rhythms.

Antarctic scientists working on a modern station do not need to worry
about cooking their meals, shopping for food or commuting to work.
The essentials of daily life are all provided.  This is a necessity of
safe working practice in Antarctica.  For, despite modern
conveniences, the dangers of Antarctica remain real, even if they are
not the constant threat the early explorers faced.  Nations conducting
Antarctic science go to great lengths to provide facilities that are
safe and practical for their inhabitants.  For every scientist
present, four or five people are there to support them, and keep them
alive and happy.

The Antarctic scientist is left to concentrate on getting their
experiment working.  This is not to say that life is without its
concerns.  In summer a base is generally overcrowded.  Work space is
limited and may need to be regularly negotiated for. Daily timetables
can be subservient to the operational needs of the station.  There is
little time for leisure, and waking hours are consumed by the
requirements of work.  With constant day light and the extremely dry
air, sleeping can be difficult.  A person can be in a state of permanent
stress.  It takes five minutes to get dressed every time one goes
outside.  Antarctica, and your immediate surroundings, become the real
world.  Life back home seems like a dream.  Indeed, for an increasing
number of people who winter-over from year to year, it does become a
dream!

Wintering is another matter and the experience can only properly be
described by someone who has done so (the author has not).  Six months
without the Sun at the Pole is a very long time, though it is not
actually dark for all that time.  Twilight lasts for a month.  When
the Moon is up it is easy to see across the ice.  But the cold is much
harder to endure in winter than it is in summer.  Skiways are not
maintained, and the daily walk to work over the snow can become a feat
of endurance.  The only company is that of the other winter-overers.
They are the only people you will see for eight months.  While this
can be a source of intense friendships, it can also lead to strong
inter-personal problems, for it is impossible to avoid a person's
company.  Many do find themselves counting out the ``days'' to the
first flight out in the Spring.

However, in one important way life has become much easier for the
winter-overer, and that is through communications.  No-one is truly
alone anymore through the winter.  The daily email torrent does not stop.  The
internet is there to browse, news from around the world continues to
arrive, phone calls can be made. Principal investigators ring in to
discuss a problem with an experiment.  Webcams let people back home
keep tabs on both experiment and experimenter.  Conference calls and
internet video conferencing are arranged among members of the research
team, scattered around the globe.  Science can be done and papers
written.  In extreme circumstances it is even possible to get out of
Antarctica in a medical emergency.

This is not to say that winter-overers don't go through periods of
depression.  In the heart of winter some still become hermits, locking
themselves away in their rooms, watching videos for much of the day,
neglecting their work.  But this happens less than it did.  The
converse is also true.  Some scientists are so involved with their
work, and in their daily correspondence with colleagues, that they
seem to slot straight back into their previous occupation when they
get off the ``Ice''.  Yet it is also true to say that for most
winter-overers it is the transition back to normal life that is the
hardest aspect to deal with, not the winter itself.

A final observation about working in Antarctica.  It is not as
expensive as might have initially been expected, at least for
small-scale experiments.  When a scientist arrives in Antarctica
everything needed to support their daily life has already been paid
for.  For the scientist, their job is to bring the working experiment
to the appropriate mainland departure point.  From there, everything
else is done by the Antarctic programs of the nation(s) they are
working with, to bring them to their destination and then support them
there.  Scientists do not pay for their deployment or living costs in
Antarctica.  The situation is akin to the user of a space telescope --
they do not have to factor in the launch costs of the spacecraft or
the operation of mission control when planning their experiment.
These costs are subsumed within the budgets of national programs.
Essentially a decision has already been made by nations to support
personnel in Antarctica and so this is budgeted for.  The processes
which determine what experiments are to be supported also determine
the available personnel slots that the national Antarctic program will
support.  If your program is good enough to be rated for deployment,
that deployment is already paid for.  You simply have to find the
funds to have your medicals, get yourself to the point of departure,
and of course build your experiment.

\subsection{Experimental Challenges}
\label{sec:challenges}
While technology makes working in Antarctica increasingly easy, it
never will be the same as in the lab back home.  Careful planning
ahead is essential.  Never try to commission an instrument for the
first time in Antarctica, always test it before deployment.  This is,
of course, wise practice anywhere, but essential in Antarctica.
Though, given the many competing demands most scientists face at their
home institutions, this is easier said than done.  It is not a total
disaster, however, if something breaks.  Stations generally have
extensive supplies of parts. Station personnel are inventive when
called upon to find something they don't have.  Machine workshops can
often be used to re-make a broken part.  Even if not, it is still
possible to order a replacement, at least in summer.  At the South
Pole items can generally be flown in within the week.

As discussed further by McGrath et al.\ (2008), unique
operational aspects of an Antarctic observatory arise from its
remoteness, the polar environment and the unusual observing cycle
afforded by long periods of darkness and daylight.  A telescope must
be designed for remote observing via satellite communications, and
must overcome both limited physical access and data transfer rates.
Commissioning and lifetime operations must deal with extended
logistics chains, continual wintertime darkness, extremely low
temperatures and frost accumulation.

One of the most difficult issues for many astronomical applications
has been the supply and maintenance of cryogens, especially liquid
helium, which must be flown in from the mainland. Even with just a 1\%
evaporation loss per day, by mid-winter a helium supply would be
exhausted, so halting sub-millimetre and CMBR experiments, for
instance. Fortunately, with the continued development of closed-cycle
systems for instruments, this should become a problem of the past.

The design challenges caused by the extreme cold are dominated by the
continued performance of lubricants, the mechanical clearances that
may change due to thermally induced dimensional changes, and the
operation of electronics designed for a room temperature environment.
However, all of these issues are quite readily overcome with proper
mechanical design, so long as it takes account of the
temperature requirements.  The high rates of change of temperature
that can take place within the boundary layer can be more problematic,
especially for components with high thermal mass and tight thermal
equilibrium requirements -- notably, the primary mirror of a
telescope.

\begin{figure*}
\begin{center}
\includegraphics[width=1.0\textwidth]{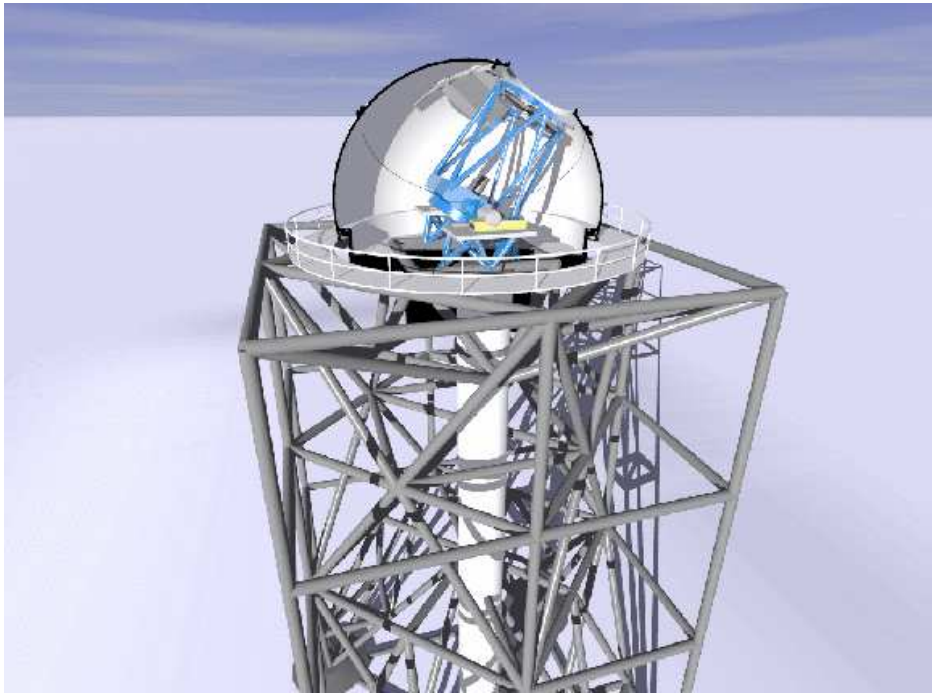}
\end{center}
\caption{A concept design for the 2.4\,m PILOT optical--IR telescope
  (Saunders et al.\ 2008a).  The design reflects a temperature and
  humidity controlled callote-style enclosure for the telescope,
  placed on top of a stiff tower that raises it above the surface
  boundary layer.  Cold air from close to the ice surface is drawn
  into the enclosure and warmed to the ambient temperature at
  telescope level, thereby lowering its humidity below the frost
  point.  Careful venting of the air to the atmosphere maintains the
  superb free air seeing. Image by Andrew McGrath, Anglo Australian
  Observatory.}
\label{fig:pilot}
\end{figure*}

In addition to the debilitating effects on image quality caused by
turbulence, the surface layer also poses an engineering challenge for
two further reasons.  The first is the large vertical temperature
gradient that can occur within it (up to $\sim 1^{\circ}$/m at surface
level and even $\sim 0.15^{\circ}$/m at 30\,m at Dome C), especially
on the most stable days in winter.  The second is the super-saturated
humidity within the layer, which readily leads to icing on exposed
surfaces.  To solve both these problems, a temperature and
humidity-controlled enclosure for a telescope may be required (e.g.\
see Saunders et al.\ 2008a).  Such an enclosure (e.g. see
Fig.~\ref{fig:pilot}) could be continuously flushed with sub-saturated
air, matched in temperature to the external air at the dome aperture.
This air is drawn from closer to the surface of the snow (where it is
colder, and so drier).  It could be heated using waste heat from the
instrumentation, resulting in its humidity falling below the
saturation point.  In addition, such a scheme could deliver excellent
dome seeing, as the temperatures could be closely matched and venting
conducted so that the external airflow suffers minimal disruption.

There is also a need to consider how to provide electrical power for
any significant astronomical facility, which might require several
tens of kW.  This might not readily be expected to be drawn from a
station diesel power supply.  However solar energy (in summer) and
wind power (the turbine mounted on a tower to place it above the
boundary layer, where a steady wind blows) could provide a clean
energy source for observatories. As discussed by McGrath et al.\
(2008), this only imposes a modest addition to the total cost of a
facility.  A clean energy source may also be necessary to gain public
support for a new facility, so minimising pollution and maintaining
the pristine condition of the environment.  In particular, it would
minimise contributions from locally produced aerosols (e.g.\ from the
exhausts of diesel generators), important for maintaining the
stability of the local sky emission.

\subsection{Antarctic Politics}
\label{sec:politics}
Antarctic politics is unique, truly in a world of its own.  While on a
national level local issues may determine how a country runs its
Antarctic program and funds its science, on an international level
Antarctic affairs are matters for co-operation.  The governance of
Antarctica is through the Antarctic Treaty\footnote{See
  http://www.scar.org/treaty/at\_text.html.}, first signed by 12
nations in 1959, and now with 45 signatories attached to it.  Of
these, 28 nations are consultative members, which means that they have
voting rights, a status attained through the conduct of sustained
scientific research in Antarctica.  29 nations operate field stations,
19 of which run the 36 year-round stations.  The Treaty bypasses
issues of sovereignty by not recognising any territorial claims made
while the Treaty is in force.  It states that Antarctica is to be used
for peaceful purposes only, with scientific investigation and
collaboration specifically encouraged.  The Antarctic Treaty is in
fact a remarkably short and simple document comprising just 14
articles.  Seven nations have laid claim to parts of Antarctica
(Argentina, Australia, Britain, Chile, France, Norway and New
Zealand), with Argentina, Britain and Chile having overlapping claims.
The USA, on the other hand, has not made a claim, but has reserved the
right to do so.  The Treaty avoids any issue over these claims by
neither recognising them nor denying them.  Any country (or indeed
person) is free to go anywhere in Antarctica without restriction,
assuming they have the means to do so.

A measure of the Antarctic Treaty's success is that there remains a
large sector of West Antarctica that has not been claimed by any
nation.  The Treaty was further strengthened in 1991 by the adoption
of the Madrid Protocol.  This defines Antarctica as a natural reserve
devoted to peace and science, and specifically forbids any mineral
activity, except for research.  It also strengthened measures for the
conservation of flora and fauna, and set up standards for waste
management.

\subsection{SCAR and the IAU}
\label{sec:iau}
SCAR, the Scientific Committee for Antarctic Research, is the peak
body for Antarctic affairs. Established in 1957, as a committee of
ICSU, the International Council for Science, SCAR is charged with the
initiation, promotion and co-ordination of scientific research in
Antarctica. Within SCAR there are 31 Full Members (countries with
active research programs in Antarctica), 4 Associate Members
(countries without independent research programs in Antarctica) and 9
Union Members (those ICSU scientific unions with an interest in
Antarctica research).  The IAU (International Astronomical Union) is
one of these Union Members.

Operating alongside SCAR are a number of bodies that have been set up
by further international treaties in order to manage the conservation
of flora and fauna in and around Antarctica, including particularly
that in the Southern Ocean.  SCAR holds an open science congress every
two years, which also serves as the major forum bringing together the
managers of national Antarctic programs.  Through a motion put forward
at its $22^{nd}$ congress in Rome in 1994, SCAR passed a resolution
recognising the scientific value of Antarctic astronomy and calling
for the development of the field.

Like the IAU, SCAR has numerous divisions within it. There are three
Standing Scientific Groups (SSGs), covering Geo Sciences, Life
Sciences and Physical Sciences.  There are also five Scientific
Research Programs (SRPs), which have a focus on international
scientific co-ordination. In 2010 Astronomy became one of these
programs -- {\it Astronomy and Astrophysics from Antarctica}
(AAA)\footnote{See www.phys.unsw.edu.au/jacara/AAA\_SRP\_webpage/. }
-- replacing the ICESTAR program (involving solar-terrestrial \&
aeronomy research).  John Storey, of the University of New South
Wales, has been appointed the first Chief Officer of AAA, which has
set for itself several goals (see Storey, 2010).  These involve the
coordination of site testing activities (including related activities
in the Arctic) and the provision of publically accessible data bases
for their data, as well as improving cooperation with other physical
sciences.  They also look to future development: defining and
prioritising science goals, creating a roadmap for developing future
astronomical facilities and stimulating international cooperation to
construct and use those facilities.  To undertake this work AAA has
been structured into four themes, covering site testing, the Arctic,
science goals and major facilities.

The International Astronomical Union also supports the development of
the Astronomy in Antarctic.  At the IAU's $21^{st}$ General Assembly,
in Buenos Aires in 1991, a working group {\it Encouraging the
  International Development of Antarctic Astronomy}\footnote{See
  www.phys.unsw.edu.au/jacara/iau.} was formed, with Peter Gillingham
of the Anglo Australian Observatory as its first chair.  The author is
the current chair.  The IAU passed a resolution at that General
Assembly encouraging the development of the then fledgling field.  The
Working Group is under the auspices of two IAU Divisions (Division IX
on Optical \& Infrared Techniques and Division X on Radio Astronomy).
There have been regular sessions on Antarctic astronomy at subsequent
IAU General Assemblies (see Gillingham (ed) 1992, Burton (ed) 2005b,
2007, 2010b), as well as several international conferences and
workshops devoted to aspects of the field.

\subsection{Funding for Antarctic Astronomy}
SCAR and the IAU serve principally as fora where scientists can come
together to discuss progress and plans for their fields of endeavour.
They do not fund Antarctic science, however, for that is the province
of nations, acting either individually or collaboratively.

The funding process itself, for deciding upon and supporting Antarctic
science, differs considerably between nations.  These have evolved
according to the Antarctic traditions and histories within individual
countries, to differing national priorities, including whether a
nation has made territorial claim to the continent, and to the
different cultures surrounding science funding.  In some countries
Antarctic science is confined mainly within their national Antarctic
agency, with both the logistics and science conducted by that agency.
In other countries, science funding is separated from the logistics of
operating the bases.  More commonly, a mix of the two exists (see
Burton, 2005).

While the peer review process might be followed for determining on
science priorities, this is often contained within pre-defined
government goals for Antarctica.  These can serve to support current
activities but make it difficult for new ones to emerge.  This has
created difficulties for furthering astronomy in Antarctica in many
nations, for astronomy is not a traditional Antarctic science. It can
be excluded from a national Antarctic program by falling outside those
goals, and so need to seek support from other funding agencies.  These
agencies are, quite often, resistant to funding research in Antarctica
as they argue that it should be funded by the national Antarctic
program.  With no agency clearly responsible for deciding upon funding
for astronomical ventures in Antarctica, it can be difficult to
develop astronomical programs in Antarctica beyond relatively low-cost
site testing activities, despite the well-recognised strength of many
of the science cases that have been made.  Those nations that have
managed to resile this conflict, between supplying the logistics and
deciding upon the science, have managed to do the best Antarctic
science, regardless of the scientific discipline.  No where is this
success more apparent than in the exotic range of science now evident
at the South Pole, nurtured slowly over three decades of planning and
pursuit.

Despite these differences between funding policies of nations,
collaborative activity forged between scientists can be a spontaneous
event, with national boundaries playing little role.  One of the
delights of working as an Antarctic scientist is the relative ease
with which international collaborations can be made.  If you can
devise an experiment that will be advantageous to conduct at the
station of another country, that country can generally be persuaded to
support you.  Of course, it needs to be logistically possible to
set-up and operate the experiment.  There also need to be scientists
from the host country supporting the project, and it will have to pass
through an assessment process to determine its relative standing among
other projects.  But once these steps are taken, the nation running
the base will take it there, and provide the logistics to install the
experiment, all without charge.  It is up to the experimenter to find
the funds and personnel to build the experiment in the first place,
but most of the operations costs become subsumed in those of the
station.  The science, of course, needs to be shared among all the
partners involved.

\section{Antarctic Astronomy Tomorrow}
\label{sec:tomorrow}
\subsection{Future Science}
Extensive science cases have been written for astronomy in Antarctica,
a diverse range of facilities proposed, funding sought for some of
them and roadmaps prepared for future developments.  Yet any words
written today, about what astronomy will be done in Antarctica
tomorrow, will almost certainly contain more speculation than fact.
This section therefore will only briefly discuss the many ideas that
are current, and try to draw together common themes that indicate
likely future directions.

Some areas of science have been well served in Antarctica, with potent
science cases leading to funded facilities, following staged
development programs which will continue over the coming decade.  This
clearly is the case with the fields of CMBR and neutrino astrophysics.
Both are characterised by a series of increasingly sensitive
experiments, though for the former these have yielded detections of
subtle effects while for the latter non-detections.  Yet both fields
have essentially followed the path expected of them by the
quantitative predictions underpinning their science.

Traditional photon astronomy has, however, not been so well served by
Antarctic astronomy, with only modest-sized facilities built so far
for observations in the infrared and sub-millimetre.  We therefore
consider here some of the science and facilities envisaged for the
thermal-IR and THz portions of these spectral regimes.

\paragraph{Infrared Astronomy} 
A series of science cases have been developed for Antarctic
facilities, focussed largely on the prospects for infrared astronomy
given the greater sensitivity of an Antarctic telescope over an
equivalent-sized mid-latitude facility (e.g.\ Burton et al.\ 1994,
Burton et al.\ 2001, Burton et al.\ 2005, Lawrence et al.\ 2009a,
2009b, 2009c).  These science cases have concentrated on
intermediate-scale facilities (i.e.\ 2\,m-class telescopes), perceived
as the most likely size for the next Antarctic IR telescope.  The
capabilities of IR facilities elsewhere, both on Earth and in space,
have improved dramatically over this period. So too has the science
focus for prospective Antarctic facilities of this scale changed as
this has occurred.  It has narrowed into the parameter space where
modest Antarctic facilities may still compete with the best facilities
available elsewhere -- that science which requires wide-field, high
spatial resolution imaging in the near-infrared and
(spectroscopically) the mid-infrared.  This science is still
impressive, however, even for a modest-size telescope.  It includes
such projects as searching for the first light in the Universe via
pair-instability supernovae and from IR-only emitting gamma-ray
bursts, as well as deep 2.4$\mu$m imaging surveys looking for the
first galaxies to evolve in the universe.  Exo-planet searches would
also be well facilitated, carried out both through the transit
technique and through micro-lensing. It would also be possible to
image the warm molecular gas of the Galaxy directly, on the arcsecond
scale, as opposed to the arcminute scale available in the millimetre,
by measuring the 17$\mu$m line of molecular hydrogen, the ortho-ground
state transition of the molecule which is sensitive to this
environment.  No Antarctic telescope has been funded yet that could
undertake any of these investigations, though several have been
proposed as we discuss further in \S~\ref{sec:future} (e.g.\ AIRO at
the South Pole, PILOT/PLT at Dome C and KDUST at Dome A).

\paragraph{Terahertz Astronomy}
\label{sec:thz}
Even at the best temperate latitude sites such as the Chajnantor
plateau in Chile there is only occasional access to the THz windows.
Only from Antarctica could substantive THz surveys be considered,
other than from space.  Two of the most important cooling lines in the
dense interstellar medium emit in the THz regime; [CII] at 1.9~THz
(158$\mu$m) and [NII] at 1.5~THz (205$\mu$m).  These lines provide
tools to probe the formation and life-cycle of interstellar clouds.  A
THz telescope has been funded for Antarctica, the 0.8\,m Stratospheric
THz Observatory (STO; Walker et al.\ 2008).  This is to be flown
from a long duration balloon scheduled for launch from McMurdo in the
2010/11 austral summer. STO aims to survey $\sim 30^{\circ}$ of the
Galactic plane for emission from these species, with a spatial scale
of $1'$. Further ahead, there are plans for much larger THz facilities
at Domes A and F, capable of extending observations to extra-galactic
sources.  The estimates made for the THz transparency (Yang et
al.\ 2010) at Dome A suggest that the [NII] line could be measured
there, with $> 28$\% transmission, for a quarter of the time.  At
Ridge A these same estimates suggest that even the [CII] line will be
accessible, for perhaps 10\% of the time.

\subsection{Future Astronomical Facilities in Antarctica}
\label{sec:future}
In this section we discuss briefly discuss some of the plans that have
been formulated for new facilities on the Antarctic plateau.

\paragraph{South Pole}
The IceCube neutrino telescope dominates the logistics of supporting
the South Pole station. Completing this facility so that it can go
into full operation is a priority for the next few years.  The
existing CMBR facilities at Pole will continue to operate, building
their data sets over several years so as to be able to address their
science goals, as outlined in \S\ref{sec:cmbr}.  This includes the
QUaD experiment on DASI (E--mode polarization mapping), BICEP (B--mode
detection) and the SPT (SZ-effect in distant galaxy clusters).  Once
this last survey is completed, the plans for the SPT are to use it for
conventional sub-millimetre astronomy, building upon programs started
with the AST/RO telescope. SPT will be capable of undertaking surveys
at the highest frequencies reached by ALMA, as well as conducting THz
observations when conditions permit.  While a 2\,m class infrared
telescope was once proposed for the South Pole (AIRO; Jackson 2001),
this was before the first measurements of seeing were obtained from
Dome C.  An infrared telescope is not currently on the agenda for the
South Pole. On the other hand, a 2\,m optical/UV telescope, ACWI --
the Antarctic Cosmic Web Imager -- is being developed.  This will be
used to search for red-shifted UV resonant line emission (Ly$\alpha$
121.6nm, CIV 155nm \& OVI 103nm; Moore et al.\ 2008b) from the
intergalactic medium at $z \sim 2-3$, in order to study large-scale
structure and the dark matter distribution (i.e.\ the Cosmic Web).
The technical driver is the extremely accurate sky subtraction that
should be attainable due to the combination of low extinction
(resulting from the low aerosols) and long duration observations at
constant zenith angle, so minimising instrumental systematics (high
image quality is not required for this application).

\paragraph{Dome C}
Astronomical activity at Dome C to date has consisted of site testing
and several small-scale prototype facilities, that will culminate with
the operation of the 80\,cm mid-infrared IRAIT telescope.
Consideration of future plans for facilities at Concordia station has
been conducted under the auspices of the European Union-funded ARENA
consortium, a network program run from 2006-09. Each of the three
yearly conferences resulted in a book summarising the deliberations
over the previous year, evolving as knowledge of the site conditions
and technical challenges grew (Epchtein \& Candidi 2007, Zinnecker,
Epchtein \& Rauer 2008, Spinoglio \& Epchtein 2010). ARENA finished
its activities by producing a roadmap for future development at Dome C
(Epchtein et al.\ 2010). This was organised through 6 working groups,
who considered the prospects at Dome C for infrared, sub-millimetre,
interferometric, time-series, CMBR and solar science, respectively.
ARENA see the development of a 2\,m class, wide-field infrared
telescope as the next stage in the development of Dome C (see Burton
et al.\ 2010), such as that based on the design study conducted for
the PILOT / PLT telescope (see Storey et al.\ 2007, Saunders et al.\
2008a, 2008b).  Funding is being sought for a Phase B study for this
concept, before construction could start.

Dome C is clearly an excellent site for many photometric monitoring
experiments, for sub-millimetre and solar telescopes, for
CMBR measurements and for infrared interferometers.  The
ARENA roadmap considers a staged development of Concordia station to
support all these types of facilities.

\paragraph{Dome A}
Although the construction of Kunlun station at Dome A has only
recently begun, China has drawn up plans for several major
astronomical facilities, under the auspices of the Chinese Center for
Antarctic Astronomy (Gong et al.\ 2010, Cui 2010).  These include a
three telescope array of 0.5\,m Schmidt telescopes (AST3) to search
for transiting exo-planets and obtain light curves of supernovae, a
4\,m infrared telescope (KDUST; Zhao et al.\ 2010) to search for
distant supernovae and distant galaxies, and a 10\,m+ diameter sub-mm
/ THz telescope. A further automated observatory (PLATO--A), with
increased power capacity, is also planned for Dome A\@.

\paragraph{Dome F}
While site testing at Dome F has barely commenced, it is clear the
site will also be exceptional for observational astronomy.  Site
testing is envisaged to continue, making use of another PLATO-style
automated observatory, as at Dome A\@.  Two prototype projects are
under construction for future deployment there, a 40\,cm infrared and
a 30\,cm THz telescope (Ichikawa 2010).  Two facilities have been
proposed after this, an ultra light weight 2.5\,m infrared telescope
(Takato et al.\ 2008) and a 10\,m class THz telescope. The infrared
telescope would undertake deep surveys for galaxies at 2.4$\mu$m, and
the THz facility examine dusty galaxies to study galaxy evolution at
high redshift.

\paragraph{Ridge A}
As argued by Saunders et al.\ (2009) this may be the best site on the
planet for THz astronomy.  No stations are as yet envisaged for Ridge
A, however, nor are any overland traverses planned to the site.  Site
quantification may have to take place through Twin Otter deployment of
an automated observatory, such as using a modified, light-weighted
version of the PLATO module.

\subsection{Final Words}
Antarctica is a land of extremes.  Descriptors such as coldest,
driest, highest, windiest and calmest can all be applied to the
continent.  It is barely a century since the first explorers ventured
into the interior.  The Antarctic plateau has been found to
provide the pre-eminent conditions for making many kinds of
astronomical observations from the Earth.  Antarctica remains a
challenging place to work, but technology now provides for ready
access to the interior of the continent, as well as for sophisticated
scientific experiments to be carried out.

Human perceptions of the continent have not, however, caught up with
recent accomplishments in Antarctica.  While the `heroic age' of
Antarctic exploration is long over, it still stirs the imagination
when people think about Antarctica.  Paradoxically, this has also
limited our ability to fully exploit the unique conditions to conduct
front-line science, as much of our thinking is mired by
pre-conceptions gleaned from the survival tales of the heroic age.
This has limited the ability of many nations to grasp the new
opportunities now available.  Overcoming human perceptions about the
environment of the Antarctic plateau is proving to be at least as
difficult as developing the new front line facilities themselves.
With the start of the third millennium, and a growing knowledge and
awareness of Antarctica, such perceptions are waning.  The challenge
for the astronomer now is to work out how to grasp the opportunity, to
conduct science that might otherwise only be tackled from space.  The
Antarctic plateau provides sites where the ultimate Earth-based
telescopes may be built, if we can find the way to build on the
pioneering endeavours of the past two decades, and establish the
necessary infrastructure development that can lead to new astronomical
facilities.


%
%

\begin{acknowledgements}
  Astronomy in Antarctica has provided a peripatetic journey for me
  for nearly two decades, one with its full share of successes and
  failures, of amazing highs and deep lows, but always with a sense of
  discovery at the frontier.  A great many people have provided
  inspiration and support along the way, on a path that has now found
  me trying to tell the factual story behind a new field of scientific
  endeavour in this review article. David Allen, Peter Gillingham and
  Russell Cannon each encouraged my youthful enthusiasm in the infant
  field while I worked at the Anglo Australian Observatory and John
  Storey gave me the opportunity to take my first steps in Antarctica
  with a job at UNSW\@.  Together with Michael Ashley, the three of us
  stumbled through some early adventures at the South Pole.  Al Harper
  and John Bally, and later John Carlstrom, made possible our
  participation in the early years of the CARA venture there, and this
  gave us the springboard to take our program to the high plateau.
  Our group has waxed and waned over the years, dictated by the
  vagaries of university funding, but with Jon Lawrence's joining our
  program managed to wind its way forward, from the Pole to Dome C,
  then Dome A and now towards Dome F, with even Ridge A now in our
  sights.  All these colleagues have contributed immeasurably to
  Astronomy in Antarctica and many of the scientific highlights
  reported here.  So too have our many support staff, postdoctoral
  fellows and research students, at UNSW, the AAO and other places
  around Australia.  Several somehow managed to winter-over in
  Antarctica along the way, an opportunity I missed in my youth!  Marc
  Duldig has provided invaluable insight into the mysterious workings
  of the Antarctic establishment.  Our continually growing list of
  international colleagues, first in the USA, then France and Italy,
  extending around the European Union through the ARENA network, and
  now to China and Japan, are all contributing to the vitality of this
  field and the science that has been accomplished, as well as that
  which may one day be possible. The referee, Roland Gredel, also
  provided insightful comments and corrected several errors. But none
  of this would have happened for me without the support and
  understanding of my dear wife Connie, who could not have imagined
  what the past twenty years could possibly bring and where it would
  lead us to.
\end{acknowledgements}


\begin{thebibliography}{}
\bibitem{ack2005} Ackermann M {\em et al.} (123 authors) (2005) Search for
  extraterrestrial point sources of high energy neutrinos with
  AMANDA-II using data collected in 2000--2002. Phy. Rev. D 71:077102

\bibitem{aga2006} Agabi A, Aristidi E, Azouit M, Fossat E, Martin F,
  Sadibekova T, Vernin J, Ziad A (2006) First whole atmosphere
  nighttime seeing measurements at Dome C, Antarctica. PASP 118:344-348

\bibitem{aha2004} Aharonian F {\em et al.} (98 authors) (2004) Calibration of
  the cameras of the HESS detector. Astroparticle Physics 22:109-125

\bibitem{ahr2004} Ahrens J {\em et al.} (131 authors) (2004) Status of the
  IceCube neutrino observatory. New Astronomy Reviews 48:519-525

\bibitem{and2000} Andr\'{e}s E {\em et al.} (67 authors from the AMANDA
  Collaboration) (2000) The AMANDA neutrino telescope: principle of
  operation and first results. Astroparticle Physics 13:1-20

\bibitem{and2001} Andr\'{e}s E {\em et al.} (119 authors from the
  AMANDA Collaboration) (2001) Observation of high energy neutrinos
  using Cherenkov detectors embedded deep in Antarctic ice. Nature
  410:441-443

\bibitem{ari2003} Aristidi E, Agabi A, Vernin J, Azouit M, Martin F, Ziad A,
  Fossat E (2003) Antarctic site testing: first daytime seeing
  monitoring at Dome C. A\&A 406:L19-L22

\bibitem{ari2005a} Aristidi A, Agabi A, Fossat E, Azouit M, Martin F,
  Sadibekova T, Travouillon T, Vernin J, Ziad A (2005a) Site testing in
  summer at Dome C, Antarctica. A\&A 444:651-659

\bibitem{ari2005b} Aristidi A, Agabi K, Azouit M, Fossat E, Vernin J,
  Travouillon T, Lawrence JS, Meyer C, Storey JWV, Halter B, Roth WL,
  Walden V (2005b) An analysis of temperatures and wind speeds above
  Dome C, Antarctica. A\&A 430:739-746

\bibitem{ari2009} Aristidi E, Fossat E, Agabi A, M\'{e}karnia D, Jeanneaux E,
  Challita Z, Ziad A, Vernin J, Trinquet H (2009) Dome C site testing:
  surface layer, free atmospheric seeing and isoplanatic angle
  statistics. A\&A 499:955-965

\bibitem{ash2005} Ashley MCB, Burton MG, Calisse PG, Phillips A, Storey JWV
  (2005) Site testing at Dome C: cloud statistics from the ICECAM
  experiment. Highlights of Astronomy 13:932:934

\bibitem{ash1996} Ashley MCB, Burton MG, Storey JWV, Lloyd JP, Bally J,
  Briggs JW, Harper DA (1996) South Pole observations of the
  near-infrared sky brightness.  PASP 108:721-723

\bibitem{ask1965} Askaryan GA (1965) Coherent radio emission from cosmic
  showers in air and in dense media. Soviet Physics JETP 21:658

\bibitem{bar2003} Barwick SW {\em et al.} (18 authors) (2003) Overview of
  the ANITA project.  SPIE 4858:265-276

\bibitem{bay1923} Bayly PGW, Stillwell FL (1923) The Adelie Land meteorite.
  Scientific Reports Australasian Antarctic Expedition (1911-1914)
  Series A, Vol 4, Part 1 Geology

\bibitem{ber1999} Bernasconi PN, Rust DM, Murphy GA, Eaton HAC (1999) High
  resolution polarimetry with a balloon-borne telescope: the Flare
  Genesis experiment. In Rimmele TR, Balasubramaniam KS, Radick RR
  (eds) High resolution solar physics: theory, observations and
  techniques.  Ast. Soc. Pacific conf. ser. 183:279-287

\bibitem{bod1995} Bodhaine BA (1995) Aerosol absorption measurements at
  Barrow, Mauna Loa and the South Pole. Journal Geophysical Research
  100:8967-8976

\bibitem{bon2009} Bonner CS, Ashley MCB, Lawrence JS, Storey JWV, Luong-Van
  DM, Bradley SG (2009) SNODAR II: probing the atmospheric boundary
  layer on the Antarctic plateau. In Masciadri E, Sarazin M (eds)
  Optical turbulence: astronomy meets meteorology. Proc. Optical
  Turbulence Characterisation for Astronomical Applications, Sardinia,
  Italy, 15-18 September 2008

\bibitem{bon2010} Bonner CS, Ashley MCB, Cui X, Feng L, Gong X,
  Lawrence JS, Luong-Van DM, Storey JWV, Wang L, Yang H, Yang J, Zhou
  X, Zhu Z (2010) Height of the atmospheric boundary layer above Dome
  A, Antarctica during 2009. PASP submitted.

\bibitem{bro1988} Bromwich DH (1988) Snow in high southern latitudes.
  Reviews Geophysics 26:149-168

\bibitem{bro2000} Brooks KJ, Burton MG, Rathborne JM, Ashley, MCB, Storey
  JWV (2000) Unlocking the Keyhole: H$_2$ and PAH emission from
  molecular clumps in the Keyhole Nebula.  MNRAS 319:95-102

\bibitem{bur1986} Burova LP, Gromov VD, Luk'yanchikova NI,
  Sholomitskii GB (1986) Low humidity and submillimeter transparency
  above the Vostok Antarctic station. Soviet Astronomy Letters
  12:339-342

\bibitem{bur2005} Burton MG (2005) Astronomy in Antarctica. In Heck A (ed)
  Organisations and Strategies in Astronomy. Astrophysics \& Space
  Library, Kluwer 5:11-37

\bibitem{bur1994} Burton MG {\em et al.} (20 authors) (1994) The scientific
  potential for astronomy from the Antarctic plateau. PASA 11:127-150

\bibitem{buretal2005} Burton MG {\em et al.} (27 authors) (2005)
  Science programs for a 2-m class telescope at Dome C, Antarctica:
  PILOT, the Pathfinder for an International Large Optical Telescope.
  PASA 22:199-235

\bibitem{bur2010} Burton MG, Burgarella D, Andersen M, Busso M, Eiroa C,
  Epchtein N, Maillard J-P, Persi P (ARENA Working Group 1) (2010) A
  wide-field, optical/infrared, 2.5\,m class telescope for Antarctica.
  In Spinoglio L, Epchtein N (eds) 3$^{rd}$ ARENA Conference on an
  Astronomical Observatory at Concordia (Dome C, Antarctica). European
  Astronomical Soc. pub. ser.  40:125-135

\bibitem{bur1993} Burton MG, Allen DA, McGregor P (1993) The potential
  of near-infrared astronomy in Antarctica. Aust. Inst. Physics
  10$^{th}$ Congress (Feb 1992). ANARE Research Notes 88:293-300

\bibitem{bur2000} Burton MG, Ashley MCB, Marks RD, Schinckel AE, Storey JWV,
  Fowler A, Merrill M, Sharp N, Gatley I, Harper DA, Loewenstein RF,
  Mrozek F, Jackson JM Kraemer KE (2000) High resolution imaging of
  photodissociation regions in NGC 6334. ApJ 542:359-366

\bibitem{bur2001} Burton MG, Storey JWV, Ashley MCB (2001) Science goals for
  Antarctic infrared telescopes. PASA 18:158-165

\bibitem{bur2005b} Burton MG (ed) (2005b) Highlights of Astronomy.
  Issue 13, pp927-976.  IAU XXV General Assembly, Special
  Session 2: Astronomy in Antarctica, Sydney, July 18-19 2003. Series
  editor O Engvold. Ast. Soc. Pacific Conf. Series

\bibitem{bur2007} Burton MG (ed) (2007) Highlights of Astronomy. Issue
  14, Volume 2, pp683-712. IAU XXVI General Assembly, Special Session
  7: Astronomy in Antarctica, Prague, August 22-23 2006. Series editor
  KA van der Hulcht. Cambridge University Press

\bibitem{bur2010b} Burton MG (ed) (2010b) Highlights of Astronomy.
  Issue 15, in press.  IAU XXVII General Assembly, Special
  Session 3: Astronomy in Antarctica, Rio de Janeiro, August 6-7 2009.
  Series editor I Corbett. Cambridge University Press

\bibitem{bus2005} Bussmann RS, Holzapfel WL, Kuo CL (2005) Millimeter
  wavelength brightness fluctuations of the atmosphere above the South
  Pole. ApJ 622:1343-1355

\bibitem{bus2010} Busso M {\em et al.} (10 authors) (2010) Science with the
  IRAIT telescope: the commissioning phase.  In Spinoglio L, Epchtein
  N (eds) 3$^{rd}$ ARENA Conference on an Astronomical Observatory at
  Concordia (Dome C, Antarctica). European Astronomical Soc. pub. ser.
  40:165-170

\bibitem{cal2004} Calisse P, Ashley MCB, Burton MG, Phillips MA, Storey JWV,
  Radford SJE, Peterson JB (2004) Sub-millimetre site testing at Dome
  C, Antarctica. PASA 21:256-263

\bibitem{car2010} Carlstrom JE {\em et al.} (40 authors) (2010) The 10 meter
  South Pole Telescope. PASP in press

\bibitem{cas2009} Castro PH {\em et al.} (32 authors) (2009) Cosmological
  parameters from the QUAD CMB polarization experiment. ApJ 701:857-864

\bibitem{cha2000} Chamberlain, MA, Ashley MCB, Burton MG, Phillips A Storey
  JWV (2000) Mid-infrared observing conditions at the South Pole. ApJ
  535:501-511

\bibitem{cha1997} Chamberlin RA, Lane AP, Stark AA (1997) The 492 GHz
  atmospheric opacity at the geographic South Pole.  ApJ 476:428-433

\bibitem{chi2010} Chiang HC {\em et al.} (30 authors) (2010) Measurement of
  cosmic microwave background polarization power spectra from two
  years of BICEP data. ApJ 711:1123-1140

\bibitem{cob1999} Coble K {\em et al.} (15 authors) (1999) Anisotropy
  in the cosmic microwave background at degree angular scales: Python
  V results. ApJ 519:L5-L8

\bibitem{cro2010} Crouzet N {\em et al.} (22 authors) (2010) ASTEP South: an
  Antarctic search for transiting exoplanets around the celestial
  South Pole. A\&A 511:A36

\bibitem{cui2010} Cui X (2010) CSTAR and future plans for Dome A.  Highlights
  of Astronomy 15: in press

\bibitem{cul2010} Culverhouse T {\em et al.} (31 authors) (2010) The QUaD
  galactic plane survey 1: maps and analysis of diffuse emission. ApJ
  submitted

\bibitem{dam2010} Dam\'{e} L, Andretta V and ARENA Solar Astrophysics working
  group members (2010) ARENA solar astrophysics working group
  reporting on Dome C exceptional potential for solar observations.
  In Spinoglio L, Epchtein N (eds) 3$^{rd}$ ARENA Conference on an
  Astronomical Observatory at Concordia (Dome C, Antarctica). European
  Astronomical Soc. pub. ser. 40:451-466

\bibitem{deber2000} de Bernardis P. {\em et al.} (36 authors) (2000) A flat
  Universe from high-resolution maps of the cosmic microwave
  background radiation. Nature 404:955-959

\bibitem{dem2005} Dempsey JT, Storey JWV, Phillips A (2005) Auroral
  contribution to sky brightness for optical astronomy on the
  Antarctic plateau. PASA 22:91-104

\bibitem{dev2009} Devlin M {\em et al.} (29 authors) (2009) Over half of the
  far-infrared background light comes from galaxies at $z \ge 1.2$.
  Nature 458:737-739

\bibitem{dic2000} Dickinson JE, Gill JR, Hart SP, Hill GC, Hinton JA,
  Lloyd-Evans J, Potter D, Pryke C, Rochester K, Schwarz R, Watson AA
  (2000) A new air-Cherenkov array at the South Pole. Nuclear
  Instruments and Methods in Physics Research A 440:114-123

\bibitem{dol2010} Dolci M {\em et al.} (19 authors) (2010) Status of the
  AMICA project: ready for Antarctic adventure.  In Spinoglio L,
  Epchtein N (eds) 3$^{rd}$ ARENA Conference on an Astronomical
  Observatory at Concordia (Dome C, Antarctica). European Astronomical
  Soc. pub. ser. 40:171-176

\bibitem{dul2002} Duldig M (2002) Cosmic ray physics and astronomy. In
  Marchant HJ, Lugg DJ \& Quilty PG (eds) Australian Antarctic
  Science: the first 50 years of ANARE. Published by Australian
  Antarctic Division pp43-71

\bibitem{epc2010} Epchtein N (ed) for the ARENA consortium (2010) A vision for
  European astronomy and astrophysics at the Antarctic station
  Concordia, Dome C. Prepared by Antarctic Research, a European
  Network for Astrophysics (ARENA) for EC-FP6 contract RICA 026150

\bibitem{epc2007} Epchtein N, Candidi M (eds) (2007) 1$^{st}$ ARENA
  conference on large astronomical infrastructure at Concordia,
  prospects and constraints for Antarctic optical/IR astronomy.
  European Astronomical Soc. pub. ser. Vol 40

\bibitem{fil2010} Filimonov K and the IceCube consortium (2010) IceCube
  neutrino observatory at the South Pole: recent results. Highlights
  of Astronomy 15: in press

\bibitem{fos2003} Fossat E, Candidi M (2003) The scientific outlook
  for astronomy and astrophysics research at the Concordia station.
  Mem. della Soc. Ast. Italiana Vol 2

\bibitem{fos2005} Fossat E (2005) The Concordia Station on the
  Antarctic plateau: the best site on Earth for the 21st century
  astronomers. J. Astrophysics \& Astronomy 26:349-357

\bibitem{fow1998} Fowler A {\em et al.} (11 authors) (1998) Abu/SPIREX: South
  Pole thermal IR experiment. SPIE 3354:1170-1178

\bibitem{gab1995} Gabriel A {\em et al.} (32 authors) (1995) Global
  oscillations at low frequency from the SOHO mission (GOLF). Solar
  Physics 162:61-99

\bibitem{gil1989} Gillingham PR (1989) Antarctic optical-infrared
    observatory, presented to a meeting on the future of Australian
    astronomy, organised by the Ast. Soc. Aust, Canberra, June 1989

\bibitem{gil1992} Gillingham P (ed) (1992) Highlights of Astronomy.
  Issue 9, pp577-602. IAU XXI General Assembly, Buenos Aires, July
  1991. Series editor J Bergeron. Kluwer

\bibitem{gil1993} Gillingham PR (1993) Antarctic astronomy: introduction and
  summary of international developments affecting Australia.
  Australian Institute Physics 10th Congress Melbourne (Feb. 1992),
  ANARE Research Notes 88:290-292

\bibitem{gon2010} Gong {\em et al.} (37 authors) (2010) Dome A site testing
  and future plans.  In Spinoglio L, Epchtein N (eds) 3$^{rd}$ ARENA
  Conference on an Astronomical Observatory at Concordia (Dome C,
  Antarctica).  European Astronomical Soc. pub. ser. 40:65-72

\bibitem{gor2009} Gorham PW {\em et al.} (43 authors) (2009) New limits on
  the ultra-high energy cosmic neutrino flux from the ANITA experiment.
  Phy. Rev. Lett. 103:051103

\bibitem{gor2010} Gorham PW {\em et al.} (37 authors) (2010) Observational
  constraints on the ultra-high energy cosmic neutrino flux from the
  second flight of the ANITA experiment.  Phy. Rev. Lett. submitted

\bibitem{gre1980} Grec G, Fossat E, Pomerantz M (1980) Solar oscillations:
  full disk observations from the geographic South Pole. Nature
  288:541-544

\bibitem{gre1983} Grec G, Fossat E, Pomerantz M (1983) Full-disk observations
  of solar oscillations from the geographic South Pole: latest
  results. Solar Physics 82:55-66

\bibitem{gre2010} Gredel R (2010) Site characterisation at Dome C: the ARENA
  work. In Spinoglio L, Epchtein N (eds) 3$^{rd}$ ARENA Conference on
  an Astronomical Observatory at Concordia (Dome C, Antarctica).
  European Astronomical Soc. pub. ser. 40:11-20

\bibitem{har1990} Harper DA (1990) Infrared astronomy in Antarctica.
  In Pomerantz, M (ed) Astrophysics in Antarctica. American Institute
  Physics conf. series 198:123-129

\bibitem{her1994} Hereld M (1994) SPIREX: near infrared astronomy from the
  South Pole. In McLean IS (ed) Infrared astronomy with arrays, the
  next generation. Astrophysics and Space Science Library 190:248-252

\bibitem{her1990} Hereld M, Rauscher BJ, Harper DA, Pernic RJ (1990) GRIM: a
  near-infrared grism spectrometer and imager. SPIE Instrumentation in
  Astronomy VII pp43-48

\bibitem{hid2000} Hidas MG, Burton MG, Chamberlain MA, Storey JWV (2000)
  Infrared and sub-millimetre observing conditions on the Antarctic
  plateau. Pub. Ast. Soc. Aust. 17:260-269

\bibitem{ich2000} Ichikawa T (2010) Future plans for astronomy at Dome Fuji.
  Highlights of Astronomy 15: in press

\bibitem{ind2005} Indermuehle BT, Burton MG, Maddison ST (2005) The history
  of astrophysics in Antarctica. PASA 2005:73-90

\bibitem{ish2010} Ishii S, Seta M, Nakai N, Nagai S, Miyagawa N, Yamauchi A,
  Motoyama H, Taguchi M (2010) Site testing at Dome Fuji for
  sub-millimeter and terahertz astronomy: 220\,GHz atmospheric
  transparency.  Polar Science 3:213-221

\bibitem{jac2001} Jackson JM, Gatley I, Bania TM, Tollestrup E, Dunham
  T (2001) The Antarctic Infrared Observatory AIRO. Bulletin American
  Astro. Soc. 33:1466

\bibitem{ken1994} Kennedy JR and the GONG Team (1994) GONG, a global network
  of automated solar telescopes. In Pyper DM, Angione RJ (eds) Optical
  Astronomy from the Earth and Moon. Ast. Soc. Pacific conf. ser.
  55:188-196

\bibitem{ken2001} Kenyon SJ, Gomez M (2001) A 3$\mu$m survey of the
  Chamaeleon I dark cloud. AJ 121:2673-2680

\bibitem{ken2006} Kenyon SL, Lawrence JS, Ashley MCB, Storey JWV, Tokovinin A,
  Fossat E (2006) Atmospheric scintillation at Dome C, Antarctica:
  implications for photometry and astrometry.  118:924-932

\bibitem{} Kenyon SL, Storey JWV (2006) A review of optical sky
  brightness and extinction at Dome C, Antarctica. PASP 118:489-502

\bibitem{kim2006} Kim S, Narayanan D (2006) [CI] 809 GHz imaging of the NGC
  6334 complex. PASJ 58:753-757

\bibitem{kov2002} Kovac JM, Leitch EM, Pryke C, Carlstrom JE, Halverson NW,
  Holzapfel WL (2002) Detection of polarization in the cosmic
  microwave background using DASI. Nature 420:772-787

\bibitem{kra2008} Kravchenko I {\em et al.} (14 authors) (2008) Status of the
  RICE experiment. In Caballero R et al. (eds) Proc. 30$^{th}$ Intl.
  Cosmic Ray Conf. Yucat\'{a}n Mexico 3:1229-1232

\bibitem{kul2005} Kulesa C, Hungerford AL, Walker CK, Zhang X, Lane AP
  (2005) Large scale CO and [CI] emission in the $\rho$ Ophiuchi
  molecular cloud. ApJ 625:194-209

\bibitem{kuo2004} Kuo CL {\em et al.} (14 authors) (2004) High-resolution
  observations of the cosmic microwave background power spectrum with
  ACBAR. ApJ 600:32-51

\bibitem{lan1998} Lane AP (1998) Submillimeter transmission at South Pole.
  In Novak G, Landsberg RH (eds) Astrophysics from Antarctica ASP
  Conf. Ser. 141:289-295

\bibitem{las2009} Lascaux F, Masciadri E, Hagelin S, Stoesz J (2009)
  Mesoscale optical turbulence simulations at Dome C. MNRAS
  398:1093-1104

\bibitem{las2010} Lascaux F, Masciadri E, Hagelin S (2010) Mesoscale optical
  turbulence simulations at Dome C: refinements. MNRAS 403:1714-1718

\bibitem{law2004a} Lawrence JS (2004) Infrared and sub-millimeter atmospheric
  characteristics of high Antarctic plateau sites. PASP 116:482-492

\bibitem{law2009} Lawrence JS, Ashley MCB, Hengst S, Luong-Van DM, Storey
  JWV, Yang H, Zhou X \& Zhu, Z (2009) The PLATO Dome A site-testing
  observatory: power generation and control systems. Rev. Sci. Inst.
  80:064501:1-10

\bibitem{law2005} Lawrence JS, Ashley MCB, Storey JWV (2005) A remote,
  autonomous laboratory for optical astronomy on the Antarctic
  plateau. Aust. Journal Electrical \& Electronic Engineering 2:1-12

\bibitem{law2004b} Lawrence JS, Ashley MCB, Tokovinin A, Travouillon T
  (2004) Exceptional astronomical seeing conditions above Dome C in
  Antarctica. Nature 431:278-281

\bibitem{law2009a} Lawrence JS {\em et al.} (43 authors) (2009a) The
  science case for PILOT I: summary and overview. PASA 26:379-396

\bibitem{law2009b} Lawrence JS {\em et al.} (13 authors) (2009b) The
  science case for PILOT II: the distant universe. PASA 26:397-414

\bibitem{law2009c} Lawrence JS {\em et al.} (23 authors) (2009c) The
  science case for PILOT III: the nearby universe. PASA 26:415-438

\bibitem{lei2002} Leitch EM, Kovac JM, Pryke C, Carlstrom JE, Halverson NW,
  Holzapfel WL, Dragovan M, Reddall B, Sandberg ES (2002) Measurement
  of polarization with the Degree Angular Scale Interferometer. Nature
  420:763-771

\bibitem{li2006} Li H, Griffin GS, Krejny M, Novak G, Loewenstein RF,
  Newcomb MG, Calisse PG, Chuss DT (2006) Results of SPARO 2003:
  mapping magnetic fields in giant molecular clouds.  ApJ 648:340-354

\bibitem{llo2002} Lloyd JP, Oppenheimer BR, Graham JR (2002) The potential
  of differential astrometric interferometry from the high Antarctic
  plateau.  Pub. Ast. Soc. Aust. 19:318-322

\bibitem{llo2004} Lloyd JP, Lane BF, Swain MR, Storey JWV, Travouillon T,
  Traub WA, Walker CK (2004) Extrasolar planet science with the
  Antarctic planet interferometer. SPIE 5170:193-199

\bibitem{loe1998} Loewenstein RF, Bero C, Lloyd JP, Mrozek F, Bally J, Theil
  D (1998) Astronomical seeing at the South Pole. In Novak G,
  Landsberg R (eds) Astrophysics from Antarctica. ASP Conf. Series
  141:296-302

\bibitem{lue2010} Lueker M. {\em et al.} (43 authors) (2010) Measurements of
  secondary cosmic microwave background anisotropies with the South
  Pole Telescope. ApJ submitted

\bibitem{lyo2003} Lyo AR, Lawson WA, Mamajek EE, Feigelson ED, Sung EO,
  Crause LA (2003) Infrared study of the $\eta$ Chamaeleontis cluster
  and the longevity of circumstellar disks. MNRAS 338:616-622

\bibitem{mae2005} Maercker M, Burton MG (2005) L--band (3.5$\mu$m) IR-excess
  in massive star formation. I: 30 Doradus. A\&A 438:663-673

\bibitem{mae2006} Maercker M, Burton MG, Wright CM (2006) L--band (3.5$\mu$m)
  IR-excess in massive star formation. II: RCW57/NGC3576. A\&A
  450:253-263

\bibitem{mar2002} Marks RD (2002) Astronomical seeing from the summits of
  the Antarctic plateau. A\&A 385:328-336

\bibitem{mar2005} Marks RD (2005) Antarctic site testing: measurement of
  optical seeing at the South Pole. PhD Thesis, University of New
  South Wales

\bibitem{mar1996} Marks RD, Vernin J, Azouit M, Briggs JW, Burton MG,
  Ashley MCB, Manigault JF (1996) Antarctic site testing:
  microthermal measurements of surface-layer seeing at the South Pole.
  118: 385-390

\bibitem{mar1999} Marks RD, Vernin J, Azouit M, Manigault JF, Clevelin C
  (1999) Measurement of optical seeing on the high Antarctic plateau.
  A\&A Supp. 134:161-172

\bibitem{mar2008} Marsden G {\em et al.} (38 authors) (2008) The
  Balloon-borne Large-Aperture Submillimeter Telescope for
  polarization: BLAST-pol. (2008) SPIE 7020:702002:1-12

\bibitem{mar2004} Martin CL, Walsh WM, Xiao K, Lane AP, Walker CK Stark AA
  (2004) The AST/RO survey of the Galactic Centre region I: the inner
  3 degrees. ApJS 150:239-262

\bibitem{mas2006} Masi S {\em et al.} (44 authors) (2006) Instrument, method,
  brightness and polarization maps from the 2003 flight of BOOMERanG.
  A\&A 458:687-716

\bibitem{mcc1962} McCracken KG (1962) The cosmic-ray flare effect 3:
  deductions regarding the interplanetary magnetic field. Journal
  Geophysical Research 67:447-458

\bibitem{mcg2008} McGrath A, Saunders W, Gillingham P, Ward D, Storey J,
  Lawrence J, Haynes R (2008) Running PILOT: operational challenges
  and plans for an Antarctic observatory. SPIE 7016:70160G:1-12

\bibitem{mck1996} McKay DS, Gibson EK, Thomas-Keprta KL, Vali H, Romanek C,
  Clemmett SJ, Chillier XDF, Maechling CR, Zare RN (1996) Search for
  past life on Mars: possible relic biogenic activity in Martian
  meteorite ALH84001. Science 273:924-930

\bibitem{moo2008a} Moore A {\em et al.} (49 authors) (2008a) Gattini: a
  multi-site campaign for the measurement of sky brightness in
  Antarctica. SPIE 7012:701226:1-10

\bibitem{moo2008b} Moore AM, Martin C, Maitless NC, Travouillon T
  (2008b) ACWI: an experiment to image the Cosmic Web from Antarctica.
  SPIE 7012:70122A:1-11

\bibitem{mor1990} Morse R, Gaidos J (1990) A South Pole facility to
  observe very high energy gamma ray sources. In Pomerantz, M (ed)
  Astrophysics in Antarctica.  American Inst. Physics conf. series
  198:24-34

\bibitem{mos2007} Mosser B, Aristidi E (2007) Duty cycle of Doppler
  ground-based asteroseismic observations. 119:127-133

\bibitem{mot2010} Motizuki Y, Takahashi K, Makishima K, Bamba A, Nakai Y,
  Yano Y, Igarashi M, Motoyama H, Kamiyama K, Suzuki K, Imamura, T
  (2010) An Antarctic ice core recording of both supernovae and solar
  cycles. Nature submitted

\bibitem{nag1975} Nagata T (ed) (1975) Yamato meteorites collected in
  Antarctic in 1969. Memoirs National Polar Institute Polar Research,
  Special Issue 5

\bibitem{net2009} Netterfield CB {\em et al.} (27 authors) (2009) BLAST: the
  mass function, lifetimes and properties of intermediate mass cores
  from a 50 square degree sub-millimetre galactic plane survey in Vela
  $l \sim 265^{\circ}$. ApJ 707:1824-1835

\bibitem{ngu1996} Nguyen HT, Rauscher BJ Severson SA, Hereld M, Harper DA,
  Loewenstein RF, Mrozek F, Pernic RJ (1996) The South Pole
  near-infrared sky brightness. PASP 108:718-720

\bibitem{nov1998} Novak G, Landsberg RH (eds) (1998) Astrophysics
  from Antarctica. Astron. Soc. Pacific Conf. Series 141

\bibitem{nov2003} Novak G, Chuss DT, Renbarger T, Griffin GS, Newcomb MG,
  Peterson JB, Loewenstein RF, Pernic D, Dotson JL (2003) First
  results from the Sub-millimeter Polarimeter for Antarctic Remote
  Observations (SPARO): evidence of large-scale toroidal magnetic
  fields in the Galactic Center. ApJL 583:L83-L86

\bibitem{obe2006} Oberst TE, Parshley SC, Stacey GJ, Nikola T, L\"ohr A,
  Harnett JI, Tothill NFH, Lane AP, Stark AA, Tucker CE (2006)
  Detection of the 205$\mu$m [NII] line from the Carina Nebula. ApJL
  652:L125-L128

\bibitem{olm2009} Olmi L {\em et al.} (37 authors) (2009) The BLAST survey of
  the Vela Molecular Cloud: physical properties of the dense cores in
  Vela-D. ApJ 707:1836-1851

\bibitem{par1957} Parsons NR (1957) Directional measurements of the daily
  variation of cosmic ray meson intensity at $\lambda = 73^{\circ}$S.
  Aust. J. Phys. 10:462-470

\bibitem{pas2008} Pascale E {\em et al.} (30 authors) (2008) The
  Balloon-borne Large Aperture Submillimeter Telescope: BLAST. ApJ
  681:400-414

\bibitem{phi1999} Phillips A, Burton MG, Ashley MCB, Storey JWV, Lloyd JP,
  Harper DA, Bally J (1999) The near-infrared sky emission at the
  South Pole in winter. ApJ 527:1009-1022

\bibitem{pet2000} Peterson JB, Griffith GS, Newcomb MG, Alvarez DL, Cantalupo
  CM, Morgan D, Miller KW, Ganga K, Pernic D, Thoma M (2000) First
  results from Viper: detection of small-scale anisotropy at 40\,GHz.
  ApJ 532:L83-L86

\bibitem{pet2003} Peterson JB, Radford SJE, Ade PAR, Chamberlin RA, O'Kelly
  MJ, Peterson KM, Schartman E (2003) Stability of the sub-millimeter
  brightness of the atmosphere above Mauna Kea, Chajnantor and the
  South Pole. PASP 115:383-388

\bibitem{pla2010} Plagge T {\em et al.} (45 authors) (2010)
  Sunyaev-Zel'dovich cluster profiles measured with the South Pole
  Telescope. ApJ submitted

\bibitem{pom1958} Pomerantz MA, Agarwal SP, Pontis VR (1958) Direct
  observation of periodic variation of primary cosmic-ray intensity.
  Phys. Rev. 109:224-225

\bibitem{pom1990} Pomerantz M (ed) (1990) Astrophysics in Antarctica.
  American Inst. Physics conf. series 198.

\bibitem{rat2005} Rathborne JM, Burton MG (2005) Results from the South Pole
  Infra-Red EXplorer Telescope. Highlights of Astronomy 13:937-944

\bibitem{rat2002} Rathborne JM, Burton MG, Brooks KJ, Cohen M, Ashley MCB,
  Storey JWV (2002) Photodissociation regions and star formation in
  the Carina Nebula. MNRAS 331:85-97

\bibitem{sau2008a} Saunders W, Gillingham P, McGrath A, Haynes R, Brzeski J,
  Storey J, Lawrence J (2008a) PILOT: a wide-field telescope for the
  Antarctic plateau. SPIE 7012:70124F:1-9

\bibitem{sau2008b} Saunders W, Gillingham P, McGrath A, Haynes R, Storey J,
  Lawrence J, Burton M, Jenkins C, Alcione M (2008b) Proposed
  instrumentation for PILOT. SPIE 7014:70144N:1-9

\bibitem{sau2009} Saunders W, Lawrence JS, Storey JWV, Ashley MCB, Kato S,
  Minnis P, Winker DM, Liu G, Kulesa C (2009) Where is the best site
  on Earth?  Domes A, B, C and F, and Ridges A and B. PASP 121:976-992

\bibitem{sev2000} Severson S (2000) Death of a comet: SPIREX observations
  of the collision of SL9 with Jupiter. PhD Dissertation, University of
  Chicago

\bibitem{sie2010} Siebenmorgen R (2010) Thermal infrared instruments for
  Antarctica: what can be gained.  In Spinoglio L, Epchtein N (eds)
  3$^{rd}$ ARENA Conference on an Astronomical Observatory at
  Concordia (Dome C, Antarctica).  European Astronomical Soc. pub.
  ser. 40:147-155

\bibitem{smi1998} Smith CH, Harper DA (1998) Mid-infrared sky brightness
  site testing at the South Pole. PASP 110:747-753

\bibitem{smi1989} Smith NJT, Gaisser TK, Hillas AM, Ogden PA, Patel M,
  Perrett JC, Pomerantz MA, Reid RJO, Stanev T, Watson AA (1989) The
  South Pole Air Shower Experiment. In Stepanian AA, Fegan DJ, Cawley
  MF (eds) Very High Energy Gamma Ray Astronomy. Proc. Workshop
  Crimea, 17-21 April 1988 pp55

\bibitem{smy1977} Smythe WD, Jackson BV (1977) Atmospheric water vapor
  at South Pole. Applied Optics 16:2041-2042

\bibitem{spe2003} Spergel DN {\em et al.} (17 authors) (2003)
  First-year Wilkinson Microwave Anisotropy Probe (WMAP) observations:
  determination of cosmological parameters. ApJS 148:175-194

\bibitem{spi2010} Spinoglio L, Epchtein N (eds) (2010) 3$^{rd}$ ARENA
  conference on an astronomical observatory at Concordia (Dome C,
  Antarctica).  European Astronomical Soc. pub. ser. Vol 40

\bibitem{sta2009} Staniszewski Z {\em et al.} (44 authors) (2009) Galaxy
  clusters discovered with a Sunyaev-Zel'dovich effect survey. ApJ
  701:32-41

\bibitem{sta1997} Stark AA, Bolatto AD, Chamberlin RA, Lane AP, Bania TM,
  Jackson JM, Lo KY (1997) First detection of 492 GHz [CI] emission
  from the Large Magellanic Cloud. ApJL 480:L59-L62

\bibitem{sta2001} Stark AA {\em et al.} (27 authors) (2001) The Antarctic
  Sub-millimeter Telescope and Remote Observatory (AST/RO). PASP
  113:567-585

\bibitem{sto2004} Storey JWV (2004) Antarctica: the potential for
  interferometry. SPIE 5491:169-175

\bibitem{sto2009} Storey JWV (2009) Astronomy and Astrophysics from
  Antarctica. Association Asia Pacific Physical Societies Bulletin.
  19:4-10

\bibitem{sto2010} Storey JWV (2010) Astronomy and Astrophysics from
  Antarctica: a new SCAR Scientific Research Program. Highlights of
  Astronomy 15: in press

\bibitem{sto1996} Storey JWV, Ashley MCB, Burton MG (1996) An automated
  astrophysical observatory for Antarctica. PASA 13:35-38

\bibitem{sto2007} Storey JWV, Ashley MCB, Burton MG, Lawrence JS
  (2007) PILOT -- the pathfinder for an international large optical
  telescope. In Epchtein N, Candidi M (eds) 1$^{st}$ ARENA conference
  on large astronomical infrastructures at Concordia, prospects and
  constraints for Antarctic optical/IR astronomy.  European
  Astronomical Soc. pub.  ser. 25:255-259

\bibitem{sto2003} Storey JWV, Ashley MCB, Lawrence JS, Burton MG (2003) Dome
  C--the best astronomical site in the world? Mem. S. A. It. Supp.
  2:13-18

\bibitem{sto2005} Storey JWV (2005) Astronomy from Antarctica.
  Antarctic Science 17:555-560

\bibitem{str2008} Strassmeier K {\em et al.} (14 authors) (2008) First
  time-series optical photometry from Antarctica. sIRAIT monitoring of
  the RS CVn binary V841 Centauri and the $\delta$-Scuti star V1034
  Centauri.  A\&A 490:287-295

\bibitem{tak2010} Takahashi Y {\em et al.} (26 authors) (2010)
  Characterisation of the BICEP telescope for high-precision cosmic
  microwave background polarimetry. ApJ 711:1141-1156

\bibitem{tak2008} Takato N, Ichikawa T, Uraguchi F, Lundock R, Murata
  C, Taniguchi Y, Motoyama H, Fukui K, Taguchi M (2008) A 2\,m class
  telescope at Dome Fuji. In Zinnecker H, Epchtein N, Rauer H (eds)
  2$^{nd}$ ARENA conference on the astrophysical science cases at Dome
  C. European Astronomical Soc. pub. ser. 33:271-274

\bibitem{tay1990} Taylor MJ (1990) Photometry of the 4686\AA\ emission
  line of Gamma 2 Velorum from the South Pole. AJ 100:1264-1269

\bibitem{tom2008} Tomasi C, Petkov B, Benedetti E, Valenziano L, Lupi
  A, Vitale V, Bonaf\'{e} U (2008) A refined calibration procedure of
  two-channel sun photometers to measure atmospheric precipitable water
  at various Antarctic sites. J. Atmospheric \& Oceanic Technology,
  25:213-229

\bibitem{tow1990} Townes GH, Melnick G (1990) Atmospheric transmission
  in the far-infrared at the South Pole and astronomical applications.
  PASP 102:357-367.

\bibitem{tra2004} Travouillon T (2004) Measurements of optical turbulence on
  the Antarctic plateau and their impact on astronomical observations.
  PhD Thesis, University of New South Wales

\bibitem{tra2003a} Travouillon T, Ashley MCB, Burton MG, Storey JWV,
  Loewenstein RF (2003a) Atmospheric turbulence at the South Pole and
  its implications for astronomy. A\&A 400:1163-1172

\bibitem{tra2003b} Travouillon T, Ashley MCB, Burton MG, Lawrence J, Storey
  JWV (2003b) Low atmospheric turbulence at Dome C: preliminary
  results. Mem. Soc. Astron. Italy 2:150-153

\bibitem{tri2008} Trinquet H, Agabi A, Vernin J, Azouit M, Aristidi E, Fossat
  E (2008) Nighttime optical turbulence vertical structure above Dome
  C in Antarctica.  PASP 120:203-211

\bibitem{tuc1993} Tucker GS, Griffin GS, Nguyen HT, Peterson JB (1993) A
  search for small-scale anisotropy in the cosmic microwave
  background. ApJL 419:L45-L48

\bibitem{val1999} Valenziano L, dall'Oglio G (1999) Millimetre astronomy from
  the high Antarctic plateau: site testing at Dome C. PASA 16:167-174

\bibitem{vanste1993} van Stekelenborg J, Gaisser TK, Perrett JC, Petrakis JP,
  Stanev TS, Beaman J, Hillas AM, Johnson PA, Lloyd-Evans J, Smith
  NJT, Watson AA (1993) Search for point sources of ultra-high energy
  $\gamma$-rays in the southern hemisphere with the South Pole Air
  Shower Experiment. Phy. Rev D 48:4504-4517

\bibitem{ver2009} Vernin J, Chadid M, Aristidi E, Agabi A, Trinquet H, van
  der Swaelmen M (2009) First single star scidar measurement at Dome
  C, Antarctica. A\&A 500:1271-1276

\bibitem{wal2005} Walden VP, Town MS, Halter B, Storey JWV (2005) First
  measurements of the infrared sky brightness at Dome C, Antarctica.
  PASP 117:300-308

\bibitem{wal1777} Wales W, Bayly W (1777) The original astronomical
  observations made in the course of a voyage towards the South Pole
  and round the world in his majesty's ships the Resolution and the
  Adventure in the years 1772-1775. Publishers Board of Longitude.
  Printers WA Strahan, London.

\bibitem{wal2008} Walker CK {\em et al.} (24 authors) (2008) The Stratospheric
  Terahertz Observatory (STO): an LDB experiment to investigate the
  life cycle of the interstellar medium. In 19$^{th}$ Intl. Symp. on
  Space Terahertz Technology. Grongingen pp28-31

\bibitem{win1991} Windhorst RA {\em et al.} (10 authors) (1991) The
  discovery of a young radio galaxy at $z=2.390$ -- probing initial
  star formation at $z$ less than approximately 3.0. ApJ 380:362-383

\bibitem{yan2010} Yang H, Kulesa CA, Walker CK, Tothill NFH, Yang J, Ashley
  MCB, Cui X, Feng L, Lawrence JS, Luong-Van DM, Storey JWV, Wang L,
  Zhou X, Zhu Z (2010) Exceptional terahertz transparency and
  stability above Dome A, Antarctica. PASP 122:490-494

\bibitem{yan2009} Yang H {\em et al.} (37 authors) (2009) The PLATO Dome A
  site-testing observatory: instrumentation and first results. PASP
  121:174-184

\bibitem{zha2001} Zhang X, Lee Y, Bolatto AP, Stark AA (2001) CO 4--3 and
  [CI] observations of the Carina molecular cloud complex. ApJ
  553:274-287

\bibitem{zha2010} Zhao G--B, Zhan H, Wang L, Fan Z, Zhang X (2010)
  Probing dark energy with the Kunlun Dark Universe Survey Telescope.
  ApJ submitted

\bibitem{zho2010} Zhou X {\em et al.} (32 authors) (2010) The first release
  of the CSTAR point source catalog from Dome A, Antarctica. PASP 122:347-353

\bibitem{zia2008} Ziad A, Aristidi E, Agabi A, Borgino J, Martin F, Fossat E
  (2008) First statistics of the turbulence outer scale at Dome C.
  A\&A 491:917-921

\bibitem{zin2008} Zinnecker H, Epchtein N, Rauer H (eds) (2008) 2$^{nd}$
  ARENA conference on the astrophysical science cases at Dome C.
  European Astronomical Soc. pub. ser. Vol 33

\bibitem{zou2010} Zou H {\em et al.} (34 authors) (2010) The sky
  brightness and transparency in $i$--band at Dome A, Antarctica. AJ
  submitted


\end{thebibliography}

%
%

\end{document}